\input harvmac
%\draftmode
\def\ev#1{\langle#1\rangle}
\def\sign{{\rm sign}}

\input amssym
\input epsf.tex

\newcount\figno
\figno=0
\def\fig#1#2#3{
\par\begingroup\parindent=0pt\leftskip=1cm\rightskip=1cm\parindent=0pt
\baselineskip=11pt
\global\advance\figno by 1
\midinsert
\epsfxsize=#3
\centerline{\epsfbox{#2}}
\vskip 12pt
{\bf Fig. \the\figno:} #1\par
\endinsert\endgroup\par
}
\def\figlabel#1{\xdef#1{\the\figno}}
\def\encadremath#1{\vbox{\hrule\hbox{\vrule\kern8pt\vbox{\kern8pt
\hbox{$\displaystyle #1$}\kern8pt}
\kern8pt\vrule}\hrule}}

% FONTS

% fraktur
\newfam\frakfam
\font\teneufm=eufm10
\font\seveneufm=eufm7
\font\fiveeufm=eufm5
\textfont\frakfam=\teneufm
\scriptfont\frakfam=\seveneufm
\scriptscriptfont\frakfam=\fiveeufm

% black board bold

\def\bb{
\font\tenmsb=msbm10
\font\sevenmsb=msbm7
\font\fivemsb=msbm5
\textfont1=\tenmsb
\scriptfont1=\sevenmsb
\scriptscriptfont1=\fivemsb
}

%\newfam\msbfam
%\font\tenmsb=msbm10
%\font\sevenmsb=msbm7
%\font\fivemsb=msbm5
%\textfont\msbfam=\tenmsb
%\scriptfont\msbfam=\sevenmsb
%\scriptscriptfont\msbfam=\fivemsb
%\def\bb{\fam\msbfam \tenmsb}

% double stroke math

\newfam\dsromfam
\font\tendsrom=dsrom10
\textfont\dsromfam=\tendsrom
\def\ds{\fam\dsromfam \tendsrom}

% bold math italics

\newfam\mbffam
\font\tenmbf=cmmib10
\font\sevenmbf=cmmib7
\font\fivembf=cmmib5
\textfont\mbffam=\tenmbf
\scriptfont\mbffam=\sevenmbf
\scriptscriptfont\mbffam=\fivembf

% bold math cal

\newfam\mbfcalfam
\font\tenmbfcal=cmbsy10
\font\sevenmbfcal=cmbsy7
\font\fivembfcal=cmbsy5
\textfont\mbfcalfam=\tenmbfcal
\scriptfont\mbfcalfam=\sevenmbfcal
\scriptscriptfont\mbfcalfam=\fivembfcal

% math script

\newfam\mscrfam
\font\tenmscr=rsfs10
\font\sevenmscr=rsfs7
\font\fivemscr=rsfs5
\textfont\mscrfam=\tenmscr
\scriptfont\mscrfam=\sevenmscr
\scriptscriptfont\mscrfam=\fivemscr

% MACROS

% bras, kets, ...

% tilde, hat, bar, ...

\def\tilde{\widetilde}

\def\bar{\overline}
\def\b{\bar}
\def\bsq#1{{{\b{#1}}^{\lower 2.5pt\hbox{$\scriptstyle 2$}}}}
\def\bexp#1#2{{{\b{#1}}^{\lower 2.5pt\hbox{$\scriptstyle #2$}}}}
\def\dotexp#1#2{{{#1}^{\lower 2.5pt\hbox{$\scriptstyle #2$}}}}

% basic math

\def\rt2{\sqrt{2}}
\def\half {{1 \over 2}}

\def\det{\mathop{\rm det}}

\def\Tr{\mathop{\rm Tr}}

\def\sign{\mathop{\rm sign}}

% bold greek characters

\font\tenbifull=cmmib10
\font\tenbimed=cmmib7
\font\tenbismall=cmmib5
\textfont9=\tenbifull \scriptfont9=\tenbimed
\scriptscriptfont9=\tenbismall

\mathchardef\bbGamma="7000
\mathchardef\bbDelta="7001
\mathchardef\bbPhi="7002
\mathchardef\bbAlpha="7003
\mathchardef\bbXi="7004
\mathchardef\bbPi="7005
\mathchardef\bbSigma="7006
\mathchardef\bbUpsilon="7007
\mathchardef\bbTheta="7008
\mathchardef\bbPsi="7009
\mathchardef\bbOmega="700A
\mathchardef\bbalpha="710B
\mathchardef\bbbeta="710C
\mathchardef\bbgamma="710D
\mathchardef\bbdelta="710E
\mathchardef\bbepsilon="710F
\mathchardef\bbzeta="7110
\mathchardef\bbeta="7111
\mathchardef\bbtheta="7112
\mathchardef\bbiota="7113
\mathchardef\bbkappa="7114
\mathchardef\bblambda="7115
\mathchardef\bbmu="7116
\mathchardef\bbnu="7117
\mathchardef\bbxi="7118
\mathchardef\bbpi="7119
\mathchardef\bbrho="711A
\mathchardef\bbsigma="711B
\mathchardef\bbtau="711C
\mathchardef\bbupsilon="711D
\mathchardef\bbphi="711E
\mathchardef\bbchi="711F
\mathchardef\bbpsi="7120
\mathchardef\bbomega="7121
\mathchardef\bbvarepsilon="7122
\mathchardef\bbvartheta="7123
\mathchardef\bbvarpi="7124
\mathchardef\bbvarrho="7125
\mathchardef\bbvarsigma="7126
\mathchardef\bbvarphi="7127

% dotted spinor indices

% bared indices

% bared spinors

% capital cal letters

\def\CJ{{\cal J}}

\def\CL{{\cal L}}
\def\CM{{\cal M}}
\def\CN{{\cal N}}
\def\CO{{\cal O}}

\def\CQ{{\cal Q}}

% double stroke symbols: unit matrix, reals, complex, quaternions, integers, natural numbers

\def\1{{\ds 1}}
\def\T{\hbox{$\bb T$}}
\def\R{\hbox{$\bb R$}}
\def\C{\hbox{$\bb C$}}

\def\Z{\hbox{$\bb Z$}}

\def\S{\hbox{$\bb S$}}

% miscellaneous objects

\noblackbox

\def\unit{\relax{\rm 1\kern-.26em I}}
\def\nada{\relax{\rm 0\kern-.30em l}}
\def\tilde{\widetilde}

\def\det{{\rm det}}

%% MACROS
\noblackbox
\def\IL{\relax{\rm I\kern-.18em L}}
\def\IH{\relax{\rm I\kern-.18em H}}
\def\IR{\relax{\rm I\kern-.18em R}}
\def\IC{\relax\hbox{$\inbar\kern-.3em{\rm C}$}}
\def\IZ{\relax\ifmmode\mathchoice
{\hbox{\cmss Z\kern-.4em Z}}{\hbox{\cmss Z\kern-.4em Z}} {\lower.9pt\hbox{\cmsss Z\kern-.4em Z}}
{\lower1.2pt\hbox{\cmsss Z\kern-.4em Z}}\else{\cmss Z\kern-.4em Z}\fi}
\def\CM {{\cal M}}

\def\CN {{\cal N}}

\def\CJ {{\cal J}}
\def\partialslash{\not{\hbox{\kern-2pt $\partial$}}}

\def\CL {{\cal L}}

\def\CO {{\cal O}}

%% MORE MACROS
\def\CM {{\cal M}}
\def\CN {{\cal N}}

\def\CO {{\cal O}}

\def\CQ {{\cal Q }}

\def\det{{\rm det}}
\def\Tr{{\rm Tr}}

\font\manual=manfnt \def\dbend{\lower3.5pt\hbox{\manual\char127}}

\def\IZ{\relax\ifmmode\mathchoice
{\hbox{\cmss Z\kern-.4em Z}}{\hbox{\cmss Z\kern-.4em Z}} {\lower.9pt\hbox{\cmsss Z\kern-.4em Z}}
{\lower1.2pt\hbox{\cmsss Z\kern-.4em Z}}\else{\cmss Z\kern-.4em Z}\fi}
\def\half {{1\over 2}}

\def\lfm#1{\medskip\noindent\item{#1}}

\def\bar{\overline}

\def\rt2{\sqrt{2}}
\def\irt2{{1\over\sqrt{2}}}

%  \slashchar puts a slash through a character to represent contraction
%  with Dirac matrices. Use \not instead for negation of relations, and use
%  \hbar for hbar.
\def\slashchar#1{\setbox0=\hbox{$#1$}           % set a box for #1
   \dimen0=\wd0                                 % and get its size
   \setbox1=\hbox{/} \dimen1=\wd1               % get size of /
   \ifdim\dimen0>\dimen1                        % #1 is bigger
      \rlap{\hbox to \dimen0{\hfil/\hfil}}      % so center / in box
      #1                                        % and print #1
   \else                                        % / is bigger
      \rlap{\hbox to \dimen1{\hfil$#1$\hfil}}   % so center #1
      /                                         % and print /
   \fi}
\def\FI{{\zeta\over 2\pi}}
\def\foursqr#1#2{{\vcenter{\vbox{
    \hrule height.#2pt
    \hbox{\vrule width.#2pt height#1pt \kern#1pt
    \vrule width.#2pt}
    \hrule height.#2pt
    \hrule height.#2pt
    \hbox{\vrule width.#2pt height#1pt \kern#1pt
    \vrule width.#2pt}
    \hrule height.#2pt
        \hrule height.#2pt
    \hbox{\vrule width.#2pt height#1pt \kern#1pt
    \vrule width.#2pt}
    \hrule height.#2pt
        \hrule height.#2pt
    \hbox{\vrule width.#2pt height#1pt \kern#1pt
    \vrule width.#2pt}
    \hrule height.#2pt}}}}
\def\psqr#1#2{{\vcenter{\vbox{\hrule height.#2pt
    \hbox{\vrule width.#2pt height#1pt \kern#1pt
    \vrule width.#2pt}
    \hrule height.#2pt \hrule height.#2pt
    \hbox{\vrule width.#2pt height#1pt \kern#1pt
    \vrule width.#2pt}
    \hrule height.#2pt}}}}
\def\sqr#1#2{{\vcenter{\vbox{\hrule height.#2pt
    \hbox{\vrule width.#2pt height#1pt \kern#1pt
    \vrule width.#2pt}
    \hrule height.#2pt}}}}

\def\figin{\epsfcheck\figin}\def\figins{\epsfcheck\figins}
\def\epsfcheck{\ifx\epsfbox\UnDeFiNeD
\message{(NO epsf.tex, FIGURES WILL BE IGNORED)}
\gdef\figin##1{\vskip2in}\gdef\figins##1{\hskip.5in}% blank space instead
\else\message{(FIGURES WILL BE INCLUDED)}%
\gdef\figin##1{##1}\gdef\figins##1{##1}\fi}
\def\DefWarn#1{}
\def\figinsert{\goodbreak\midinsert}
\def\ifig#1#2#3{\DefWarn#1\xdef#1{fig.~\the\figno}
\writedef{#1\leftbracket fig.\noexpand~\the\figno}%
\figinsert\figin{\centerline{#3}}\medskip\centerline{\vbox{\baselineskip12pt \advance\hsize by
-1truein\noindent\footnotefont{\bf Fig.~\the\figno:\ } \it#2}}
\bigskip\endinsert\global\advance\figno by1}
%\WittenHF
\lref\WittenHF{
  E.~Witten,
  ``Quantum Field Theory and the Jones Polynomial,''
Commun.\ Math.\ Phys.\  {\bf 121}, 351 (1989).
%%CITATION = IASSNS-HEP-88-33%%
}
%\AharonyBX
\lref\AharonyBX{
  O.~Aharony, A.~Hanany, K.~A.~Intriligator, N.~Seiberg and M.~J.~Strassler,
  ``Aspects of N = 2 supersymmetric gauge theories in three dimensions,''
  Nucl.\ Phys.\  B {\bf 499}, 67 (1997)
  [arXiv:hep-th/9703110].
  %%CITATION = NUPHA,B499,67;%%
}

%\HitchinEA
\lref\HitchinEA{
  N.~J.~Hitchin, A.~Karlhede, U.~Lindstrom and M.~Rocek,
  ``Hyperkahler Metrics and Supersymmetry,''
Commun.\ Math.\ Phys.\  {\bf 108}, 535 (1987).
%%CITATION = STOCKHOLM-ITP-86-8%%
}

%\AganagicUW
\lref\AganagicUW{
  M.~Aganagic, K.~Hori, A.~Karch and D.~Tong,
  ``Mirror symmetry in (2+1)-dimensions and (1+1)-dimensions,''
JHEP {\bf 0107}, 022 (2001).
[hep-th/0105075].
%%CITATION = hep-th/0105075%%
}

%\AffleckAS
\lref\AffleckAS{
  I.~Affleck, J.~A.~Harvey and E.~Witten,
  ``Instantons and (Super)Symmetry Breaking in (2+1)-Dimensions,''
Nucl.\ Phys.\ B {\bf 206}, 413 (1982).
%%CITATION = PRINT-82-0478 (PRINCETON)%%
}
%\ClossetVG
\lref\ClossetVG{
  C.~Closset, T.~T.~Dumitrescu, G.~Festuccia, Z.~Komargodski and N.~Seiberg,
  ``Contact Terms, Unitarity, and F-Maximization in Three-Dimensional Superconformal Theories,''
JHEP {\bf 1210}, 053 (2012).
[arXiv:1205.4142 [hep-th]].
%%CITATION = arXiv:1205.4142%%
}
%\CalliasKG
\lref\CalliasKG{
  C.~Callias,
  ``Index Theorems on Open Spaces,''
Commun.\ Math.\ Phys.\  {\bf 62}, 213 (1978).
%%CITATION = MIT-CTP-694%%
}
%\NekrasovUH
\lref\NekrasovUH{
  N.~A.~Nekrasov and S.~L.~Shatashvili,
  ``Supersymmetric vacua and Bethe ansatz,''
Nucl.\ Phys.\ Proc.\ Suppl.\  {\bf 192-193}, 91 (2009).
[arXiv:0901.4744 [hep-th]].
%%CITATION = arXiv:0901.4744%%
}
%\IntriligatorEX
\lref\IntriligatorEX{
  K.~A.~Intriligator and N.~Seiberg,
  ``Mirror symmetry in three-dimensional gauge theories,''
Phys.\ Lett.\ B {\bf 387}, 513 (1996).
[hep-th/9607207].
%%CITATION = hep-th/9607207%%
}
%\GatesNR
\lref\GatesNR{
  S.~J.~Gates, M.~T.~Grisaru, M.~Rocek and W.~Siegel,
  ``Superspace Or One Thousand and One Lessons in Supersymmetry,''
Front.\ Phys.\  {\bf 58}, 1 (1983).
[hep-th/0108200].
%%CITATION = hep-th/0108200%%
}
%\PolyakovFU
\lref\PolyakovFU{
  A.~M.~Polyakov,
  ``Quark Confinement and Topology of Gauge Groups,''
Nucl.\ Phys.\ B {\bf 120}, 429 (1977).
%%CITATION = NORDITA-76/33%%
}
%\AganagicUW
\lref\AganagicUW{
  M.~Aganagic, K.~Hori, A.~Karch and D.~Tong,
  ``Mirror symmetry in (2+1)-dimensions and (1+1)-dimensions,''
JHEP {\bf 0107}, 022 (2001).
[hep-th/0105075].
%%CITATION = hep-th/0105075%%
}
%\IntriligatorCP
\lref\IntriligatorCP{
  K.~A.~Intriligator and N.~Seiberg,
  ``Lectures on Supersymmetry Breaking,''
Class.\ Quant.\ Grav.\  {\bf 24}, S741 (2007).
[hep-ph/0702069].
%%CITATION = hep-ph/0702069%%
}
%\IvanovFN
\lref\IvanovFN{
  E.~A.~Ivanov,
  ``Chern-Simons matter systems with manifest N=2 supersymmetry,''
Phys.\ Lett.\ B {\bf 268}, 203 (1991).
}
%\IntriligatorRX
\lref\IntriligatorRX{
  K.~A.~Intriligator, N.~Seiberg and S.~H.~Shenker,
  ``Proposal for a simple model of dynamical SUSY breaking,''
Phys.\ Lett.\ B {\bf 342}, 152 (1995).
[hep-ph/9410203].
%%CITATION = hep-ph/9410203%%
}
%\ClossetVP
\lref\ClossetVP{
  C.~Closset, T.~T.~Dumitrescu, G.~Festuccia, Z.~Komargodski and N.~Seiberg,
  ``Comments on Chern-Simons Contact Terms in Three Dimensions,''
[arXiv:1206.5218 [hep-th]].
%%CITATION = arXiv:1206.5218%%
}
%\DoreyKQ
\lref\DoreyKQ{
  N.~Dorey, D.~Tong and S.~Vandoren,
  ``Instanton effects in three-dimensional supersymmetric gauge theories with matter,''
JHEP {\bf 9804}, 005 (1998).
[hep-th/9803065].
%%CITATION = hep-th/9803065%%
}
%\KapustinPK
\lref\KapustinPK{
  A.~Kapustin, E.~Witten and ,
  ``Electric-Magnetic Duality And The Geometric Langlands Program,''
Commun.\ Num.\ Theor.\ Phys.\  {\bf 1}, 1 (2007).
[hep-th/0604151].
%%CITATION = hep-th/0604151%%
}
%\BuchbinderEM
\lref\BuchbinderEM{
  I.~L.~Buchbinder, N.~G.~Pletnev and I.~B.~Samsonov,
  ``Effective action of three-dimensional extended supersymmetric matter on gauge superfield background,''
JHEP {\bf 1004}, 124 (2010).
[arXiv:1003.4806 [hep-th]].
%%CITATION = arXiv:1003.4806%%
}
%\HitchinEA
\lref\HitchinEA{
  N.~J.~Hitchin, A.~Karlhede, U.~Lindstrom and M.~Rocek,
  ``Hyperkahler Metrics and Supersymmetry,''
Commun.\ Math.\ Phys.\  {\bf 108}, 535 (1987).
%%CITATION = STOCKHOLM-ITP-86-8%%
}
%\AganagicUW
\lref\AganagicUW{
  M.~Aganagic, K.~Hori, A.~Karch and D.~Tong,
  ``Mirror symmetry in (2+1)-dimensions and (1+1)-dimensions,''
JHEP {\bf 0107}, 022 (2001).
[hep-th/0105075].
%%CITATION = hep-th/0105075%%
}
%\JafferisNS
\lref\JafferisNS{
  D.~Jafferis and X.~Yin,
  ``A Duality Appetizer,''
[arXiv:1103.5700 [hep-th]].
%%CITATION = arXiv:1103.5700%%
}
%\BashkirovVY
\lref\BashkirovVY{
  D.~Bashkirov,
  ``Aharony duality and monopole operators in three dimensions,''
[arXiv:1106.4110 [hep-th]].
%%CITATION = arXiv:1106.4110%%
}
%\KlebanovTD
\lref\KlebanovTD{
  I.~R.~Klebanov, S.~S.~Pufu, S.~Sachdev and B.~R.~Safdi,
  ``Entanglement Entropy of 3-d Conformal Gauge Theories with Many Flavors,''
JHEP {\bf 1205}, 036 (2012).
[arXiv:1112.5342 [hep-th]].
%%CITATION = arXiv:1112.5342%%
}
%\KomargodskiPC
\lref\KomargodskiPC{
  Z.~Komargodski and N.~Seiberg,
  ``Comments on the Fayet-Iliopoulos Term in Field Theory and Supergravity,''
JHEP {\bf 0906}, 007 (2009).
[arXiv:0904.1159 [hep-th]].
%%CITATION = arXiv:0904.1159%%
}
\lref\IStwo{K. Intriligator and N. Seiberg, {\bf will discuss it here}}
%\HwangJH
\lref\HwangJH{
  C.~Hwang, H.~-C.~Kim and J.~Park,
  ``Factorization of the 3d superconformal index,''
[arXiv:1211.6023 [hep-th]].
%%CITATION = arXiv:1211.6023%%
}
%\WittenBS
\lref\WittenBS{
  E.~Witten,
  ``Toroidal compactification without vector structure,''
JHEP {\bf 9802}, 006 (1998).
[hep-th/9712028].
%%CITATION = hep-th/9712028%%
}
%\WeinbergUV
\lref\WeinbergUV{
  S.~Weinberg,
  ``Nonrenormalization theorems in nonrenormalizable theories,''
Phys.\ Rev.\ Lett.\  {\bf 80}, 3702 (1998).
[hep-th/9803099].
%%CITATION = hep-th/9803099%%
}
%\SeibergAJ
\lref\SeibergAJ{
  N.~Seiberg and E.~Witten,
  ``Monopoles, duality and chiral symmetry breaking in N=2 supersymmetric QCD,''
Nucl.\ Phys.\ B {\bf 431}, 484 (1994).
[hep-th/9408099].
%%CITATION = hep-th/9408099%%
}
%\WittenYC
\lref\WittenYC{
  E.~Witten,
  ``Phases of N=2 theories in two-dimensions,''
Nucl.\ Phys.\ B {\bf 403}, 159 (1993).
[hep-th/9301042].
%%CITATION = hep-th/9301042%%
}
%\SmilgaUY
\lref\SmilgaUY{
  A.~V.~Smilga,
  ``Once more on the Witten index of 3d supersymmetric YM-CS theory,''
JHEP {\bf 1205}, 103 (2012).
[arXiv:1202.6566 [hep-th]].
%%CITATION = arXiv:1202.6566%%
}
%\HenningsonVB
\lref\HenningsonVB{
  M.~Henningson,
  ``Ground states of supersymmetric Yang-Mills-Chern-Simons theory,''
JHEP {\bf 1211}, 013 (2012).
[arXiv:1209.1798 [hep-th]].
%%CITATION = arXiv:1209.1798%%
}
%\AcharyaDZ
\lref\AcharyaDZ{
  B.~S.~Acharya and C.~Vafa,
  ``On domain walls of N=1 supersymmetric Yang-Mills in four-dimensions,''
[hep-th/0103011].
%%CITATION = hep-th/0103011%%
}
%\JafferisUN
\lref\JafferisUN{
  D.~L.~Jafferis,
  ``The Exact Superconformal R-Symmetry Extremizes Z,''
JHEP {\bf 1205}, 159 (2012).
[arXiv:1012.3210 [hep-th]].
%%CITATION = arXiv:1012.3210%%
}
%\ClossetRU
\lref\ClossetRU{
  C.~Closset, T.~T.~Dumitrescu, G.~Festuccia and Z.~Komargodski,
  ``Supersymmetric Field Theories on Three-Manifolds,''
[arXiv:1212.3388 [hep-th]].
%%CITATION = PUPT-2432%%
}
%\BergmanNA
\lref\BergmanNA{
  O.~Bergman, A.~Hanany, A.~Karch and B.~Kol,
  ``Branes and supersymmetry breaking in three-dimensional gauge theories,''
JHEP {\bf 9910}, 036 (1999).
[hep-th/9908075].
%%CITATION = hep-th/9908075%%
}
%\PoppitzKZ
\lref\PoppitzKZ{
  E.~Poppitz and M.~Unsal,
  ``Chiral gauge dynamics and dynamical supersymmetry breaking,''
JHEP {\bf 0907}, 060 (2009).
[arXiv:0905.0634 [hep-th]].
%%CITATION = arXiv:0905.0634%%
}
%\OhtaIV
\lref\OhtaIV{
  K.~Ohta,
  ``Supersymmetric index and s rule for type IIB branes,''
JHEP {\bf 9910}, 006 (1999).
[hep-th/9908120].
%%CITATION = hep-th/9908120%%
}
%\SeibergPQ
\lref\SeibergPQ{
  N.~Seiberg,
  ``Electric - magnetic duality in supersymmetric nonAbelian gauge theories,''
Nucl.\ Phys.\ B {\bf 435}, 129 (1995).
[hep-th/9411149].
%%CITATION = hep-th/9411149%%
}
%\AharonyDHA
\lref\ARSW{
  O.~Aharony, S.~S.~Razamat, N.~Seiberg and B.~Willett,
  ``3d dualities from 4d dualities,''
[arXiv:1305.3924 [hep-th]].
%%CITATION = arXiv:1305.3924%%
}

%\IntriligatorID
\lref\IntriligatorID{
  K.~A.~Intriligator and N.~Seiberg,
  ``Duality, monopoles, dyons, confinement and oblique confinement in supersymmetric SO(N(c)) gauge theories,''
Nucl.\ Phys.\ B {\bf 444}, 125 (1995).
[hep-th/9503179].
%%CITATION = hep-th/9503179%%
}
\lref\DineFI{M. Dine, ``Fields, Strings, and Duality: TASI 96," eds. C. Efthimiou and B. Greene, (World Scientific, Singapore, 1997).}
%\CveticXN
\lref\CveticXN{
  M.~Cvetic, T.~W.~Grimm and D.~Klevers,
  ``Anomaly Cancellation And Abelian Gauge Symmetries In F-theory,''
JHEP {\bf 1302}, 101 (2013).
[arXiv:1210.6034 [hep-th]].
%%CITATION = arXiv:1210.6034%%
}
%\WittenDS
\lref\WittenDS{
  E.~Witten,
  ``Supersymmetric index of three-dimensional gauge theory,''
In *Shifman, M.A. (ed.): The many faces of the superworld* 156-184.
[hep-th/9903005].
%%CITATION = hep-th/9903005%%
}
%\KatzTH
\lref\KatzTH{
  S.~H.~Katz and C.~Vafa,
  ``Geometric engineering of N=1 quantum field theories,''
Nucl.\ Phys.\ B {\bf 497}, 196 (1997).
[hep-th/9611090].
%%CITATION = hep-th/9611090%%
}
%\AffleckAS
\lref\AffleckAS{
  I.~Affleck, J.~A.~Harvey and E.~Witten,
  ``Instantons and (Super)Symmetry Breaking in (2+1)-Dimensions,''
Nucl.\ Phys.\ B {\bf 206}, 413 (1982).
%%CITATION = PRINT-82-0478 (PRINCETON)%%
}
%\StrasslerHY
\lref\StrasslerHY{
  M.~J.~Strassler,
  ``Confining phase of three-dimensional supersymmetric quantum electrodynamics,''
In *Shifman, M.A. (ed.): The many faces of the superworld* 262-279.
[hep-th/9912142].
%%CITATION = hep-th/9912142%%
}
%\BorokhovIB
\lref\BorokhovIB{
  V.~Borokhov, A.~Kapustin and X.~-k.~Wu,
  ``Topological disorder operators in three-dimensional conformal field theory,''
JHEP {\bf 0211}, 049 (2002).
[hep-th/0206054].
%%CITATION = hep-th/0206054%%
}

%\MaldacenaSS
\lref\MaldacenaSS{
  J.~M.~Maldacena, G.~W.~Moore and N.~Seiberg,
  ``D-brane charges in five-brane backgrounds,''
JHEP {\bf 0110}, 005 (2001).
[hep-th/0108152].
%%CITATION = hep-th/0108152%%
}
%\PantevRH
\lref\PantevRH{
  T.~Pantev and E.~Sharpe,
  ``Notes on gauging noneffective group actions,''
  arXiv:hep-th/0502027.
  %%CITATION = HEP-TH/0502027;%%
}
%\PantevZS
\lref\PantevZS{
  T.~Pantev and E.~Sharpe,
  ``GLSM's for gerbes (and other toric stacks),''
  Adv.\ Theor.\ Math.\ Phys.\  {\bf 10}, 77 (2006)
  [arXiv:hep-th/0502053].
  %%CITATION = 00203,10,77;%%
}
%\CaldararuTC
\lref\CaldararuTC{
  A.~Caldararu, J.~Distler, S.~Hellerman, T.~Pantev and E.~Sharpe,
  ``Non-birational twisted derived equivalences in abelian GLSMs,''
  arXiv:0709.3855 [hep-th].
  %%CITATION = ARXIV:0709.3855;%%
}
%\SeibergQD
\lref\SeibergQD{
  N.~Seiberg,
  ``Modifying the Sum Over Topological Sectors and Constraints on Supergravity,''
JHEP {\bf 1007}, 070 (2010).
[arXiv:1005.0002 [hep-th]].
%%CITATION = arXiv:1005.0002%%
}
%\BanksZN
\lref\BanksZN{
  T.~Banks and N.~Seiberg,
  ``Symmetries and Strings in Field Theory and Gravity,''
Phys.\ Rev.\ D {\bf 83}, 084019 (2011).
[arXiv:1011.5120 [hep-th]].
%%CITATION = arXiv:1011.5120%%
}
%\HellermanFV
\lref\HellermanFV{
  S.~Hellerman and E.~Sharpe,
  ``Sums over topological sectors and quantization of Fayet-Iliopoulos parameters,''
Adv.\ Theor.\ Math.\ Phys.\  {\bf 15}, 1141 (2011).
[arXiv:1012.5999 [hep-th]].
%%CITATION = arXiv:1012.5999%%
}

%\SeibergNZ
\lref\SeibergNZ{
  N.~Seiberg and E.~Witten,
  ``Gauge dynamics and compactification to three-dimensions,''
In *Saclay 1996, The mathematical beauty of physics* 333-366.
[hep-th/9607163].
%%CITATION = hep-th/9607163%%
}
%\CollieIZ
\lref\CollieIZ{
  B.~Collie and D.~Tong,
  ``The Partonic Nature of Instantons,''
JHEP {\bf 0908}, 006 (2009).
[arXiv:0905.2267 [hep-th]].
%%CITATION = arXiv:0905.2267%%
}
%\DunneQY
\lref\DunneQY{
  G.~V.~Dunne,
  ``Aspects of Chern-Simons theory,''
[hep-th/9902115].
%%CITATION = hep-th/9902115%%
}
%\OlmezAU
\lref\OlmezAU{
  S.~Olmez and M.~Shifman,
  ``Revisiting Critical Vortices in Three-Dimensional SQED,''
Phys.\ Rev.\ D {\bf 78}, 125021 (2008).
[arXiv:0808.1859 [hep-th]].
%%CITATION = arXiv:0808.1859%%
}
\lref\Shamirthesis{
 I.~Shamir,
 ``Aspects of three dimensional Seiberg duality,''
 M. Sc. thesis submitted to the Weizmann Institute of Science, April 2010.
 }
%\ClossetVP
\lref\ClossetVP{
  C.~Closset, T.~T.~Dumitrescu, G.~Festuccia, Z.~Komargodski and N.~Seiberg,
  ``Comments on Chern-Simons Contact Terms in Three Dimensions,''
[arXiv:1206.5218 [hep-th]].
%%CITATION = arXiv:1206.5218%%
}
%\KaoGF
\lref\KaoGF{
  H.~-C.~Kao, K.~-M.~Lee and T.~Lee,
  ``The Chern-Simons coefficient in supersymmetric Yang-Mills Chern-Simons theories,''
Phys.\ Lett.\ B {\bf 373}, 94 (1996).
[hep-th/9506170].
%%CITATION = hep-th/9506170%%
}

%\KoroteevRB
\lref\KoroteevRB{
  P.~Koroteev, M.~Shifman, W.~Vinci and A.~Yung,
  ``Quantum Dynamics of Low-Energy Theory on Semilocal Non-Abelian Strings,''
Phys.\ Rev.\ D {\bf 84}, 065018 (2011).
[arXiv:1107.3779 [hep-th]].
%%CITATION = arXiv:1107.3779%%
}

%\FestucciaWS
\lref\FestucciaWS{
  G.~Festuccia and N.~Seiberg,
  ``Rigid Supersymmetric Theories in Curved Superspace,''
JHEP {\bf 1106}, 114 (2011).
[arXiv:1105.0689 [hep-th]].
%%CITATION = arXiv:1105.0689%%
}

%\FendleyVE
\lref\FendleyVE{
  P.~Fendley and K.~A.~Intriligator,
  ``Scattering and thermodynamics of fractionally charged supersymmetric solitons,''
Nucl.\ Phys.\ B {\bf 372}, 533 (1992).
[hep-th/9111014].
%%CITATION = hep-th/9111014%%
}
%\FendleyDM
\lref\FendleyDM{
  P.~Fendley and K.~A.~Intriligator,
  ``Scattering and thermodynamics in integrable N=2 theories,''
Nucl.\ Phys.\ B {\bf 380}, 265 (1992).
[hep-th/9202011].
%%CITATION = hep-th/9202011%%
}
%\GatesQN
\lref\GatesQN{
  S.~J.~Gates, Jr. and H.~Nishino,
  ``Remarks on the N=2 supersymmetric Chern-Simons theories,''
Phys.\ Lett.\ B {\bf 281}, 72 (1992).
%%CITATION = UMDEPP-92-127%%
}
%\ZupnikRY
\lref\ZupnikRY{
  B.~M.~Zupnik and D.~G.~Pak,
  ``Topologically Massive Gauge Theories In Superspace,''
Sov.\ Phys.\ J.\  {\bf 31}, 962 (1988).
}
%\IntriligatorAU
\lref\IntriligatorAU{
  K.~A.~Intriligator and N.~Seiberg,
  ``Lectures on supersymmetric gauge theories and electric - magnetic duality,''
Nucl.\ Phys.\ Proc.\ Suppl.\  {\bf 45BC}, 1 (1996).
[hep-th/9509066].
%%CITATION = hep-th/9509066%%
}
%\WittenDF
\lref\WittenDF{
  E.~Witten,
  ``Constraints on Supersymmetry Breaking,''
Nucl.\ Phys.\ B {\bf 202}, 253 (1982).
%%CITATION = PRINT-82-0163 (PRINCETON)%%
}
%\WilczekCY
\lref\WilczekCY{
  F.~Wilczek and A.~Zee,
  ``Linking Numbers, Spin, and Statistics of Solitons,''
Phys.\ Rev.\ Lett.\  {\bf 51}, 2250 (1983).
%%CITATION = NSF-ITP-83-148%%
}
%\WilczekDU
\lref\WilczekDU{
  F.~Wilczek,
  ``Magnetic Flux, Angular Momentum, and Statistics,''
Phys.\ Rev.\ Lett.\  {\bf 48}, 1144 (1982).
%%CITATION = NSF-ITP-81-117%%
}
%\MezincescuGB
\lref\MezincescuGB{
  L.~Mezincescu and P.~K.~Townsend,
  ``Semionic Supersymmetric Solitons,''
J.\ Phys.\ A A {\bf 43}, 465401 (2010).
[arXiv:1008.2775 [hep-th]].
%%CITATION = arXiv:1008.2775%%
}
%\MorrisonFR
\lref\MorrisonFR{
  D.~R.~Morrison and M.~R.~Plesser,
  ``Summing the instantons: Quantum cohomology and mirror symmetry in toric varieties,''
Nucl.\ Phys.\ B {\bf 440}, 279 (1995).
[hep-th/9412236].
%%CITATION = DUKE-TH-94-78%%
}
%\WessCP
\lref\WessCP{
  J.~Wess and J.~Bagger,
  ``Supersymmetry and supergravity,''
Princeton, USA: Univ. Pr. (1992) 259 p.
}
%\deBoerKR
\lref\deBoerKR{
  J.~de Boer, K.~Hori and Y.~Oz,
  ``Dynamics of N=2 supersymmetric gauge theories in three-dimensions,''
Nucl.\ Phys.\ B {\bf 500}, 163 (1997).
[hep-th/9703100].
%%CITATION = hep-th/9703100%%
}
%\GoldhaberKN
\lref\GoldhaberKN{
  A.~S.~Goldhaber, A.~Rebhan, P.~van Nieuwenhuizen and R.~Wimmer,
  ``Quantum corrections to mass and central charge of supersymmetric solitons,''
Phys.\ Rept.\  {\bf 398}, 179 (2004).
[hep-th/0401152].
%%CITATION = hep-th/0401152%%
}
%\DumitrescuIU
\lref\DumitrescuIU{
  T.~T.~Dumitrescu and N.~Seiberg,
  ``Supercurrents and Brane Currents in Diverse Dimensions,''
JHEP {\bf 1107}, 095 (2011).
[arXiv:1106.0031 [hep-th]].
%%CITATION = arXiv:1106.0031%%
}
%\LeePM
\lref\LeePM{
  B.~-H.~Lee and H.~Min,
  ``Quantum aspects of supersymmetric Maxwell Chern-Simons solitons,''
Phys.\ Rev.\ D {\bf 51}, 4458 (1995).
[hep-th/9409006].
%%CITATION = HYUPT-94-04%%
}
%\GiveonZN
\lref\GiveonZN{
  A.~Giveon and D.~Kutasov,
  ``Seiberg Duality in Chern-Simons Theory,''
Nucl.\ Phys.\ B {\bf 812}, 1 (2009).
[arXiv:0808.0360 [hep-th]].
%%CITATION = arXiv:0808.0360%%
}
%\AharonyGP
\lref\AharonyGP{
  O.~Aharony,
  ``IR duality in d = 3 N=2 supersymmetric USp(2N(c)) and U(N(c)) gauge theories,''
Phys.\ Lett.\ B {\bf 404}, 71 (1997).
[hep-th/9703215].
%%CITATION = hep-th/9703215%%
}
%\LeeEQ
\lref\LeeEQ{
  C.~-k.~Lee, K.~-M.~Lee and H.~Min,
  ``Selfdual Maxwell Chern-Simons solitons,''
Phys.\ Lett.\ B {\bf 252}, 79 (1990).
%%CITATION = CU-TP-478%%
}
%\HananyEA
\lref\HananyEA{
  A.~Hanany and D.~Tong,
  ``Vortex strings and four-dimensional gauge dynamics,''
JHEP {\bf 0404}, 066 (2004).
[hep-th/0403158].
%%CITATION = hep-th/0403158%%
}
%\HananyHP
\lref\HananyHP{
  A.~Hanany and D.~Tong,
  ``Vortices, instantons and branes,''
JHEP {\bf 0307}, 037 (2003).
[hep-th/0306150].
%%CITATION = hep-th/0306150%%
}
%\KapustinHPK
\lref\KapustinHPK{
  A.~Kapustin and B.~Willett,
  ``Wilson loops in supersymmetric Chern-Simons-matter theories and duality,''
[arXiv:1302.2164 [hep-th]].
%%CITATION = arXiv:1302.2164%%
}
%\IntriligatorPU
\lref\IntriligatorPU{
  K.~A.~Intriligator and S.~D.~Thomas,
  ``Dynamical supersymmetry breaking on quantum moduli spaces,''
Nucl.\ Phys.\ B {\bf 473}, 121 (1996).
[hep-th/9603158].
%%CITATION = hep-th/9603158%%
}
%\IzawaPK
\lref\IzawaPK{
  K.~-I.~Izawa and T.~Yanagida,
  ``Dynamical supersymmetry breaking in vector - like gauge theories,''
Prog.\ Theor.\ Phys.\  {\bf 95}, 829 (1996).
[hep-th/9602180].
%%CITATION = hep-th/9602180%%
}
%\BeemMB
\lref\BeemMB{
  C.~Beem, T.~Dimofte and S.~Pasquetti,
  ``Holomorphic Blocks in Three Dimensions,''
[arXiv:1211.1986 [hep-th]].
%%CITATION = arXiv:1211.1986%%
}
%\ShifmanDR
\lref\ShifmanDR{
  M.~Shifman and A.~Yung,
  ``NonAbelian string junctions as confined monopoles,''
Phys.\ Rev.\ D {\bf 70}, 045004 (2004).
[hep-th/0403149].
%%CITATION = hep-th/0403149%%
}

%\AharonyHDA
\lref\AharonyHDA{
  O.~Aharony, N.~Seiberg and Y.~Tachikawa,
  ``Reading between the lines of four-dimensional gauge theories,''
[arXiv:1305.0318 [hep-th]].
%%CITATION = WIS-03-13-APR-DPPA%%
}

%\AuzziFS
\lref\AuzziFS{
  R.~Auzzi, S.~Bolognesi, J.~Evslin, K.~Konishi and A.~Yung,
  %``NonAbelian superconductors: Vortices and confinement in N=2 SQCD,''
Nucl.\ Phys.\ B {\bf 673}, 187 (2003).
[hep-th/0307287].
%%CITATION = hep-th/0307287%%
}
\lref\WittenNV{
  E.~Witten,
  ``Supersymmetric index in four-dimensional gauge theories,''
Adv.\ Theor.\ Math.\ Phys.\  {\bf 5}, 841 (2002).
[hep-th/0006010].
%%CITATION = hep-th/0006010%%
}

%\LeeYC
\lref\LeeYC{
  B.~-H.~Lee, C.~-k.~Lee and H.~Min,
  ``Supersymmetric Chern-Simons vortex systems and fermion zero modes,''
Phys.\ Rev.\ D {\bf 45}, 4588 (1992).
%%CITATION = SNUTP-92-04%%
}
%\WillettGP
\lref\WillettGP{
  B.~Willett and I.~Yaakov,
  ``N=2 Dualities and Z Extremization in Three Dimensions,''
[arXiv:1104.0487 [hep-th]].
%%CITATION = arXiv:1104.0487%%
}
%\WardIJ
\lref\WardIJ{
  R.~S.~Ward,
  ``Slowly Moving Lumps In The Cp**1 Model In (2+1)-dimensions,''
Phys.\ Lett.\ B {\bf 158}, 424 (1985).
%%CITATION = Print-85-0135 (DURHAM)%%
}
%\KacGW
\lref\KacGW{
  V.~G.~Kac and A.~V.~Smilga,
  ``Vacuum structure in supersymmetric Yang-Mills theories with any gauge group,''
In *Shifman, M.A. (ed.): The many faces of the superworld* 185-234.
[hep-th/9902029].
%%CITATION = hep-th/9902029%%
}
%\WittenNV
\lref\WittenNV{
  E.~Witten,
  ``Supersymmetric index in four-dimensional gauge theories,''
Adv.\ Theor.\ Math.\ Phys.\  {\bf 5}, 841 (2002).
[hep-th/0006010].
%%CITATION = hep-th/0006010%%
}
%\JackiwPR
\lref\JackiwPR{
  R.~Jackiw, K.~-M.~Lee and E.~J.~Weinberg,
  ``Selfdual Chern-Simons solitons,''
Phys.\ Rev.\ D {\bf 42}, 3488 (1990).
%%CITATION = CU-TP-473%%
}
%\HindmarshYY
\lref\HindmarshYY{
  M.~Hindmarsh,
  ``Semilocal topological defects,''
Nucl.\ Phys.\ B {\bf 392}, 461 (1993).
[hep-ph/9206229].
%%CITATION = hep-ph/9206229%%
}
%\VachaspatiDZ
\lref\VachaspatiDZ{
  T.~Vachaspati and A.~Achucarro,
  ``Semilocal cosmic strings,''
Phys.\ Rev.\ D {\bf 44}, 3067 (1991).
}
%\PasquettiFJ
\lref\PasquettiFJ{
  S.~Pasquetti,
  ``Factorisation of N = 2 Theories on the Squashed 3-Sphere,''
JHEP {\bf 1204}, 120 (2012).
[arXiv:1111.6905 [hep-th]].
%%CITATION = arXiv:1111.6905%%
}
%\OlmezAU
\lref\OlmezAU{
  S.~Olmez and M.~Shifman,
  ``Revisiting Critical Vortices in Three-Dimensional SQED,''
Phys.\ Rev.\ D {\bf 78}, 125021 (2008).
[arXiv:0808.1859 [hep-th]].
%%CITATION = arXiv:0808.1859%%
}
%\LeeseFN
\lref\LeeseFN{
  R.~A.~Leese and T.~M.~Samols,
  ``Interaction of semilocal vortices,''
Nucl.\ Phys.\ B {\bf 396}, 639 (1993).
%%CITATION = DAMTP-92-40%%
}
%\IntriligatorAN
\lref\IntriligatorAN{
  K.~A.~Intriligator,
  ``Fusion residues,''
Mod.\ Phys.\ Lett.\ A {\bf 6}, 3543 (1991).
[hep-th/9108005].
%%CITATION = hep-th/9108005%%
}
%\WittenXI
\lref\WittenXI{
  E.~Witten,
  ``The Verlinde algebra and the cohomology of the Grassmannian,''
In *Cambridge 1993, Geometry, topology, and physics* 357-422.
[hep-th/9312104].
%%CITATION = hep-th/9312104%%
}
%\HananyVM
\lref\HananyVM{
  A.~Hanany and K.~Hori,
  ``Branes and N=2 theories in two-dimensions,''
Nucl.\ Phys.\ B {\bf 513}, 119 (1998).
[hep-th/9707192].
%%CITATION = hep-th/9707192%%
}
%\DoreyRB
\lref\DoreyRB{
  N.~Dorey and D.~Tong,
  ``Mirror symmetry and toric geometry in three-dimensional gauge theories,''
JHEP {\bf 0005}, 018 (2000).
[hep-th/9911094].
%%CITATION = hep-th/9911094%%
}
%\PisarskiYJ
\lref\PisarskiYJ{
  R.~D.~Pisarski and S.~Rao,
  ``Topologically Massive Chromodynamics in the Perturbative Regime,''
Phys.\ Rev.\ D {\bf 32}, 2081 (1985).
%%CITATION = FERMILAB-PUB-85-066-T%%
}
%\TongKY
\lref\TongKY{
  D.~Tong,
  ``Dynamics of N=2 supersymmetric Chern-Simons theories,''
JHEP {\bf 0007}, 019 (2000).
[hep-th/0005186].
%%CITATION = hep-th/0005186%%
}
%\GaiottoQI
\lref\GaiottoQI{
  D.~Gaiotto and X.~Yin,
  ``Notes on superconformal Chern-Simons-Matter theories,''
JHEP {\bf 0708}, 056 (2007).
[arXiv:0704.3740 [hep-th]].
%%CITATION = arXiv:0704.3740%%
}
%\GukovSN
\lref\GukovSN{
  S.~Gukov, E.~Witten and ,
  ``Rigid Surface Operators,''
Adv.\ Theor.\ Math.\ Phys.\  {\bf 14} (2010).
[arXiv:0804.1561 [hep-th]].
%%CITATION = arXiv:0804.1561%%
}
%\DimofteJU
\lref\DimofteJU{
  T.~Dimofte, D.~Gaiotto and S.~Gukov,
  ``Gauge Theories Labelled by Three-Manifolds,''
[arXiv:1108.4389 [hep-th]].
%%CITATION = arXiv:1108.4389%%
}
%\BorokhovCG
\lref\BorokhovCG{
  V.~Borokhov, A.~Kapustin and X.~-k.~Wu,
  ``Monopole operators and mirror symmetry in three-dimensions,''
JHEP {\bf 0212}, 044 (2002).
[hep-th/0207074].
%%CITATION = hep-th/0207074%%
}
%\BeniniMF
\lref\BeniniMF{
  F.~Benini, C.~Closset and S.~Cremonesi,
  ``Comments on 3d Seiberg-like dualities,''
  JHEP {\bf 1110}, 075 (2011)
  [arXiv:1108.5373 [hep-th]].
  %%CITATION = JHEPA,1110,075;%%
}
%\KapustinHA
\lref\KapustinHA{
  A.~Kapustin and M.~J.~Strassler,
  ``On mirror symmetry in three-dimensional Abelian gauge theories,''
JHEP {\bf 9904}, 021 (1999).
[hep-th/9902033].
%%CITATION = hep-th/9902033%%
}
%\ElitzurNR
\lref\ElitzurNR{
  S.~Elitzur, G.~W.~Moore, A.~Schwimmer and N.~Seiberg,
  ``Remarks on the Canonical Quantization of the Chern-Simons-Witten Theory,''
Nucl.\ Phys.\ B {\bf 326}, 108 (1989).
%%CITATION = IASSNS-HEP-89/20%%
}
%\IntriligatorUE
\lref\IntriligatorUE{
  K.~Intriligator, H.~Jockers, P.~Mayr, D.~R.~Morrison and M.~R.~Plesser,
  ``Conifold Transitions in M-theory on Calabi-Yau Fourfolds with Background Fluxes,''
[arXiv:1203.6662 [hep-th]].
%%CITATION = arXiv:1203.6662%%
}
\newbox\tmpbox\setbox\tmpbox\hbox{\abstractfont }
\Title{\vbox{\baselineskip12pt \hbox{UCSD-PTH-12-17}}}
{\vbox{\centerline{Aspects of 3d $\CN =2$ Chern-Simons-Matter Theories}}}
\smallskip
\centerline{Kenneth Intriligator$^1$ and  Nathan Seiberg$^2$}
\smallskip
\bigskip
\centerline{$^1${\it Department of Physics, University of
California, San Diego, La Jolla, CA 92093 USA}}
\medskip
\centerline{$^2${\it School of Natural Sciences, Institute for
Advanced Study, Princeton, NJ 08540 USA}}

\bigskip
\vskip 1cm

\noindent We comment on various aspects of the the dynamics of 3d $\CN =2$ Chern-Simons gauge theories and their possible phases.  Depending on the matter content, real masses and FI parameters, there can be non-compact Higgs or Coulomb branches, compact Higgs or Coulomb branches, and isolated vacua.  We compute the Witten index of the theories, and show that it does not change when the system undergoes a phase transition.  We study aspects of monopole operators and solitons in these theories, and clarify subtleties in the soliton collective coordinate quantization.  We show that solitons are compatible with a mirror symmetry
exchange of Higgs and Coulomb branches, with BPS solitons on one branch related to the modulus of the other.  Among other results, we show how to derive Aharony duality from Giveon-Kutasov duality.

\bigskip

\Date{May 2013}

\newsec{Introduction}

The use of holomorphy has shed light on the dynamics of $\CN=1$ gauge theories in four dimensions \IntriligatorAU, giving insights into the dynamics and the role of electric magnetic duality.    Following the original papers \refs{\AffleckAS\AharonyBX - \deBoerKR}, there has been progress in finding a similar picture in 3d $\CN=2$ theories (and interconnections with string theory), see e.g.\ \refs{\KapustinHA\DoreyRB \TongKY \DimofteJU \BeniniMF\IntriligatorUE-\CveticXN}, and many others. The 3d theories exhibit several interesting elements that are not present in 4d:
\item{1.} Abelian gauge groups can have non-trivial IR dynamics.
\item{2.} There are real parameters: masses and FI terms\foot{Real Fayet-Iliopoulos terms are also present in 4d.  But since they are associated with $U(1)$ gauge theories, which are not asymptotically free, they cannot be present in the UV Lagrangian.  Moreover, FI terms cannot be generated by the dynamics \refs{\DineFI\WeinbergUV - \KomargodskiPC} and hence cannot be present in the IR Lagrangian.  Therefore, their impact on the dynamics is less interesting than in 3d.}, which do not reside in (background)  chiral superfields.  So the power of holomorphy does not help to control them.
\item{3.} Chern-Simons parameters; they are quantized and cannot be continuously varied.
\item{4.} Coulomb branches, associated with expectation values of vector superfields; the chiral operators thus include monopole operators \refs{\AffleckAS, \AharonyBX, \KapustinHA, \BorokhovIB, \BorokhovCG}.

The presence of real parameters has deep consequences: it allows for interesting phase transitions. This differs from 4d SUSY theories, where  all the parameters are background chiral superfields, so complex.  Then, if supersymmetry is unbroken, there are no phase transitions \refs{\SeibergAJ, \IntriligatorAU}. There are no first order transitions, since the vacuum energy is zero, and any second or higher order transitions are associated with the expectation value of  a chiral superfield order parameter.
Such regions are subspaces of complex codimension, e.g.\ a  point in the complex plane that can be in a new phase (e.g.\ interacting SCFT vs.\  IR free). Since it is always possible to move around such regions, there cannot be walls separating different IR phase regions in 4d.  But there can be in 3d.

We will here encounter the following phases\foot{There is also the confining phase, as in \PolyakovFU, for  3d theories with ${\cal N}\leq 1$ SUSY; see \StrasslerHY\ for discussion of the confining phase in ${\cal N}=2$ SQED.}
\item{1.} Higgs.    This includes non-compact or compact moduli spaces, or isolated points.
\item{2.} Coulomb. This includes both non-compact and compact (as in \DoreyRB) possibilities.
\item{3.} Topological: isolated vacua of theories with Chern-Simons terms $k_{eff}\neq 0$.

\noindent
Most of our examples will be based on $U(1)$ and $SU(2)$ gauge theories, but they can be easily generalized to more complicated gauge groups.

We will discuss, compare, and make distinctions between, several different notions:
\item{1.} The supersymmetric vacua and their moduli.
\item{2.} Vortices.  (Particles, localized at points in space.)
\item{3.} Monopole operators. (Localized at points in spacetime.)

\noindent
These notions are inter-related.  The Coulomb branch of say a $U(1)$ gauge theory can be labeled by expectation values of a chiral superfield $X$.   On the Higgs branch, there can be vortex particle states, which can be BPS for non-zero FI term.  In \AharonyBX\ it was suggested that the Coulomb branch $\ev{X}$ can be related to a condensate of massless vortices.  This is motivated by mirror symmetry \IntriligatorEX, which exchanges the Higgs and Coulomb branches of dual theories: the vortices can be the quanta of dual, charged matter.   More generally, we can consider monopole operators, which are UV operators, independent of the vacuum, which insert a flux unit at a point in spacetime.  The above notions will be discussed, and distinguished, extensively in this work.

\subsec{Moduli spaces of vacua}

The theories in flat spacetime have a rich variety of possible vacua.  For generic real parameters, 3d $\CN =2$ theories have isolated vacua, with a mass gap. In some cases, supersymmetry is spontaneously and dynamically broken.

In general, upon tuning the real parameters to special values, there can be moduli spaces of supersymmetric vacua, which can be non-compact or compact.  Consider, for example, a $U(1)$ gauge theory, with some charged matter content.  The gauge field strength and superpartners can be written in terms of a real, linear multiplet
\eqn\sigmais{\Sigma \equiv  -{i\over 2} \epsilon ^{\alpha \beta} \overline{D}_\alpha D_\beta V=\sigma +\dots + \half \bar\theta \gamma ^\mu  \theta F^{\nu \rho }\epsilon _{ \mu \nu\rho }\equiv 2\pi {\cal J}_J;}
where $\sigma$ is the real scalar in the vector multiplet and $\dots$ includes the photino terms. Here ${\cal J}_J$ is the current supermultiplet for the conserved, $U(1)_J$ global symmetry, with charge
\eqn\qjis{q_J=\int d^2 x j_J^0=\int  {F\over 2\pi}=c_1(F) \in \Z.}
The linear multiplet \sigmais\ satisfies $D^2\Sigma =\bar D^2\Sigma =0$.

When there is a Coulomb branch, the fields in \sigmais\ are massless, and the real linear multiplet \sigmais\ can be dualized to a chiral superfield $X$, with $\bar D_\alpha X=0$ or, as an operator statement, $[\bar Q_\alpha , X]=0$.  The Coulomb modulus $X$ carries $q_J$ charge $\pm 1$, so $\ev{X}\neq 0$ spontaneously breaks $U(1)_J$.  The field $X$ can also acquire charges under the other global symmetries, as can be seen in terms of one-loop induced Chern-Simons terms \AharonyBX.

\subsec{Monopole operators}

A monopole operator changes the boundary conditions of the gauge field,  inserting a unit of $q_J$  charge at a point in spacetime.   It is a disorder operator, like the 't Hooft operator in 4d: it is most naturally defined by how it affects the fields in the functional integral; see e.g.\  \refs{\KapustinPK, \GukovSN} for analogous discussion for line and surface operators.   As a UV operator, the definition of the monopole operator is independent of the IR choice of vacuum.   Nevertheless, the monopole operator is not-unrelated to the Coulomb modulus $X$, which labels the vacuum, and which also carries $q_J$ charge.  Also, there is a connection to vortices.  When the monopole operator acts on the vacuum, it creates particle states with  $q_J$ charge, like the vortices discussed above.

It had been anticipated \AharonyBX\ that the Coulomb branch moduli can be given a mirror dual \IntriligatorEX\ interpretation as the Higgsing moduli  associated with a condensate of vortices.  This notion was made much more precise in the works \refs{\KapustinHA, \BorokhovIB, \BorokhovCG}.  The latter works focus on the theory on a spatial sphere $\S^2$, which eliminates some technical challenges. The physics is quite different between the flat space theory and the theory on a sphere,  and we will here discuss aspects of the monopole operators in non-compact spacetime, where it is interesting to understand the connection with the Coulomb branch moduli space of vacua (which are only superselection sectors on non-compact spacetime).

Inserting a monopole operator of charge $q_J$ at a spacetime point $x_0$ introduces a magnetic source there, with flux $q_J$ on any (Euclidean) $\S^2$ surrounding $x_0$, as in \qjis.
For a chiral monopole operator, this leads to
\eqn\sigmader{D^2\Sigma =0, \qquad \bar D^2 \Sigma = q_J 2\pi \delta ^{(3)} (x-x_0)\theta ^2,}
(with the roles of $D^2\Sigma$ and $\bar D^2 \Sigma$ reversed for the anti-chiral monopole operator).    This defines the  monopole operator in the UV, independent of the choice of vacuum.  Indeed,  \sigmader\ implies that the Coulomb branch coordinate $\sigma = \Sigma |$ is pushed to an asymptotic region close to the monopole,
\eqn\sigmapush{\sigma \to {q_J \over 2r}, \qquad\hbox{for}\qquad r\to 0,} with
$r$ the Euclidean spacetime distance to  the insertion point\foot{This behavior also entered in the analysis of \BorokhovCG\ on the theory on $\S^2\times \R$. The $1/r$ there fits with the operator dimension $\Delta (\Sigma )$ and one can rescale  $r\to 1$.}.
Here, \sigmapush\ gives the relation between the microscopic operator, whose definition is independent of the vacuum, and the Coulomb branch moduli space.  Moreover, the quantum numbers of the monopole operator are determined from the various Chern-Simons terms in this asymptotic region.

A Coulomb branch exists precisely when the monopole operator is gauge neutral.  When it is not, one can form a neutral composite of the monopole operator and the charged matter fields, but they do not have the correct quantum numbers to give a Coulomb branch.  See also \BashkirovVY\ for some discussion of monopole operators and the chiral ring.

\subsec{Vortices}

Though the elementary fields have charge \qjis\ $q_J=0$, they can  lead to solitonic particles with charge $q_J\neq 0$.  A classic example is a 3d $\CN =2$ version of the Abelian Higgs model, which on the Higgs branch has $q_J\neq 0$ Abrikosov-Nielsen-Olsen (ANO) vortices, with BPS mass given by the FI parameter, $m=|Z|$, $Z=q_J\zeta$.    There had been some issues in literature as to whether there are two or four states in BPS vortex SUSY representations, whether their spins are half-integral or $\pm {1\over 4}$, and issues related to the zero modes, so we will give a detailed discussion.   Some of the issues connect with the distinction between the monopole operators, which are chiral operators, vs.\  the vortex BPS states.

We note that, while chiral operators $\CO$ and BPS states are both annihilated by half the supercharges, they are annihilated by {\it different} supercharges (linear combinations).  Acting with a chiral operator on the vacuum gives a tower of energy states, and the BPS state is found by projecting to the lowest energy state in the tower.  The basic BPS representation is two-dimensional, and it becomes four-dimensional upon including its CPT conjugates.

In theories with non-minimal matter content, there are additional Bose and Fermi zero modes of the soliton or vortex solutions\foot{For the monopole operator on $\S^2$, the $\sigma \neq 0$ in \sigmapush\ lifts the matter fields, so there are then no zero modes \BorokhovCG.  Indeed, even the operators in the same supermultiplet have different operator dimension, so they have different energies on $\S^2$.}, which have log-IR-divergent norm when the theory is on non-compact space.  This was first noted in \WardIJ\ for solitons in the 3d, non-SUSY $CP^1$ sigma model, and it also occurs in the Abelian-Higgs UV completion, and more generally in theories with more than one charged field.

We will here give an interpretation\foot{An interpretation discussed in \CollieIZ\ and references cited therein is that these new parameters are the relative locations of partonic constituents of the solitons or instantons, which are frozen by strong binding forces, as evident from the mirror dual \IntriligatorEX\ variables and gauge fields.}  of these non-normalizable zero modes, which does not rely on the presence of vortices: the modes are present also in the $q_J=0$ sector.      As we will further discuss,  such zero modes  should be interpreted as non-dynamical parameters, labeling different superselection sectors of the quantum field theory, analogous to vacuum moduli.   For the non-normalizable Fermi zero modes, our prescription is to quantize them, which leads to a Fock space of states.    Because we take space non-compact, this Fock space gives multiple, disconnected Hilbert spaces.  As we discuss, these zero modes affect the vortex quantum numbers, and we will see how vortex quantum numbers match those of the monopole operator, thanks in part to the zero mode contributions.

 \subsec{Generic real masses: a mass gap.}

3d $\CN =2$ theories with generic real masses, Chern-Simons terms, and FI parameters have a mass gap, with isolated vacua.  These could be isolated SUSY vacua, or SUSY could be broken.  One of the most basic diagnostics of the dynamics is the Witten index $\Tr (-1)^F$ \WittenDF.
The Witten index of a theory in $D$ spacetime dimensions can be computed by considering the theory on $\R \times \T^{D-1}$.  Let us contrast this torus-index with the sphere-indices, on $\R \times \S^{D-1}$ or $\S^1\times \S^{D-1}$, which have been of recent interest.  The torus-index counts  SUSY vacua, whereas the sphere-index (also called the super-conformal index\foot{This is a misnomer, as the sphere index is not restricted to superconformal theories, see \FestucciaWS.}) index counts the BPS operators.  Indeed, the theory on a torus can have multiple vacua, whereas the theory on a sphere always has a unique vacuum (the identity operator, in the operator/state correspondence).  In theories with non-compact moduli spaces of vacua the Witten index is ill-defined.   This is not an issue for the sphere index.

The index\foot{In what follows, ``index" refers to the Witten (or torus) index.  Also, $G_k$ denotes gauge group $G$ with Chern-Simons coefficient $k$. } of 3d SUSY $G_k$ Yang-Mills-Chern-Simons theories, without matter, has been studied, starting with \WittenDS. But the index of theories with matter has not yet appeared in the literature\foot{It was observed via branes and the s-rule that certain 3d CS theories  break supersymmetry \refs{\BergmanNA, \OhtaIV}, and that $\CN =2$ $U(N_c)$ has SUSY vacua if $N_f\geq N_c-|k|$ \GiveonZN.  } (as far as we are aware).   We here compute the index of $\CN =2$ theories with matter, for non-zero real masses.  With massless matter, the index is generally  ill-defined.  By turning on the real parameters, the space of vacua can be made compact,  which leads to a well-defined, finite index, which we compute\foot{Note that theories related by dimensional reduction generally do {\it not} have the same index.   Susy vacua can move in or out from infinity, as the radius $R$ of the $\S _R^1$ compactification circle of dimensional reduction varies between $R=0$ and $R=\infty$. E.g.\ 4d $\CN =1$ SYM has $h=C_2(G)$ SUSY vacua \refs{\WittenDF, \WittenBS, \KacGW}, whereas 3d $\CN =2$ SYM (with $k=0$) has a runaway, with no SUSY vacua.  The $h$ 4d SUSY vacua are lost, at infinity in field space.}.

As we vary the parameters there can be phase transitions, where massless moduli arise. If the moduli are compact, the index remains well-defined and cannot change through the transition. We will see examples of this, e.g.\ changing the sign of a FI parameter can open up a compact $CP^N$ moduli space.
We will also see examples where there are phase transitions where a non-compact moduli space arises.  Since the index is ill-defined at such a transition, it could  in principle differ on either side, with additional vacua coming in from infinity, or disappearing to infinity\foot{See e.g.\ \refs{\IntriligatorPU, \IntriligatorCP}, and references cited therein, for 4d examples where that happens when SUSY is broken.}.    We show that the index of 3d $\CN =2$ theories actually does not change through any  transitions, even in the non-compact case.

As a general argument as to why the index does not change, consider compactifying the 3d theory on $\S^1_R$ of radius $R$ (such a compactification is needed anyway to define the index).  The low-energy theory is a 2d $\CN =(2,2)$ theory with massless matter,  and a tower of massive matter fields from the Fourier modes with non-zero momentum on the $\S _R^1$.  In that theory, the real parameters of the 3d theory become complex parameters, background expectation values of twisted-chiral superfields\foot{Specifically, $m_R\to \tilde m_C$ and $\zeta\to \zeta _c=\zeta +i\theta$, where $\theta$ is the 2d theta term $i\theta F$ and the additional parameters in $m_C$ are twists by global symmetry phases when circling the $\S^1$.}.  Holomorphy constrains the 2d twisted-chiral superfield parameters, much as in 4d, so any phase transitions are of complex codimension, and can thus be avoided \WittenYC.  This proves that the Witten index of 3d $\CN =2$ theories cannot change as we vary the real parameters.   Even though this proof is complete, it is nice to verify it explicitly, as we will in several examples.  We will see that the  fact that the index does not change looks highly non-trivial, as the vacua are rearranged from the 3d perspective as we cross the phase transition.

We find that the index of a general $\CN =2$ $G_k$ theory, with generic real parameters, is
\eqn\indexintro{\Tr (-1)^F=J_G(k'), \qquad k'=|k|-h+\half \sum _f T_2(r_f),}
where $J_G(k')$ is the number of primary operators of $G_{k'}$ WZW theory.   The $\sum _f$ in \indexintro\ is a contribution from the matter fields, $Q_f$, in representations $r_f$ of $G$, with $T_2(r_f)$ the quadratic index of the representation (the number of $\psi _{r_f}$ Fermion zero modes in a 4d instanton).  When there is no matter, the result \indexintro\ follows from the argument in \WittenDS: the $-h$ shift in \indexintro\ comes from integrating out the gauginos (with
$h=\half T_2(adj)$ the dual Coxeter number, and this contribution is twice that in  \WittenDS, since here we consider $\CN =2$ rather than $\CN =1$), and connection between the index of the 3d Chern-Simons theory and the 2d WZW theory follows from \refs{\WittenDS, \WittenHF}.  The matter contribution in \indexintro, shifting $k'$ by $+\half \sum _f T_2(r_f)$, is a new result, that will be explained here.  For example,  for  $U(1)_k$ gauge theory
with  matter chiral superfields $Q_i$ of charge $n_i$, our result for the index
is \eqn\indexnf{\Tr (-1)^F=|k|+\half \sum _i n_i^2, }
(with a modification for chiral theories with small $|k|$).  The matter shift of $k'$ could roughly be anticipated from the shift $k\to k_{eff}$ from integrating out the real-massive matter fields, though it will be seen that the details are much richer than this rough argument.

If $k'<0$ in \indexintro, then the theory dynamically breaks supersymmetry for generic real parameter deformations.  For example, that is the case for $SU(N_c)$ or $U(N_c)$ with $N_f$ fundamental flavors if $k-h+N_f<0$.  Though supersymmetry is broken for generic real parameter deformations, there are generally moduli spaces of supersymmetric vacua for tuned, non-generic real parameters. This will be illustrated in the examples.

If $k'=0$ in \indexintro, it follows that there is a unique supersymmetric groundstate
\eqn\indexconf{\Tr (-1)^F=1: \qquad G_k \ \hbox{gauge theory with}\ |k|-h+\half \sum _f T_2(r_f)=0 }
since $G_{k'=0}$ WZW theory has only the identity operator.    As explained in \WittenDS\ (for pure $\CN =1$ SUSY Chern-Simons theory without matter, but the same explanation applies here), confinement of electric flux can occur only if the center of $G$ acts trivially on the states, and that requires $k'=0$.   So simply confining theories must have $k'=0$.

Dual theories must have matching index.  This gives another test of dualities, in addition to matching the moduli spaces, discrete parity anomaly matching for the global symmetries in theories with\foot{When $k\neq 0$, $P$ is broken.  While the parity anomalies must still match, the check becomes unconstraining, as global Chern-Simons coefficients can then have more general RG running
\refs{\BeniniMF, \ClossetRU, \ClossetVP}.} $k=0$, and the more detailed checks of matching of the $\S^2\times \S^1$ partition function and $\S^3$ sphere indices.
For example, it was found in \AharonyBX\ that $SU(N_c)_0$ and $U(N_c)_0$, with $N_f=N_c$, has a simple dual description, to a theory of chiral superfields, without gauge fields.
The matching of $\Tr (-1)^F$ gives an  immediate check, and classification of possible generalizations.  Such dual theories, without gauge fields, all have  a unique SUSY vacuum upon adding real masses
\eqn\indexconf{\Tr (-1)^F=1: \qquad\hbox{no gauge fields, all matter with real masses.}}
So we see from \indexconf\ that only $k'=0$ theories can have a simple dual description, without gauge fields.  This matches the observation mentioned above, that the argument of \WittenDS\ shows that $k'=0$ is a necessary condition for simple confinement.
The condition \indexconf\ is indeed satisfied e.g.\ for $SU(N_c)_k$ and $U(N_c)_k$ when $|k|+N_f=N_c$, agreeing with the known cases in \refs{\AharonyBX, \GiveonZN} of theories with simple duals, without gauge fields.

\subsec{Aharony Duality and Giveon-Kutasov Duality}

We discuss some aspects of Aharony duality \AharonyGP.  The electric theory is $U(N_c)_0$, with $N_f$ flavors, and the dual theory is $U(N_f-N_c)_0$ with $N_f$ flavors and some added singlets with superpotential terms.  The unusual aspect of this duality is how the electric vs.\  magnetic monopole operators are dualized, with the dual monopole operators explicitly appearing in the dual Lagrangian. This peculiarity does not occur in 4d duals. For example, consider 4d $\CN =1$ electric-magnetic duality \SeibergPQ, e.g.\ for the case of $SO(N)$ theories \IntriligatorID: the magnetic monopoles of the electric theory map to the fundamental matter fields $q_i$ of the dual, not to monopoles in the dual.  The difference in 3d can be anticipated from the fact that monopole operators of $U(N_c)$ carry a conserved {\it global} charge $q_J$ \qjis.

For theories with non-zero Chern-Simons terms, Giveon and Kutasov \GiveonZN\ proposed an interesting duality, between $U(N_c)_k$ with $N_f$ flavors and $U(|k|+N_f-N_c)_{-k}$ with $N_f$  flavors.   Here $k\neq 0$ eliminates the monopole operators.  Correspondingly, this duality is more conventional than Aharony duality, in that the dual theory does not contain monopole operators in the Lagrangian.   It has been shown \refs{\GiveonZN, \BeniniMF}\ that, starting from Aharony duality in the UV,  there is a deformation that gives an RG flow to Giveon-Kutasov duality in the IR.  So Giveon-Kutasov duality can be derived as a consequence of Aharony duality.

We here discuss a reversed RG flow, from Giveon-Kutasov duality in the UV to Aharony duality in the IR.  In this way, we {\it derive} the peculiar monopole operator coupling of Aharony duality as an output, rather than assuming it as an input.  There has been much recent evidence for both Aharony and Giveon-Kutasov duality from considering the theories on spheres, see e.g.\ \refs{\BashkirovVY, \BeniniMF, \HwangJH}, and our connection here between the two dualities in flat spacetime provides additional evidence.  Adding generic real masses to both sides of either Giveon-Kutasov, or Aharony, duals gives another check: the massive matter fields can be integrated out, and the low-energy Chern-Simons theories are related by level-rank duality.  In particular, their indices from \indexintro, match.

Our methods are also useful for upcoming work on relating 4d and 3d dualities \ARSW.

\subsec{Outline}

The organization of this paper is as follows.  In section 2, we give some background material on $U(1)$ gauge theories and the Higgs, Coulomb, and topological vacua. In section 3, we discuss general aspects of monopole operators and vortices, and the distinction between them.  In section 4, we discuss examples of $U(1)$ gauge theories with Higgs or Coulomb moduli spaces, which can be either non-compact or compact (in this context, compact Coulomb branches  first appeared in \DoreyRB).  In section 5, we discuss some generalities about the Witten index \WittenDF, in general spacetime dimension, and we recall the results of \WittenDS\ for the index of 3d SUSY Chern-Simons theories without matter.

In section 6, we discuss the Witten index of 3d $\CN =2$ gauge theory, with gauge group $U(1)$, with Chern-Simons term $k$, and general matter content,  with generic real mass $m_i$ and FI term $\zeta$.   In section 7, we compare the $\Tr (-1)^F$ index of some dual theories.  In section 8, we discuss $SU(2)_k$ gauge theory with various matter content.  In section 9, we discuss aspects of general $SU(N_c)_k$ and $U(N_c)_k$ theories.   In section 10, we discuss a RG flow from Giveon-Kutasov duality in the UV to Aharony duality in the IR.

In appendix A, we collect some details about our conventions, the SUSY algebra and Lagrangians, vacua, and aspects of BPS vortices and zero modes.  In appendix B, we recall how the Coulomb branch is dualized via a Legendre transform between the linear multiplet and chiral superfields \refs{\HitchinEA, \deBoerKR}, and how this can  imply non-renormalization of the K\"ahler form \AganagicUW.  In appendix C, we comment about modes with log-divergent norm.  In appendix D, we provide some details about the derivation of $\Tr (-1)^F$ for $U(1)_k$ with matter, explicitly verifying that the result is independent of variations of the real parameters.

\newsec{3d, ${\cal N}=2$ SUSY $U(1)$ gauge theories}

Consider a $U(1)$ gauge theory with classical Chern-Simons term $k$, Fayet-Iliopoulos term $\zeta$, and matter fields $Q_i$ of charge $n_i$, and real mass $m_i$.
Consistency  for compact $U(1)$ (as opposed to $\R$) requires all $n_i\in \Z$ and
\eqn\consk{k + \half \sum_i n_i^2 \in \Z; \qquad\hbox{equivalently,} \qquad  k + \half \sum_i n_i \in \Z~~.}
The unbroken global symmetry for generic $\zeta$ and $m_i$ is $U(1)_R\times U(1)_J\times \prod _i U(1)_i \ /U(1)$, where $U(1)_R$ is an R-symmetry,  $U(1)_J$ is associated with the topological current \sigmais, $U(1)_i$ rotates $Q_i$ by a phase, and the $/U(1)$ is for the gauged $U(1)$.  The parameters $m_i$ and $\zeta$ can be thought of as background values of scalars in classical gauge fields for these global symmetries\foot{Therefore, if $m_i\neq 0$, any added $W_{tree}$, e.g.\ complex mass terms, must not break the $U(1)_i$ symmetry.  Doing so would lead to explicit supersymmetry breaking.}. Since one linear combination of the $U(1)_i$ generators is gauged, one linear combination of $m_i$ and $\zeta$ is redundant, and can be absorbed in shifts of the scalar $\sigma =\Sigma |$ in the dynamical gauge superfield \sigmais,
  \eqn\ginva{\sigma\to \sigma + \delta \sigma, \qquad
\zeta\to \zeta - k \delta \sigma, \qquad   m_i\to m_i - n_i \delta \sigma~~.}

It is important that the parameters $m_i$ and $\zeta$ are bottom components of linear superfields.  Therefore, they are renormalized at most at one loop.  Nevertheless, the actual masses of fields with nonzero real masses are renormalized at all orders, because of K\"ahler potential renormalization.  We will continue to refer to the parameters $m_i$ as real masses, even though they are not precisely masses of particles.

The semi-classical effective potential of the theory, with $W_{tree}=0$,  is\foot{ Our normalization of the FI term has an unconventional factor of $1/2\pi$ (and an opposite sign); this is natural in the context of 3d field theory, where  $\zeta$  can be regarded as coming from a mixed Chern-Simons term $k_{GJ}=1$ between the $U(1)$ gauge field and a background field $\Sigma _J|=\zeta$ for the topological $U(1)_J$ global symmetry.  This will eliminate many factors of $2\pi$ in later expressions and is also natural in terms of the mirror symmetry map \IntriligatorEX\ $m\leftrightarrow \zeta$.}
\eqn\claspot{V_{s.c.}= {e^2_{eff}\over 32\pi ^2}\left( \sum_i 2\pi n_i  |Q_i|^2 -\zeta _{eff} -k \sigma _{eff}\right)^2 + \sum_i (m_i + n_i\sigma)^2  |Q_i|^2}
($Q_i$ includes the $Z_i$ renormalization factor). The
 effective real mass of $Q_i$  for $\sigma \neq 0$ is
\eqn\effmass{m_i(\sigma)=m_i + n_i\sigma,}
so $Q_i$ is massless at $\sigma =\sigma _{Q_i}$, with $m_i(\sigma _{Q_i})=0$:
\eqn\sigmaiis{\sigma _{Q_i} \equiv -m_i/n_i.}
The quantum-corrected $k_{eff}$ and $\zeta _{eff}$ include renormalization from integrating out matter $Q_i$ with real masses $m_i(\sigma)$ but, we emphasize, they  must be piecewise field independent constants, modulo discontinuous jumps at the $\sigma _{Q_i}$ \sigmaiis.   Indeed, we have
\eqn\zetakeff{\eqalign{
&\zeta_{eff}=\zeta + \half \sum_i n_i m_i \sign(m_i(\sigma))\cr
&k_{eff}=k + \half \sum_i n_i^2 \sign(m_i(\sigma)) \cr
&\zeta_{eff}+ k_{eff}\sigma =\zeta +k \sigma + \half \sum_i n_i|m_i(\sigma)|\equiv F (\sigma)~~.
}}

These expressions are similar to the ones in \refs{\DoreyRB,\TongKY}.   Condition \consk\ ensures that all $k_{eff}\in \Z$; this is required for consistency. There are no higher order perturbative corrections to \zetakeff: $k_{eff}$ must remain quantized, and $\zeta_{eff}$ is a component of a background linear superfield.  Higher order perturbative corrections cannot maintain these properties.  The function $F(\sigma)$ in \zetakeff\ is the combination appearing in the potential \claspot, and is continuous: the jumps at $\sigma _{Q_i}$ are in the slopes of $F(\sigma)$.  Using  \zetakeff, the discontinuities of $k_{eff}$ and $\zeta _{eff}$ for $\sigma$ just above and below $\sigma _{Q_i}$ for generic $m_i$ are
\eqn\kzetajumps{k_{eff,i}^+-k_{eff,i}^-=n_i^2\sign (n_i)=n_i|n_i|, \qquad \hbox{and}\qquad \zeta _{eff,i}^+-\zeta _{eff,i}^-=n_i m_i \sign (n_i)=|n_i|m_i.}

The semi-classical vacua  of \claspot\ can be found via solutions of
\eqn\condos{\sum_i 2\pi n_i |Q_i|^2 =F(\sigma), \qquad\hbox{and}\qquad m_i(\sigma) Q_i=0, \quad\hbox{for all $i$~~.}}
We refer to the SUSY vacua solutions of \condos\ as follows:
\lfm{1.} ``Higgs:" vacua with some $\ev{Q_i}\neq 0$, and thus $\sigma = \sigma _{Q_i}$, as in \sigmaiis.  With non-generic real masses and opposite sign matter, there can be non-compact Higgs moduli spaces of vacua, similar to those of 4d theories. With generic real masses, on the other hand, the Higgs vacua are isolated, with a mass gap, and only exist if the sign of $F(\sigma _{Q_i})$ in \condos\ coincides with that of $n_i$.  For later use, we define
\eqn\higgsmaybe{s_i\equiv \Theta (n_i F(\sigma _{Q_i})), \qquad\hbox{where}\qquad \Theta (x)=\cases{1 & for $x>0$\cr 0 & for $x<0$,}}
so, for generic real masses, there is a Higgs vacuum at $\sigma _{Q_i}$ iff $s_i=1$.

\lfm{2.} ``Coulomb branch:"  vacua with $\ev{Q_i}=0$, and thus $F(\sigma)=0$ in \condos.  We reserve the name ``Coulomb branch" for the case where there is a continuous (possibly compact) moduli space of such vacua, i.e.\  a range of $\sigma$ where $\zeta _{eff}=k_{eff}=0$.

\lfm{3.} ``Topological vacua:" there is a low-energy $U(1)_{k_{eff}}$, with all matter massive and integrated out, $\ev{Q_i}=0$.  Such vacua are isolated, with a mass gap for $k_{eff}\neq 0$, located at the zeros of $F(\sigma)$,
\eqn\sigmacapi{\sigma _{I}=-{\zeta_{eff} (\sigma _{I})\over k_{eff}(\sigma _I)}, \qquad\hbox{i.e.} \qquad F(\sigma _I)\equiv \zeta + k\sigma _I+\half \sum _i n_i |m_i(\sigma _I)|=0,}

When there is a Coulomb branch, it is convenient to dualize  the linear multiplet $\Sigma$ to a chiral superfields $X_\pm$, with $U(1)_J$ charge \qjis\ $q_J=\pm 1$ \refs{\AharonyBX,  \deBoerKR} (see appendix B for a review).   Sometimes the $X_\pm$ fields are related to monopole operators,  though we will here make the distinction between the Coulomb vacuum modulus and the monopole operator.  Semi-classically, for a $U(1)$ gauge theory, or for $SU(2)\to U(1)$,  these are given by\foot{The relation is as follows.  The standard normalization for the $SU(2)$ Lagrangian, broken to $U(1)$ on the Coulomb branch, gives $\CL =-{1\over 4g^2} \sum _{a=1}^3 F_{\mu \nu}^a F^{a\mu \nu}\to -{1\over g^2}F_{\mu \nu}F^{\mu \nu}\equiv -{1\over 4e^2}F_{\mu \nu}F^{\mu \nu}$, so $g^2_{SU(2)}=4e^2_{U(1)}$.  Here $F_{\mu \nu}\equiv \half  F_{\mu \nu}^{a=3}$, and $A_\mu \equiv \half A_\mu ^{a=3}$, with the $\half $ introduced to couple to the fundamental with $U(1)\subset SU(2)$ charges $\pm 1$ rather than $\pm \half$; likewise, $\sigma _{U(1)}=\half \sigma _{SU(2)}$. },
\eqn\scmon{U(1):\ X_{\pm}\sim e^{\pm (2\pi \sigma/e_{eff}^2  +ia) }, \qquad \hbox{or}\qquad SU(2):\ Y\sim e^{4\pi \sigma /g_{eff}^2+ia},}
with $a$ the $2\pi$ periodic  scalar dual  to the photon, $F_{\mu \nu} =e_{eff}^2 \epsilon _{\mu \nu \sigma}\partial ^\sigma a/2\pi $, and $a$ can also be regarded as the massless Nambu-Goldstone boson \AffleckAS\ for spontaneously broken $U(1)_J$ global symmetry for $X_\pm \neq 0$.  With real masses, the Coulomb branch can become quite intricate, with non-compact and/or compact regions, as depicted in Fig.\ 1. Monopole operators push $\sigma$ to $\pm \infty$ and hence are associated with the non-compact regions ($X_{1-}$ and $X_{4+}$ in Fig.\ 1) on the ends. The compact Coulomb regions are also parameterized by chiral superfields (e.g.\ $X_{1+}$ and $X_{2-}=1/X_{1+}$ in Fig.\ 1); they do not correspond to monopole operators in the UV theory.  Instead, these compact regions lead to Skyrmions, which we will describe below.

\fig{Quantum moduli space of $U(1)$ with flavors with ($CP$ symmetric, $m_{Q_i}=-m_{\tilde Q_i}$) real masses. $M_i^i$ are mesons parameterizing Higgs branches, $X_{1-}$ and $X_{4+}$ parameterize noncompact Coulomb branches and the remaining $X_{i\pm}$ (with $X_{1+}=1/X_{2-}$, etc.) parameterize the compact regions of the Coulomb branch.  For more details, see below.}
{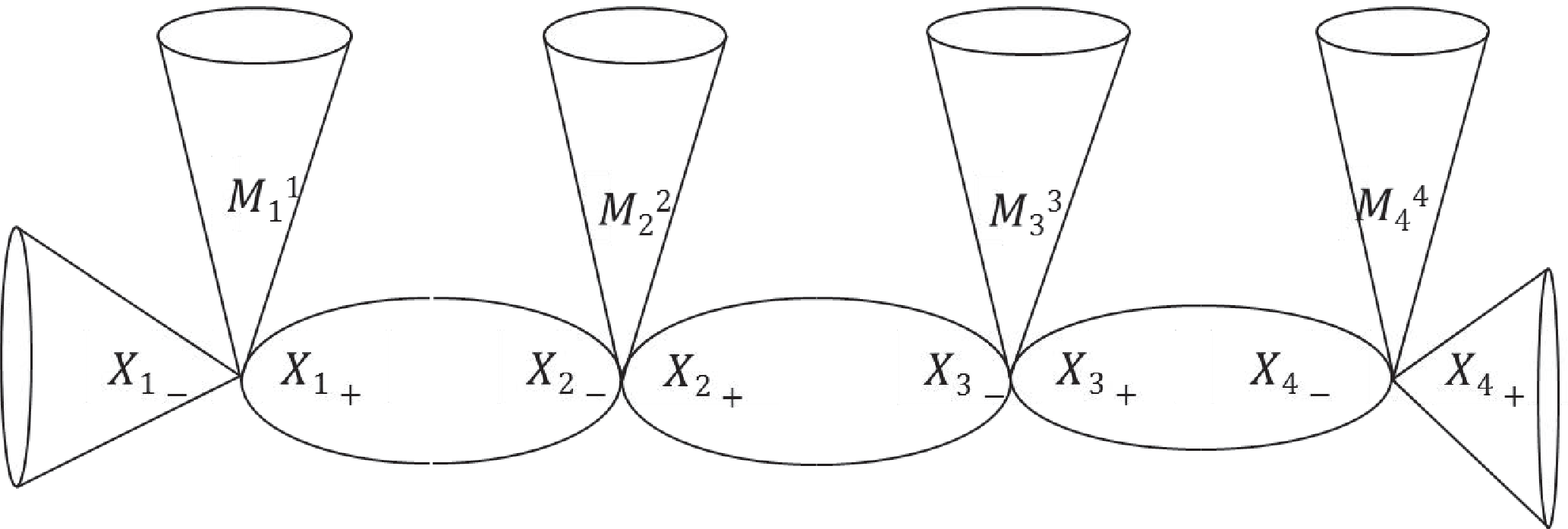}{15 truecm}
\figlabel\Onenf

As usual, it is useful to tabulate the charges of the chiral superfields under the symmetries.  When $k_{eff}\neq 0$, fields with $q_J$ charge, in particular the fields $X_\pm$, become charged under the $U(1)_{G}$ gauge group.  Likewise, $X_\pm$ get charges under the global $U(1)_i$ and $U(1)_R$  symmetries\foot{The $U(1)_R$ symmetry given here is simply a particular choice; one can obtain other conserved R-symmetries by mixing with the $U(1)_i$ global symmetries.  In particular, the superconformal R-symmetry is determined as in \refs{\JafferisUN,\ClossetVG} (including also any accidental global $U(1)$ symmetries).}, via their induced mixed Chern-Simons terms with the $U(1)_G$ gauge field \AharonyBX
\eqn\kgaugefla{k_{G,i}=\half n_i  \sign (n_i \sigma + m_i) , \qquad k_{G,R}=-\half \sum _i n_i  \sign (n_i \sigma +m_i)}
 from integrating out the $Q_i$ matter.  In particular, for the asymptotic regions $\sigma \to \pm \infty$, which we associate with operators $X_\pm$, respectively, this gives the charges
\eqn\onengen{
\vbox{\offinterlineskip\tabskip=0pt
\halign{\strut\vrule#
%%%%%%%%%%%%%%%%%%
&~$#$~\hfil\vrule
&~$#$~\hfil\vrule
&~$#$~\hfil\vrule
&~$#$~\hfil\vrule
&~$#$~\hfil\vrule
&~$#$~\hfil%\vrule
&\vrule#
\cr
%%%%%%%%%%%%%%%%%
\noalign{\hrule}
& & U(1)_G & U(1)_j & U(1)_R &  U(1)_J \cr
\noalign{\hrule}
%%%%%%%%%%%%%%%%%%
&  Q_i        &\; n_i & \;\delta _{ij} &\;0& \quad 0   \cr
&X_\pm  &\;  \mp (k\pm k_c)&\; -\half  |n_j|  &\; \half \sum _i |n_i|  & \; \pm 1 \cr
}\hrule}}
where the
asymptotic values of $k_{eff}(\sigma)$ for $\sigma \to \pm \infty$ are:
\eqn\kinfinity{k_{eff}(\sigma =\pm \infty)=k\pm k_c, \qquad k_c\equiv \half \sum _i n_i |n_i|;}
$k_c$ vanishes in theories with vector-like matter, where charge conjugation $C$: $ n_i\to -n_i$ is a symmetry.
The charges in \onengen\ are computed by going to $\sigma = \pm \infty$, which is only possible when there is a non-compact Coulomb branch, i.e.\   $k=-k_c$ (or $k=+k_c$), so $k_{eff}=0$ in the asymptotic region \kinfinity, and $X_+$ (or $X_-$) is gauge neutral.  The charges \onengen\ should thus be understood as being applicable only when the $U(1)_G$ charge of $X$ vanishes.
For $k_c=0$, both $X_+$ and $X_-$ can be neutral, while for $k_c\neq 0$ at most one can be neutral, corresponding to a non-compact Coulomb branch.

Another condition for the $X_\pm$ Coulomb branch to be unlifted is that the asymptotic value of the FI parameter obtained from \zetakeff,
\eqn\zetainfinity{\zeta _{eff}(\sigma =\pm \infty)=\zeta \pm  \zeta _c \qquad \zeta _c\equiv \half \sum _i |n_i| m_i}
must vanish, $\zeta \to \mp \zeta _c$.  This can equivalently be understood from the $X_\pm$ quantum numbers in \onengen: since $m_j$ is a background for $U(1)_j$ and $\zeta$ is a background for $U(1)_J$, the real mass of $X_\pm$ (its central charge $Z$) is determined by the charges in \onengen\ to be
\eqn\mxpm{m_{X_\pm}= \sum _j m_j (-\half |n_j|)\pm \zeta,}
so $X_\pm$ is massless only if $k=\pm k_c$ and $\zeta \to \mp \zeta _c$.

The chiral monopole operator is related to an insertion of the operator $X_\pm$, so it must have the global quantum numbers of $X_\pm$ in \onengen, in cases where there is a non-compact Coulomb branch.   In \BorokhovCG, the global quantum numbers of the monopole chiral operator were determined, for the example of $U(1)$ with $N_f$ vector-like flavors, by instead considering the energy spectrum of the theory on  $\S^2$, with the BPS monopole gauge field background, computing the vacuum charges by heat kernel regularization.  The charges in \onengen\ reproduce the results of \BorokhovCG\ via the simpler (but conceptually equivalent) method of using the induced Chern-Simons terms \kgaugefla\ of the theory on non-compact space.

For a $U(1)_k$ theory with $N_+$ matter fields $Q_i$ of charge $+1$, and $N_-$ fields $\tilde Q_{\tilde i}$ of charge $-1$, the global symmetry  is $SU(N_+)\times SU(N_-) \times U(1)_A \times U(1)_R\times U(1)_J$, with charges
\eqn\onensqed{
\vbox{\offinterlineskip\tabskip=0pt
\halign{\strut\vrule#
%%%%%%%%%%%%%%%%%%
&~$#$~\hfil\vrule
&~$#$~\hfil\vrule
&~$#$~\hfil\vrule
&~$#$~\hfil\vrule
&~$#$~\hfil\vrule
&~$#$~\hfil%\vrule
&\vrule#
\cr
%%%%%%%%%%%%%%%%%
\noalign{\hrule}
&  &  SU(N_+) &  SU(N_-) &  U(1)_A& U(1)_R &  U(1)_J &\cr
\noalign{\hrule}
%%%%%%%%%%%%%%%%%%
&  Q^i        & \; {\bf  N_+}   &\; {\bf  1}  &\; 1    &\;0& \quad 0   &\cr
& \tilde Q_{\tilde i} & \; {\bf  1}   & \; {\bf  \bar N_-} & \; 1   & \;0 &  \quad 0   &\cr
\noalign{\hrule}
%%%%%%%%%%%%%%%%%%
& M^i_{\tilde i}=Q^i \tilde Q_{\tilde i} & \; {\bf  N_+}  & \; {\bf  \bar N_-}       & \;2 & \; 0  &   \quad  0   &\cr
&X_\pm & \; {\bf  1} & \; {\bf  1} &\; -\half  (N_++N_-) &\; \half  (N_++N_-) & \; \pm 1 &\cr
}\hrule}}
The charges of $X_\pm$ are as in \onengen\ and \refs{\AharonyBX, \IntriligatorUE}\ and follow from \onengen, via the induced mixed Chern-Simons terms $k_{GA}=\half (N_++N_-)\sign (\sigma)=-k_{GR}$ between the gauge $U(1)_G$ and $U(1)_A$ and $U(1)_R$,
with $\sign (\sigma)=\pm 1$ for $X_\pm$.   As in \onengen, the operators $X_\pm$ are only gauge neutral if $k_{eff}=0$; if not, $X_\pm$ are removed from the chiral ring of spin zero, gauge invariant operators and, correspondingly, there is no Coulomb branch.   Using  \zetakeff,
\eqn\krenn{k_{eff}=k + \half(N_+-N_-)\sign(\sigma),}
for $m_i=m^{\tilde i}=0$.   So there is a non-compact half-Coulomb branch for $k=\pm \half (N_+-N_-)$, upon taking  $\zeta \to 0$, parameterized by $X_\mp$.

As a soon-useful aside,  we comment on the case when the  charges $n_i$ all have a common integer factor, i.e.\  all $n_i=n\tilde n_i$, with $n$ and $\tilde n_i$ integers.  This  motivates the rescaling
\eqn\rescaling{V\to \tilde V\equiv nV, \quad e^2\to \tilde e^2 \equiv n^2 e^2, \quad n_i \to \tilde n_i\equiv n_i/n, \quad k\to \tilde k\equiv k/n^2, \quad \zeta \to \tilde \zeta \equiv \zeta /n.}
If the gauge group is non-compact, i.e.\  $\R$ instead of $U(1)$, the original and rescaled theories would be physically identical.  For compact $U(1)$, with $\tilde k$ an integer, the  rescaled theory \rescaling\ is a  $\Z _n$ orbifold of the original theory.  Indeed, even if all $\ev{Q_i}\neq 0$, the original $U(1)$ gauge group is not fully broken, $U(1)\to \Z_n$.  If  the original theory has $q_J=\int F/2\pi \in \Z$, the rescaled theory \rescaling\ has $\tilde q_J=\int \tilde F/2\pi \in n \Z$.  The scalar dual to the photon  of the rescaled and original theories are related as $\tilde a = n a /n^2=a/n$ (the $1/n^2$ is from $\tilde e^2$), so while $a\cong a+2\pi$, the periodicity of $\tilde a $ is $2\pi /n$,  i.e.\  it is a $\Z _n$ orbifold.

\newsec{Aspects of vortices and monopole operators}

As discussed in the introduction, we are here interested in vortices and their connection with monopole operators in flat spacetime.  Many aspects of such operators were elucidated in \refs{\BorokhovIB, \BorokhovCG}, focussing on the theory on $\S^2\times \R$ for technical reasons.
Here we focus on flat $\R ^{1,2}$ spacetime.  A key physical difference is that theories on  non-compact space  can have multiple vacua, including continuous moduli spaces, which are superselection sectors.  We will  discuss some  additional superselection parameters.  Theories on compact space, e.g.\ $\S^2\times \R$, instead have a unique vacuum, with classical moduli integrated over\foot{Moreover, as in \FestucciaWS, to have unbroken supercharges on $\S^2\times \R$ requires a  background $U(1)_R$ gauge field $b^\mu$ leading to $\Delta \CL = -\half b^\mu J^{FZ}_\mu$, here with $b_0\sim -1/r$.  For the Euclidean theory $\S^2\times \R_E$ or $\S^2\times \S^1$, these backgrounds are imaginary  \FestucciaWS.}.

To insert a monopole operator, say at $x=0$, one removes a small ball from $\R ^3$ around the origin and imposes  boundary conditions on the field strength, as in \qjis,  around the hole,
 setting the charged matter fields to zero there, $Q_i=0$.  In our supersymmetric context, we are interested in monopole operators that are chiral, annihilated by half the supercharges.  In superspace, this chiral condition is written as \sigmader\ for $\Sigma$, which requires that $\sigma =\Sigma |$ satisfies
\eqn\approxdeq{-\partial _E^2 \sigma = 2\pi q_J \delta^{(3)}(x)}
with $\partial _E^2$ the 3d Euclidean Laplacian. So the solution near the hole is
\eqn\sigmanear{\sigma \approx  {q_J\over 2r}  ~~,}
 with $r$ the Euclidean radial distance to the hole.   Indeed,  in order for the configuration to be chiral, annihilated by half of the supercharges, requires the dimensional reduction of the 4d condition that the field strength be self-dual, which gives
 \eqn\sigmabc{r^2 \partial_r \sigma \Big|_{r\to 0}= - \half  q_J, }
 so the solution near the hole is again given by \sigmanear.

 As the hole becomes small and $r \to 0$, the field $\sigma$ at the hole, given by \sigmanear\ has $\sigma \to  \infty \sign (q_J)$.    The monopole operator insertion thus forces
the field $\sigma$ nearby to be out in an asymptotic infinity region  on the Coulomb branch.   This is intuitively satisfying: it explains why this operator is associated with the Coulomb branch.  The inserted operator of charge $q_J$ is chiral, $X_+^{q_J}$  for $q_J>0$, or $X_-^{|q_J|}$ for $q_J<0$.  Anti-chiral operators have opposite sign in \approxdeq, \sigmanear\ and \sigmabc; they are $(X_-^\dagger)^{q_J}$ for $q_J>0$ and $(X_+^\dagger)^{|q_J|}$ for $q_J<0$.

\subsec{Chiral operators vs.\  BPS states}

Monopole operators are chiral operators, whereas vortices are BPS states.  We here emphasize the general distinction between chiral operators vs.\  BPS states.

 The supersymmetry algebra is (see appendix A for details)
\eqn\zenop{\{Q_\alpha, \bar Q_\beta\} = 2\gamma^\mu_{\alpha\beta} P_\mu + 2i \epsilon_{\alpha\beta} Z,}
where the $SL(2,\R)$ Lorentz spinor has $\overline{(Q_\alpha)}= \bar Q_\alpha$. The energy $P^0=-P_0$ is positive.  The central element $Z$ gets contributions only  from the global (not gauge, see \AharonyBX\ for discussion), non-R, $U(1)$ charges and associated real parameters
\eqn\zisi{Z=\sum _i q_i m_i,}
where $q_i$ is the charge under the global $U(1)_i$ symmetry, and the sum includes the $U(1)_J$ global symmetry, with $m_J=\zeta$ the FI parameter.

It is convenient to define
\eqn\Qpmd{Q_\pm=\half(Q_1\pm i Q_2), \qquad \bar Q_\pm=\overline{Q_\mp}=\half(\bar Q_1\pm i \bar Q_2),}
such that $Q_\pm$ and $\bar Q_\pm$ have spin $\pm\half$.  In terms of these, the algebra \zenop\ is
\eqn\zenopm{\{Q_\pm, \bar Q_\pm\} =\pm i P_1 + P_2 , \qquad
\{Q_\pm, \bar Q_\mp\} =  P^0 \pm Z.
}

A chiral operator $\CO (x^\mu)$, at spacetime point $x^\mu$, preserves half the supercharges\foot{Note that we cannot pick  $\CO$ to be annihilated by one of the $Q$'s and one of the $\bar Q$'s: that would be  incompatible with Lorentz invariance.},
\eqn\opeq{[\bar Q_\alpha, \CO(x)]=0.}
The two $Q$'s puts the operator $\CO (x^\mu)$ in a four dimensional representation
\eqn\fourdr{\CO\qquad ;\qquad \Psi_\alpha(\CO)=[Q_\alpha, \CO] \qquad ; \qquad ~F(\CO)=\{Q,[Q,\CO]\}}
(If $\CO$ has spin, it is more complicated.)  If $\CO$ is complex, CPT means that this representation appears twice.   The equation of motion can reduce the representation, and can make it 2d.  For example: the bottom component of a free chiral superfield (even with real mass) satisfies the Klein-Gordon equation, the Fermions satisfy the Dirac equation, and $F=0$.

When $Z=0$, the representation is completely analogous to 4d $\CN =1$.   On the other hand, when a chiral operator $\CO$ is charged under a global non-R $U(1)_i$ symmetry, with real parameter $m_i$ non-zero, then $\CO$ creates states with $Z\neq 0$ in \zisi.  The representation is still four-dimensional \fourdr, but the algebra acts on this representation differently.

For a massive particle state,  we can go to the rest frame, $P_1=P_2=0$, $P^0=m$.
A BPS state with $Z>0$ is annihilated by $Q_-$ and its complex conjugate $\bar Q_+$ in \zenopm,
\eqn\bpsa{Z>0: \qquad  Q_-|BPS\rangle = \bar Q_+|BPS\rangle =0.}
The remaining two supercharges make a two-dimensional representation
\eqn\bpsp{Z>0: \qquad |BPS\rangle = \pmatrix{|a\rangle \cr |b\rangle }, \qquad \bar Q_-|a\rangle =0, \qquad |b\rangle = Q_+|a\rangle.}
The CPT conjugates have $Z<0$ and are thus anti-BPS states, with
\eqn\abpsa{Z<0: \qquad \bar Q_-|\overline{BPS}\rangle = Q_+|\overline{BPS}\rangle =0}
and with $Q_-$ and $\bar Q_+$ forming a two-dimensional representation
\eqn\tdhcpt{Z<0: \qquad |\overline{BPS}\rangle = \pmatrix{|\bar a\rangle  \cr |\bar b\rangle}, \qquad  Q_-|\bar a\rangle=0, \quad   |\bar b\rangle = \bar Q_+|\bar a \rangle .}
The R-charges and spins of these states are\foot{The state and its CPT conjugate have the same spin ($P$ and $T$ both flip the sign).  For example, for Chern-Simons-Maxwell theory, the spectrum includes a single kind of particle with spin ${\rm sign}(k)$, which is its own CPT conjugate.}
\eqn\spectrum{
\vbox{\offinterlineskip\tabskip=0pt
\halign{\strut\vrule#
%%%%%%%%%%%%%%%%%%
&~$#$~\hfil\vrule
%&~$#$~\hfil\vrule
&~$#$~\hfil\vrule
&~$#$~\hfil\vrule
&~$#$~\hfil\vrule
&~$#$~\hfil\vrule
&~$#$~\hfil%\vrule
&\vrule#
\cr
%%%%%%%%%%%%%%%%%
\noalign{\hrule}
&  & U(1)_R &  U(1)_{spin} & \; Z\cr
\noalign{\hrule}
%%%%%%%%%%%%%%%%%%
&  |a\rangle      &\;r& \quad s  & \;>0\cr
&|b\rangle  &\; r-1  & \; s+\half & \;>0 \cr
&  |\bar a\rangle      &\;-r& \quad s & \;<0  \cr
&|\bar b\rangle  &\; -r+1  & \; s+\half & \;<0\cr
}\hrule}}

The simplest example is a free chiral superfield with real mass $m$. The real mass $m$ couples to a global $U(1)$ symmetry under which the chiral superfield has charge $+1$. $Z$ is the product of this charge and $m$.  For positive $m$ we have two states with charge $+1$: a scalar and a Fermion with spin $+\half$ and their CPT conjugate states with charge $-1$: a scalar and a Fermion with spin $+\half$.  This corresponds to the states with $s=0$ in \spectrum.

To clarify the relation between the chiral operator $\CO$ \opeq\ and the BPS states, consider the state $\CO|0\rangle$.  While  $\CO|0\rangle$ is annihilated by $\bar Q_\alpha$, it is {\it not} a BPS state: it is not annihilated by any $Q_\alpha$.  The BPS states \bpsp\ are obtained by projecting to the lowest energy state in  $\CO|0\rangle$:
\eqn\BPSlimit{|a\rangle = \lim_{\tau \to \infty} e^{-(H-Z)\tau}\CO|0\rangle \qquad |b\rangle = Q_+|a\rangle,}
where the global charges $q_i$  of the operator $\CO$ must be such that \zisi\ has $Z>0$ for the limit \BPSlimit\ to give a non-zero state.  Similarly, upon projection to the lowest energy state,
anti-chiral operators with $Z<0$ can create anti-BPS states (the CPT conjugates of \BPSlimit)
\eqn\aBPSlimit{|\bar a\rangle = \lim_{\tau \to \infty} e^{-(H+Z)\tau}\overline\CO|0\rangle \qquad |\bar b\rangle = \overline Q_+|a\rangle.}

\subsec{BPS states from solitons}

In appendix A.3, we summarize aspects of BPS vortices for general $U(1)_k$ gauge theory with matter.  There are (anti) BPS field configurations when the FI parameter $\zeta \neq 0$, which are particles with $m=|Z|$, with $Z=q_J\zeta$.   The BPS field configurations  are annihilated by $Q_-$ and $\bar Q_+$ \bpsp, which acting on the Fermions give the BPS differential equations for the bosonic fields.  The (anti) BPS equations indeed require $Z>0$ ($Z<0$).  The vortex-soliton with charge $q_J=\pm 1$ always has a single normalizable zero mode supermultiplet.  The bosonic component gives the position of the BPS particle in the spatial plane.  Its Fermionic superpartner  is quantized and gives the BPS doublet \bpsp.

For $U(1)$ with a single matter field, these are the only zero modes.  With additional matter fields, there are restrictions on the vacuum in order to admit BPS configurations: essentially,  the vacuum must have a single field having non-zero expectation value.  In such configurations, each matter field has a zero mode supermultiplet.  The matter field with expectation value gives a normalizable zero mode with the same interpretation as above, giving the BPS particle position, and the doublet structure.  The additional matter fields contribute non-normalizable zero modes, whose interpretation is more subtle.  These issues are discussed further in appendix C.

A Chern-Simons term $k$ affects the Gauss' law constraint, giving
\eqn\Gaussk{{1\over e^2}\oint _\infty E_i dx^i \equiv q_{Gauss}=q_{elec}+k q_J,}
where $q_{elec}$ is the charge of the matter fields.  In the Higgs phase, $E_i/e^2\to 0$ in the IR, so $q_{Gauss}=0$ and thus vortices have electric charge (which is anyway screened) when $k\neq 0$,
\eqn\eleck{q_{elec}=-kq_J.}
This affects the spin of the vortices as in \refs{\WilczekDU, \LeePM}
\eqn\spinshift{\Delta s= -\half k q_J^2 . }

We will show in the next section that, when a Coulomb branch exists, the quantum numbers of a vortex coincide with those of the monopole operator in \onengen.  The Coulomb branch exists precisely when there is a spin zero, neutral vortex state, whose condensation can be interpreted as giving the Coulomb branch as a mirror Higgs branch.

\newsec{Examples of $U(1)$ gauge theories with moduli, and vortices}

We here discuss some examples with compact and non-compact Higgs or Coulomb branches, and aspects of the connection between the Coulomb branch and moduli operators.   For general $U(1)_k$ theory, with matter fields $Q_i$ of charges $n_i$, the theory with generic real parameters $m_i$ and $\zeta$ has only isolated vacua, with a mass gap; that case will be discussed in section 6, where we compute the Witten index.  In the present section, we discuss tuned values of $k$, $m_i$ and $\zeta$, that lead to moduli spaces.

As we noted after \kinfinity, the asymptotic regions $\sigma \to \pm \infty$ of the Coulomb branch, corresponding to $X_\pm$, only exist if $k_{eff}$ there vanishes, i.e.\  if $k=\mp k_c$.  As seen from \condos, a Coulomb branch, where all $Q_i=0$, requires $F(\sigma)=0$, i.e.\  both $k_{eff}=0$ and $\zeta _{eff}=0$, and using \zetakeff\ this requires a particular value of both $k$ and $\zeta$, to get $F(\sigma \to \pm \infty)=0$:
\eqn\cnoncom{\hbox{non-compact Coulomb,}\  X=X_\pm : \qquad\hbox{for}\qquad k=\mp k_c, \quad \hbox{and}\quad \zeta = \mp \half \sum _i |n_i| m_i.}
  Semi-classically, the Coulomb branch $X_+$ in \cnoncom\ has $\sigma \in (\sigma _{Q_i}^{max}, +\infty)$, where $\sigma ^{max}_{Q_i}$ is the largest of the $\sigma _{Q_i}$ in \sigmaiis, and likewise $X_-$ has $\sigma \in (-\infty, \sigma _{Q_i}^{min})$.   There will instead be a compact Coulomb branch if $k$ and $\zeta$ are tuned such that the function $F(\sigma)$ in \zetakeff\ vanishes in an intermediate region for $\sigma$, in between two $\sigma _{Q_i}$ values.

There will be a non-compact Higgs branch moduli space if two (or more) $\sigma _{Q_i}$, with opposite sign $n_i$, are tuned to coincide.  There will be a compact Higgs branch if two (or more) $\sigma _{Q_i}$, with the same sign $n_i$, are tuned to coincide and $\zeta$ is chosen such that $\sign (F(\sigma _{Q_i}))=\sign (n_i)$, to satisfy \condos.

We now illustrate these moduli spaces in simple, characteristic examples.

\subsec{$U(1)_k$ with one field, $Q$, of integer charge $n>0$}

For the $n=1$ case, we refer to this theory as $N_f=\half$ ($N_f=1$ has $Q$ and its charge-conjugate, $\tilde Q$).  The quantization condition \consk\ is $k+\half n\in \Z$; so $n=1$ has $k+\half \in \Z$, hence $k\neq 0$.    As we will discuss, the general $n$ case is a $\Z _n$ orbifold of the $n=1$ case.

\bigskip
\centerline{\bf Vacua}

The theory has a single real parameter, which we take to be $\zeta$ (using \ginva\ to set $m_Q=0$).    The effective CS and FI terms are \zetakeff
\eqn\kren{k_{eff}=k+\half n^2\sign (\sigma), \qquad \zeta _{eff}=\zeta.}
For generic $k$ and $\zeta$, the theory has a mass gap, with isolated supersymmetric vacua.  As in \cnoncom, there is a Coulomb branch solution of \condos\ if $k_{eff}=\zeta _{eff}=0$, i.e.\
\eqn\coulombi{\hbox{non-compact Coulomb,}\  X=X_\pm : \qquad\hbox{for}\qquad k=\mp \half n^2, \quad \hbox{and}\quad \zeta = 0.}
There are conserved $U(1)_R$ and $U(1)_J$ global symmetries with charges as in \onengen: the Coulomb modulus $X$ has $R(X)=\half n$, and $q_J(X)=-\sign (k)$ (we take $R(Q)=0$, as in \onengen, so $\ev{Q}\neq 0$ does not break $U(1)_R$).

The $n=1$ case of \coulombi, with $|k|=\half$,  is dual  to a single, free chiral superfield \refs{\DimofteJU, \BeniniMF, \BeemMB}, which fits with $R(X)=\half$.   The duality requires that the K\"ahler potential for $X$ be smooth at the origin (i.e. in the region $|X|^2\ll e^2$)
\eqn\effK{K(|X|) \sim \cases{
e^2 \Big[\log (|X|)\Big]^2 & $|X| \to \infty $\cr
|X|^2 & $|X| \to 0$ ~~}}
where the $U(1)_J$ symmetry implies that $K$ only depends on $|X|$ and we included the large $|X|$ dependence that's known from the perturbative $U(1)$ limit.  Since the duality maps the global symmetry $U(1)_J\to U(1)_X$, it maps $\zeta\to m_X$: the FI parameter $\zeta$ of the $U(1)$ theory to the real mass parameter $m_X$ for the free superfield $X$ of the dual. For
$\zeta =m_X=0$,  the dual  theories have the same, non-compact moduli space. Turning on $\zeta \neq 0$, or $m_X\neq 0$ in the dual, gives $\Tr (-1)^F=1$ supersymmetric vacuum.

The Coulomb branch theory \coulombi\ for integer $n>1$ is, as discussed after \rescaling, a $\Z _n$ orbifold of the theory with $n=1$.    The Coulomb coordinate can be written in terms of $\tilde X$ which has the quantum numbers of $X^{1/n}$, e.g.\ it has the $U(1)_R$ charge of a free field, but with a $\Z _n$ phase identification.   The local theory for $n>1$ is free, but there are twisted sector, codimension two, line-operator fields associated with the orbifold.

\bigskip
\centerline{\bf Vortices and monopole operators}

We take $Q$ to have $U(1)$ charge $n=1$. Taking $\zeta >0$, there is a Higgs vacuum, $|Q_1|^2=\zeta /2\pi $, breaking $U(1)$.  This is the $\CN=2$ version of the classic Abelian Higgs model, which has ANO vortices and has been much discussed (with some debates) in the literature; see e.g.\ \refs{\GoldhaberKN, \LeePM,  \MezincescuGB}. The necessary presence of the Chern-Simons term $k\neq 0$, since $k+\half \in \Z$, has important effects.  As in the analogous case of 2d LG theories  \refs{\FendleyVE, \FendleyDM} the BPS solitons are doublets,  and the spins are half-integral rather than semionic.

  The fields for the static soliton with winding number $q_J$ have the standard vortex asymptotics far from the core \LeeEQ,
\eqn\qbcs{\lim_{z\to \infty} Q_1 = \sqrt{{\zeta \over 2\pi}}e^{iq_J\theta }+\dots \qquad A_\theta =q_J+\dots, \qquad\hbox{(large $|z|$)}}
where $z=x+iy=|z|e^{i\theta}$ is the space coordinate, and   the  $\dots$ are higher order in $1/|z|$.  The deviations from the asymptotic behavior \qbcs\ is here exponentially suppressed:
since the theory is in the Higgs phase, the magnetic flux is confined by the Meissner effect to a thin flux tube, with core size set by the photon mass.

For the $\CN =2$ theory, the self-duality equations \LeeEQ\ are
the BPS soliton equations \LeePM.   The vortex with $U(1)_J$ charge $q_J=1$ has a zero mode which forms one chiral superfield particle degree of freedom, with complex Bose component associated with the position of the vortex in the spatial plane.  The complex Fermi zero mode superpartner is $\delta _+$, and complex conjugate $\delta _-$, associated with the broken supersymmetries $Q_+$ and $\bar Q_-$.  The $\delta_+$ zero mode has R-charge $+1$ and spin $-\half$ and its complex conjugate, from $\bar \delta_-$, has the opposite charges.  The quantization of $\delta _+$ and $\bar \delta _-$ leads to the BPS doublet of states  $|a\rangle$ and $|b\rangle$ in \spectrum,   with charges (using \spinshift)
\eqn\SUSYzmc{
\vbox{\offinterlineskip\tabskip=0pt
\halign{\strut\vrule#
%%%%%%%%%%%%%%%%%%
&~$#$~\hfil\vrule
&~$#$~\hfil\vrule
&~$#$~\hfil\vrule
&~$#$~\hfil\vrule
&~$#$~\hfil
&\vrule#
\cr
%%%%%%%%%%%%%%%%%
\noalign{\hrule}
&  &U(1)_{spin} &   U(1)_J& U(1)_R  \cr
\noalign{\hrule}
%%%%%%%%%%%%%%%%%%
&  |a\rangle       & \;-\half k - {1\over 4}  &\; 1    &\; {1\over 2}\cr
&  |b\rangle   & \; -\half k +{1\over 4}    & \; 1  & \; -{1\over 2}\cr
\noalign{\hrule}
}}}
Because we consider compact $U(1)$ (as needed for vortices), the $k$ quantization condition \consk\ is $k\in \Z +\half$. So the spin in \SUSYzmc\ is always half-integral or integral, not semionic.  The solitons \SUSYzmc\ have BPS mass $m=|\zeta|$.

We have seen that the theory with $k=-\half$ has a gauge-neutral operator $X_+$, which is a chiral superfield with zero spin and $R=\half$.  For $\zeta =0$, $X_+$ is massless and there is an associated non-compact Coulomb branch.   For $k=-\half$, we can indeed identify $|a\rangle$ with the projection \BPSlimit\ of $X_+|0\rangle$, since $|a\rangle$ has the same quantum numbers \SUSYzmc\ as $X_+$ for $k=-\half$.  The state $|b\rangle$ in this case is similarly obtained from the Fermionic superpartner $\Psi (X_+)|0\rangle$.    The conjugate states $|\bar a\rangle$ and $|\bar b \rangle$ are likewise created by the anti-chiral operator $X_+^\dagger$ and its Fermionic partner.  For $k=+\half$, on the other hand, the state $|b\rangle$ in \SUSYzmc\ has spin 0, with $R=-\half$.  In this case, the  state $|b\rangle$ has the quantum numbers to be identified with the projection of $X_-^\dagger |0\rangle$, and $|a\rangle$ is created by the Fermionic parter of $X_-^\dagger$.

For $|k|\neq \pm \half$, the vortex has nonzero charge (charge is anyway screened in the Higgs phase) \eleck\ and spin $s\neq 0$ \SUSYzmc.   Therefore, they cannot condense without spoiling rotational symmetry.  This fits with the fact seen from \onengen\ that $X_\pm$ are only gauge neutral operators if $|k|=|k_c|=\half$, so the Coulomb branch only exists in that case.

\subsec{$U(1)_k$ with $N$ fields $Q_i$ of charge $n_i=+1$}

The quantization condition \consk\ here requires $k+\half N\in \Z$.  Using \ginva, we take  $m_i$ in the Cartan subalgebra of the $SU(N)$ global symmetry, $\sum _i m_i=0$.

\bigskip
\centerline{\bf Vacua}

There are Higgs vacua, for any $k$, at $\sigma _{Q_i}=-m_i$ if $\zeta _{eff}>0$.
For generic $m_i$, these are isolated SUSY vacua, with a mass gap.  Upon tuning two or more $m_i$ to coincide, they instead lead to a compact Higgs branch moduli spaces of vacua.   Taking all $m_i=0$, the theory with $\zeta >0$, and any $k$ has a compact Higgs branch $\CM _H=CP^{N-1}$.

Now consider Coulomb vacua.  Non-compact Coulomb branches \cnoncom\ $X_\pm$ exist for $k=\mp \half N$, and $\zeta =0$. For $|k|<\half N$, and $m_i$ chosen appropriately and $\zeta$ appropriately tuned, there is a compact $CP^1$ Coulomb branch, for $\sigma$ between the neighboring values of $\sigma_{Q_i}=-m_i$ where $k_{eff}=0$. Consider e.g.\ $N=2$ fields, with $m_1=-m_2=m>0$.
Then \zetakeff\ gives
\eqn\exkeff{F(\sigma)=\zeta _{eff}+k_{eff}\sigma = \zeta +k\sigma +\half |\sigma +m|+\half |\sigma -m|.}
For $k =1$ (or $k=-1$) and $\zeta =0$, there is a non-compact Coulomb branch \cnoncom\  with $\sigma <-m$ ($\sigma >m$).  For $k=0$, at $\zeta =-m$, there is a compact Coulomb branch, $|\sigma |<m$, giving $\CM _C\cong CP^1$ upon including the compact scalar $a$ dual to the photon \DoreyRB.   The local theory on this $CP^1$ Coulomb branch is everywhere smooth.  In particular, at $\sigma = -m$ (or $\sigma = m$), the metric is smooth as in \effK, as the low-energy theory there is $U(1)_{k_{low}}$, with $k_{low}=- \half$ and one charged field $Q$.

For generic $m_i$, the non-compact or compact Coulomb branches mentioned above are all smooth, and free.  Upon tuning two or more $m_i$ to coincide, the low-energy theory has a singularity associated with additional massless degrees of freedom.  To illustrate it, consider taking all $m_i=0$, so \zetakeff\ gives  $k_{eff}=k+\half N \sign (\sigma )$ and $\zeta _{eff}=\zeta$.  There is a non-compact Coulomb branch $X_+$ (or $X_-$) for $k=-\half N$ ($k=\half N$) and $\zeta =0$, with $\sigma \geq 0$ ($\sigma \leq 0$).  The theory at the origin, $X=0$, is singular for $N>1$, i.e.\  there are additional degrees of freedom there; for general $N$ and $k$, it  is an interacting SCFT.

\bigskip
\centerline{\bf Solitons (for $m_i=0$)}

For $\zeta >0$ and any $k$, the vacuum is on the compact Higgs branch $\CM _H=CP^{N-1}$; by an $SU(N)$ rotation, we take the vacuum to be $Q_i|=\sqrt{\zeta /2\pi}\delta _{i, 1}$.  This theory has BPS vortices, where $Q_1$ and $A_\theta$ have the same field configuration solution as in the $N=1$ case, with asymptotic behavior \qbcs\  far from the core.  The solution has the same, associated, normalizable chiral superfield zero mode, giving the particle's spatial position and super-doublet structure \bpsp.  The matter superfields, $Q_{i>1}$, provide $N-1$ additional chiral superfield zero modes, satisfying $D_z Q_i=0$, with charge $n_i=1$ coupling to the winding gauge configuration $A_\theta$ of the vortex.

As discussed in appendix C, these additional zero modes are non-normalizable.  The bosonic components are frozen, superselection sector parameters. The question then arises whether or not to quantize the non-normalizable Fermion zero modes.  As we discuss in appendix C, the Fermi components are to be quantized, but map between different Hilbert spaces, so they affect the vortices' quantum numbers, but not the number of states.

This $U(1)$ linear sigma model gauge theory is the UV completion of the low-energy $CP^{N-1}$ non-linear sigma model.  The $U(1)$ vortices are, correspondingly, the UV completion of the Skyrmionic lump solutions \WardIJ\ of the low-energy $CP^{N-1}$ sigma model, associated with field configurations that wrap the non-trivial two-cycles in the moduli space,  $\Sigma _2 \supset \CM _H$, with  size $\int _{\Sigma _2}\omega = \zeta$ calibrated by K\"ahler form $\omega$, around the two dimensional space surrounding the soliton.  Indeed, there are BPS solutions that interpolate between the ANO vortices and the $CP^{N-1}$ Skyrmionic lumps; see e.g.\ \refs{\VachaspatiDZ\HindmarshYY- \LeeseFN}.

Following the prescription in appendix C, the  upshot is that there are  BPS vortices as in \spectrum, with
$U(1)_R$ charge $r=\half N$ and $U(1)_{spin}$, $s=\half (k\pm \half N)$.  When $k=\mp \half N$, we find the spin $0$ states, matching the quantum numbers of $X_+$ and $X_-^\dagger$, which are neutral, matching the existence of non-compact Coulomb branches for $|k|= \half N$.

When there is a compact Coulomb branch, we can likewise consider BPS Skyrmions which wrap the compact Coulomb vacua.  Consider e.g.\ the $N=2$ case with real masses $m$ \exkeff, where for $k=0$ there is a compact Coulomb branch for $|\sigma|\leq m$, $\CM _C\cong CP^1$.      The BPS mass of the Skyrmion is given by the size of the $CP^1$, which is calibrated by a K\"ahler form which is non-renormalized, see \refs{\deBoerKR, \AganagicUW}\ and appendix B.  So the BPS Skyrmion mass is
\eqn\mSkyrm{m_{BPS}= \int \Omega = \int {1\over 4\pi} d\sigma _i da^i= |m|.}
We will discuss an analogous setup in the next subsection, which will be more interesting, in that the Coulomb branch Skyrmions map to the quanta of the Higgs branch meson operator.  In the present case, this does not happen, as there is no non-compact Higgs branch and, correspondingly, no gauge-invariant analog of the meson chiral operator, because here $Q_1$ and $Q_2$ both have positive charge.  We identify the topological charge of the Skyrmion in the  present example with a $U(1)\subset SU(2)_F$; this interpretation is consistent with \mSkyrm.

\subsec{$U(1)_k$ with $N_f=1$ flavor of matter, $Q$ and $\tilde Q$, of charge $\pm 1$}

We take $m_Q=m_{\tilde Q}= m\geq 0$.   The function $F(\sigma )$ is
\eqn\exkeffii{F(\sigma)=k_{eff}\sigma +\zeta _{eff}=\zeta +k\sigma +\half |\sigma +m|-\half |\sigma -m|.}

\bigskip
\centerline{\bf Vacua}

For $m=0$, there is a non-compact Higgs branch moduli space $\CM _H$ of SUSY vacua, labeled by arbitrary $\ev{M}$ with $M=Q\tilde Q$, for all $k$ and $\zeta$.  The metric on $\CM _H$ depends on $\zeta$, e.g.\ it can be singular at $M=0$ for $\zeta=0$, where there is an interacting SCFT for all $k$.   For $m\neq 0$, $\CM _H$  is lifted, and there is instead an isolated supersymmetric Higgs vacuum at $\sigma = -m$ (or $\sigma =+m$), for $\zeta$ positive (or negative).  The low-energy theory
there  is $U(1)_{k_{eff}}$ with light field $Q$ (or $\tilde Q$) and $k_{eff}=k-\half$ (or $k_{eff}=k+\half$) from integrating out the other field.

Now consider the topological or Coulomb vacua, where \exkeffii\ gives $F(\sigma)=0$.  For $k\neq 0,-1$, there are no Coulomb vacua, only isolated topological vacua, for any $m$.
For $k=0$, there are non-compact Coulomb branches where $\sigma \to \pm \infty$, labeled by $X_\pm$,  for $\zeta = \mp m$.  The $X_+$ non-compact Coulomb branch starts at $\sigma =m$, where $Q$ is massless and $\tilde Q$ can be integrated out, leading to $U(1)_{-\half}$ there: this is the theory of subsection 4.1, which is dual to a smooth, free field theory $X$.  Likewise, the $X_-$ non-compact Coulomb branch, for $\sigma <-m$, comes from the CP conjugate theory, $U(1)_\half$, with massless field $\tilde Q$ at $\sigma =-m$.  Upon tuning $m\to 0$, $X_\pm$, merge together and the theory at the origin is an interacting SCFT, dual to $W=MX_+X_-$ \AharonyBX.

For $k=-1$, there is a compact Coulomb branch for $\zeta =0$, with $F(\sigma)=0$ for  $|\sigma|\leq m$.  Including the dual photon, this $\CM _C\cong CP^1$.   The $CP^1$ is everywhere smooth; in particular, the local theory at $\sigma =\pm m$ is the $U(1)_{k=\mp \half}$ theory, dual to a free field.

\bigskip
\centerline{\bf Solitons}

There is a Higgs-branch BPS vortex for non-zero $\zeta$ and any $k$, $m$.  For $m\neq 0$, where the Higgs vacua are isolated with a mass gap, the vortex is essentially that of the low-energy $N_f=\half$ theory \qbcs.  For $\zeta$ positive (negative), the light field $Q$ (or $\tilde Q)$ has winding expectation value \qbcs\ in the low-energy theory at $\sigma =-m$ (or $\sigma =-m$).  The other field has real mass, and does not contribute any additional Bose or Fermi zero modes.

For $m=0$, for any $k$, there is a BPS vortex for any non-zero $\zeta$, but only for one, special vacuum on the noncompact Higgs branch.  For $\zeta >0$ it is the vacuum with $|Q|^2= \zeta /2\pi$, with $\tilde Q=0$.  The BPS soliton has $Q=Q_1$ and gauge field as in the  $N_f=\half$ theory, \qbcs.   There is  the same, normalizable Bose and Fermion zero modes as the $N_f=\half$ theory, associated with the super-translations of the soliton core.  There is a complex, non- normalizable Bose and Fermion zero mode associated with $\tilde Q$.  Following the prescription in appendix C, this mode gives superselection sectors, which contributes to the vortex quantum numbers in the spectrum \spectrum,  but not its number of states.

There is also a Coulomb branch Skyrmion for $k=-1$, $\zeta =0$, where $\CM _C\cong CP^1$, since $\pi _2(\CM _C)=\Z$.  As discussed in the previous subsection, for Higgs-branch Skyrmions, $q_J$ is given by the $\pi _2(\CM _H)$ winding number.  We have here a mirror-dual situation, where now $\pi _2(\CM_C)$ gives the charge $q_M=\half q_A$ under the global $U(1)$ symmetry which assigns charge 1 to $M$ (so $\half$ to $Q$ and $\tilde Q$).  In particular, the basic Skyrmion wrapping the $\CM _C$ Coulomb branch has the same quantum numbers as the operator $M$, so can potentially be identified as the BPS state associated with the chiral operator $M$.  The BPS Skyrmion has mass given as in \mSkyrm.  It becomes massless as $m\to 0$, and it can there condense and be interpreted in terms of the non-compact Higgs branch which exists for $m=0$.

\subsec{$U(1)_k$ with $N_+$ matter fields $Q_i$ of charge $+1$, and $N_-$ matter fields of charge $-1$.}

The CS term satisfies \consk, $k + \half (N_++N_-)\in \Z$.  Taking the real masses to vanish,
the global symmetries are as in \onensqed.  There is a non-compact Coulomb branch,  parameterized by $X_\pm$, for $k=\pm \half (N_+-N_-)$, and $\zeta =0$. There is a Higgs branch $\CM _H$, of complex dimension ${\rm dim} _C \CM _H=N_++N_--1$.  For $N_+N_-=0$, $\CM_H$ is compact and only exists for $\zeta$ of correct sign.  For $N_+N_-\neq 0$,  the Higgs branch is non-compact, and can be parameterized by $M^i_{\tilde i}= Q^i\tilde Q_{\tilde i}$ with rank 1,
\eqn\Mcons{\epsilon_{ij...}\epsilon^{\tilde i\tilde j...}M^i_{\tilde i}M^j_{\tilde j} =0 ~.}
When $\zeta _{eff}\to 0$, the Higgs branch can touch the origin, $Q_i=\tilde Q_{\tilde i}=0$, where it can connect with the Coulomb branch, when it exists.  For $\zeta _{eff}\neq 0$, the Higgs branch fields are away from the origin, and there can be BPS solitons on special points or subspaces of $\CM _H$.

We can turn on real masses $m_i$ for $Q^i$, and $m^{\tilde i}$ for $\tilde Q_{\tilde i}$, in the Cartan subalgebra of $SU(N_+)\times SU(N_-)\times U(1)_A$.   Let us consider the case $m_{Q_i}=m_{\tilde Q_{\tilde i}}=m$, i.e.\  a background value for $U(1)_A$ only; by using $P$, we can take $m>0$ without loss of generality.   Using \zetakeff, we find
\eqn\kzetax{\eqalign{k_{eff}&=k+\half N_+ \sign (m+\sigma)+\half N_-\sign (m-\sigma), \cr
\zeta_{eff}&=\zeta +\half mN_+ \sign (m+\sigma)-\half m N_-\sign (m-\sigma).}}
There is a Coulomb branch when $k_{eff}=\zeta _{eff}=0$.  So there is a  non-compact Coulomb branch $X_+$, for $\sigma >m$, if $k=-\half (N_+-N_-)$ and $\zeta =-\half m (N_++N_-)$, as then $k_{eff}=\zeta _{eff}=0$.  Likewise, there is a non-compact Coulomb branch $X_-$, for $\sigma <-m$, if $k=\half (N_+-N_-)$ and $\zeta =\half m (N_++N_-)$.  The global charges of $X_\pm$ are as in \onensqed.

There is also a compact Coulomb branch, with $|\sigma |<m$, if $k=-\half (N_++N_-)$, and $\zeta = -\half m (N_+-N_-)$.   The low-energy theory at $\sigma =-m$, after integrating out the massive $\tilde Q_{\tilde i}$ fields, has $N_+$ massless fields $Q^i$ of charge $+1$, and $k_{low}=-\half N_+$.   This low-energy theory is smooth at $\sigma =-m$ only for $N_+=1$, otherwise there are additional degrees of freedom.  The low-energy theory at $\sigma =+m$ likewise has $N_-$ massless $\tilde Q_{\tilde i}$ fields of charge $-1$, with $k_{low}=-\half N_-$, and is smooth there if $N_-=1$.

\subsec{$U(1)$ with $N_+=N$ fields $Q_i$ of charge $+1$, and $N_-=1$ field\foot{These theories arise  via $M$ theory on local fourfolds with  $G$ flux \IntriligatorUE.} $\tilde Q$ of charge $-1$.}

These theories are a special case of those of the previous subsection, and the discussion of the Coulomb branches there directly apply here.  The nice aspect of this class is that the Higgs branch
simplifies, as it is simply given $N$ unconstrained moduli $M_i=Q_i \tilde Q$.
The classical metric on ${\cal M}_H$ can be found from
\eqn\kclis{K_{cl}=Q^\dagger _i e^V Q^i +\tilde Q^\dagger e^{-V}\tilde Q-\zeta V}
upon integrating out $V$, which gives
\eqn\kclii{K_{cl}=\sqrt{\zeta ^2+4M_i^\dagger M^i} +\zeta \log (\sqrt{4 M_i^\dagger M^i+\zeta ^2}-\zeta ) .}
For $\zeta =0$, $K_{cl}=2\sqrt{M_i^\dagger M_i}$ is singular at the origin, while for $\zeta \neq 0$ $\CM _H$ is smoothed out.

 For $\zeta <0$, $\CM _H$ is topologically trivial, $\CM _H\cong \C ^N$.  For $\zeta >0$, on the other hand,  $c_1(\CM _H)\neq 0$ and there is a K\"ahler form that calibrates a two-cycle in $\CM _H$, with volume $\sim \zeta $, similar to the Fubini-Study K\"ahler potential of $CP^{N-1}$.  These two cases can be seen from the behavior of \kclii\ near the origin:
\eqn\kexp{K_{cl}(M_i^\dagger M ^i\ll \zeta ^2)\approx \cases{{1\over \zeta }M_i^\dagger M^i+\zeta \log (M_i^\dagger M^i) & $\zeta >0$\cr {1\over |\zeta |}M_i^\dagger M^i & $\zeta <0$}}
with both cases smooth at the origin for $\zeta\neq 0$.  Far from the origin, \kclis\ gives
\eqn\kexp{K_{cl}(M_i^\dagger M^i\gg \zeta ^2)\approx 2\sqrt{M_i^\dagger M^i}+\half \zeta \log ( M_i^\dagger M^i).}
The $\zeta \log(M_i^\dagger M^i)$ term can be eliminated by a K\"ahler transformation for $N=1$, but leads to a non-zero contribution to the K\"ahler metric for $N>1$.  Quantum effects will of course locally modify the classical metric, but we expect that the full quantum metric on $\CM _H$ has the same topology, with a non-trivial two-cycle, of size $\sim \zeta$ for $\zeta >0$.  So the $\zeta >0$ case has BPS Skyrmion, similar to those of the $CP^{N-1}$ sigma model.

\subsec{$U(1)_0$ with $N_f=N_+=N_-=2$, with nonzero real masses}

We recall this case \AharonyBX\ to illustrate the possibility of a moduli space with multiple components, both compact and non-compact. Taking $m_{Q_1}=-m_{\tilde Q_1}=-m_{Q_2}=m_{\tilde Q_2}=m>0$, and $\zeta =0$, the moduli space has two singular points, $\sigma =\pm m$, each of which has a low-energy $U(1)_{k=0}$ with $N_f=1$ light flavor, i.e.\ the $W=MX_+X_-$ SCFT.    The moduli space is similar to Fig. 1.  The theory is dual to \AharonyBX\
\eqn\wex{W=-M_1^1 X_{1+}X_{1-}-M_2^2 X_{2+}X_{2-} +\lambda (X_{1+}X_{2-}-1),}
with $\lambda$ a Lagrange multiplier.  $X_{1-}$ and $X_{2+}$ parameterize the Coulomb branches that look like non-compact cones. The compact $\CM _C\cong CP^1$ region in between has local coordinates $X_{1+}$ and $X_{2-}$ near the two points where the $CP^1$ intersects the non-compact Higgs and Coulomb branches ($M_1^1$ and $X_{1-}$, or $M_2^2$ and $X_{2+}$, respectively).

Since the low-energy theory at $\sigma =-m$ and $\sigma =+m$ are each a copy of $U(1)_0$ with $N_f=1$, each has corresponding Higgs branch BPS vortices for $\zeta \neq 0$.  As $\zeta \to 0$, the vortices on the $M_1^1$ Higgs branch can condense to give the $X_{1\pm }$ Coulomb branch moduli.  Likewise, vortices on the $M_2^2$ branch can condense to give the $X_{2\pm}$ moduli.   On the other hand, the chiral monopole operators are defined via \sigmader\ independent of the vacuum, which leads to   \sigmapush, $\sigma \to \pm \infty$,  so always corresponds to non-compact Coulomb regions; specifically here, $X_{1-}$ or $X_{2+}$.

We can also consider Skyrmion solitons wrapping the compact $CP^1$ part of the Coulomb branch, though the local theory at $\sigma =\pm m$ is not smooth, but contains the additional degrees of freedom of the interacting SCFT there.  This soliton has the quantum numbers of $M_1^2$ or $M_2^1$.

\newsec{Generalities about the Witten index}

We here recall a few standard aspects about the Witten index \WittenDF, $I=\Tr (-1)^F$ of SUSY theories in $D$ spacetime dimensions. The index  can be computed by reducing the theory to SQM on $\T^{D-1}\times \R$, or from the partition function on Euclidean $\T^D$, with periodic boundary conditions for the Fermions in every $\S^1$.   The index is ill-defined if the theory on $\R ^{1,D-1}$ has a continuous, non-compact moduli space of vacua.  Such vacua are superselection sectors of the theory in $D\geq 3$ non-compact spacetime dimensions, but are integrated over when the theory is put on $\T^{D-1}$, leading to an ill-defined $\Tr (-1)^F$.

\subsec{The index for 3d SUSY CS theories with no matter}

Consider first pure,  $\CN =0$ Chern-Simons theory, with gauge group $G$ and Chern-Simons coefficient $k$, without the Maxwell term, gauginos, or matter, on a spatial torus $\T^2$.  This theory has a moduli space of flat connections that are zero energy states and, upon quantization, are related \refs{\WittenHF, \ElitzurNR} to the conformal blocks of the 2d $G_k$ WZW theory.  The number of vacua is thus $J_G(k)\equiv$ the number of conformal blocks of $G_k$ WZW theory.
This is also the number of vacua of the theory with the Maxwell term added, since the
 Maxwell term is irrelevant in the IR compared with the Chern-Simons term.

The $\CN =1$ supersymmetric version of this theory, without matter, was analyzed in \WittenDS.  The result is that the vacua counted by $J_G(k)$ count the supersymmetric groundstates, once one accounts for a shift of the level from integrating out the gauginos\foot{There is also a $+h$ contribution to the WZW level from  integrating out gluons, but it was stated in \WittenDS\ that such
bosonic contributions are already accounted for, without shifting $k'$.   This subtlety was debated and analyzed in detail in  \SmilgaUY\ and references therein.  The upshot is that the original result of \WittenDS\ is correct: only the Fermionic shifts to $k'$ should be included. See also  \HenningsonVB.}:
\eqn\indexnomat{\Tr (-1)^F\equiv I(k)=J_G(|k'|), \qquad |k'_{\CN =1}|= |k|-\half h ,\qquad |k'_{\CN =2}|=|k|-h,}
($h=N$ for $SU(N)$).  The index of pure ${\cal N}=2$ gauge theory without matter is then
\eqn\indexnom{\Tr (-1)^F=J_G(|k|-h),}
with supersymmetry broken\foot{For $\CN =1$, SUSY is broken for $|k|<\half h$ \WittenDS.}  for $|k|<h$.

For $U(1)_k$, the number of conformal blocks is $J_{U(1)}(|k|)=|k|$, as seen by bosonizing the $U(1)$, with $k$ vertex operators $V_q=e^{iq\phi /\sqrt{k}}$, $q=0,\dots ,k-1$.   So \indexnom\ gives
\eqn\indexci{U(1)_k\qquad\hbox{with no matter}: \qquad \Tr (-1)^F=\cases{ |k| & for $k\neq 0$\cr {\rm ill~defined}& for $k=0$.}}
For $k\neq 0$, there is a mass gap and the single classical SUSY vacuum at $\sigma =0$ acquires a topological multiplicity of $|k|$.   For $k=0$, the index is ill-defined, because of the unlifted Coulomb branch.

For $SU(N_c)_k$, the number of $SU(N_c)_{k'}$ conformal blocks is
\eqn\jksun{J_{SU(N_c)}(k')={(N_c+k'-1)!\over (N_c-1)! k'!}.}
For example, $J_{SU(2)}(k')=k'+1$, from the $SU(2)_{k'}$ representations with $j=0, \dots \half k'$.
Accounting for the  shift \indexnom, this gives for the index
\eqn\sunind{\CN =2 \qquad SU(N_c)_k:  \qquad I(k)=\cases {{(|k|-1)! \over (N_c-1)! (|k|-N_c)!}& for $|k|\geq N_c$\cr 0\ \hbox{(gapped)} & for $0<|k|<N_c$\cr
0 \ \hbox{(runaway)} & for $k=0$.}}
For $|k|=N_c$, strictly speaking, the above argument does not work ($k'=0$ incorrectly suggests
the absence of a gap), but the result \sunind\ is nevertheless applicable, giving the correct (as we will show) result that  $\Tr (-1)^F=1$ in this case.  For $0<|k|<N_c$
the vanishing index (along with our results in the following sections) is consistent with supersymmetry being  broken in a stable vacuum, with a massless Goldstino, and a mass gap for the other fields.  For $k=0$ there is the non-gapped runaway, with no stable vacuum, as in the $SU(2)$ case \AffleckAS.

\newsec{Computing the Witten index for $U(1)_k$ with matter}

We now compute the Witten index for $U(1)$ gauge theory, with CS term $k$, with matter fields $Q_i$ of charge $n_i$, and generic real parameters, $m_i$ and $\zeta$.   The vacuum solutions of \condos\ are then isolated, with a mass gap: the Higgs and topological vacua, which contribute
\eqn\types{\eqalign{U(1)_{k_{eff}}\ \hbox{``topological"}:\qquad \Tr (-1)^F |_{\sigma = \sigma _I}&=|k_{eff}(\sigma _I)| \cr
\ev{Q_i}\neq 0 \ \hbox{``Higgsed"}: \qquad \Tr (-1)^F|_{\sigma =\sigma _{Q_i}}&=  s_i n_i^2,}}
see \higgsmaybe.
 The topological vacua, at $\sigma = \sigma _I$ where $F(\sigma _I)=0$ \sigmacapi, have $|k_{eff}|$ vacua, the topological multiplicity  \indexci\ of the low-energy $U(1)_{k_{eff}}$ theory with no matter.   The Higgsed vacua, at $\sigma = \sigma _{Q_i}$ \sigmaiis\ where $m_{Q_i}(\sigma)=0$,
 exist if $s_i=1$ in \higgsmaybe, as then $\zeta_{eff}$ has the right sign for  $\ev{Q_i}\neq 0$.
Now $\ev{Q_i}$ breaks $U(1)\to \Z _{n_i}$, with $n_i$ the $U(1)$ charge of $Q_i$, and the associated $\Z _{n_i}$ orbifold leads to a Higgs vacua multiplicity\foot{In $D$ spacetime dimensions, a $\Z _n$ orbifold gives a $n^{D-1}$ multiplicity, from the twists on  $\T ^{D-1}$.} of $n_i^2$.

 The total index is obtained by summing the contributions \types\ over all the vacua
 \eqn\indexch{\Tr (-1)^F=\sum _{\sigma _{I}} |k_{eff}(\sigma _{I})|+ \sum _{\sigma _{Q_i}} s_i n_i^2.}
The individual terms \types\ generally depend on the real parameters, $m_i$ and $\zeta$, but the sum \indexch\ independent of the real parameters.   As we vary $m_i$ and $\zeta$, the vacua $\sigma _I$ and  $\sigma _{Q_i}$ can move around and collide, leading to phase transitions.  We have already given a general argument, in the introduction,  that $\Tr (-1)^F$ must be invariant under deformations of the real parameters $m_i$ and $\zeta$, since they become complexified upon compactifying on $\S^1_R$.

 We will explicitly verify the $m_i$ and $\zeta$ independence of \indexch, and evaluate the sum, in appendix D. The final result is (in terms of the critical $k_c$ from \kinfinity)
\eqn\indexgen{\Tr (-1)^F=|k|+\half \sum _i n_i^2\qquad\hbox{for}\qquad |k|\geq |k_c|,}
\eqn\indexgenn{\Tr (-1)^F=|k_c|+\half \sum _i n_i^2\qquad\hbox{for}\qquad |k|\leq |k_c|.}
These results apply whenever the index is well-defined, i.e.\ whenever there is no non-compact moduli space of vacua.  Non-compact moduli spaces occur on the Higgs branch, if two $\sigma _{Q_i}$, with opposite sign charges $n_i$, are tuned to coincide.  Such moduli spaces also  occur on the Coulomb branch if $k=\pm k_c$ and tuned $\zeta$, as in \cnoncom.  It is interesting that the results \indexgen\ and \indexgenn\ are independent of $m_i$ and $\zeta$ even in cases where they cross through $\Tr (-1)^F$ ill-defined locations, since vacua could have there moved in or out from infinity.

As an illustrative example of \indexch\ yielding \indexgen\ or \indexgenn , consider $U(1)_k$ with $N$ fields $Q_i$ of charge $+1$.  It suffices to set all real masses $m_i=0$,  with the FI parameter $\zeta$ varying.    If $\zeta >0$, there is a compact Higgs branch moduli space of SUSY vacua, ${\cal M}_H\cong CP^{N-1}$, with index given by the Euler character, while for $\zeta <0$ there are no Higgs vacua, so
\eqn\higgsn{\Tr (-1)^F|_{\rm Higgs}=\half (1+\sign (\zeta))\chi (CP^{N-1})=\half N  (1+\sign (\zeta)).}
Topological vacua exist at $\sigma _I=-\zeta/k_{eff}$ if $\sign (\sigma _I)$ fits with the sign of the shift in $k_{eff}$:
\eqn\cone{U(1)_{k\pm \half N}: \qquad \sigma _I= -{\zeta \over k\pm \half N}, \qquad \hbox{if}\qquad  \mp {\zeta\over k\pm \half N}>0.}
For $|k|\geq \half N$, one or the other of the solutions in \cone\ exist,  depending on $\sign (\zeta)$.  For $|k|<N/2$, both exist for $\zeta <0$, and neither exists for $\zeta >0$.  In sum, this gives
\eqn\indexcn{\Tr (-1)^F|_{\rm topological}=\sum |k_{eff}|=\cases {|k|-\half N \sign (\zeta) & for $|k|\geq \half N$\cr \half N (1-\sign (\zeta)) & for $|k|\leq \half N$}.}
The separate contributions \higgsn\ and \indexcn\ depend on $\sign (\zeta )$.  Indeed, at $\zeta =0$ the topological and Higgs vacua locations collide, $\sigma _{Q_i}=\sigma _I=0$, so the individual vacua can there become rearranged.  Their total is properly independent of all real parameters, with $\sign (\zeta)$ canceling between \higgsn\ and \indexcn:
 $\Tr (-1)^F=|k|+\half N$ for $|k|\geq \half N$, or $\Tr (-1)^F=N$ for $|k|\leq \half N$, agreeing with \indexgen\ and \indexgenn.  For $k=\pm k_c=\pm \half N$, and $\zeta =\zeta _{crit}=0$, there is an unlifted Coulomb branch, so $\Tr (-1)^F$ is ill-defined. Again, we note that $\Tr (-1)^F$ nevertheless takes the same value for $\zeta >\zeta _{crit}$ and $\zeta <\zeta _{crit}$.

Finally, note that vector-like matter, i.e.\ pairs $Q_i$ and $\tilde Q_i$ with charges $\pm n_i$, do not cancel in the index \indexgen\ and \indexgenn.  This might seem strange, since one could decouple such matter with a complex mass term $W=m_C Q_i\tilde Q_i$, and $\Tr (-1)^F$ is invariant under localized, continuous deformations.
 The point, again, is that real vs.\  complex masses are mutually exclusive if SUSY is not explicitly broken.  Since  \indexgen\ and \indexgenn\ apply  for
 generic real masses, they cannot be continuously connected with $m_C\neq 0$, while keeping $\Tr (-1)^F$ well defined.  The only option is to go through $m_i=m_C=0$, where the theory has an unlifted Higgs branch moduli space of vacua, so $\Tr (-1)^F$ is ill-defined, and vacua can move in or out from infinity upon taking  $m_C\neq 0$.  The existence of real masses is special to odd spacetime dimensions, which fits with the index contribution scaling as $n_i^{D-1}$, so vector-like matter contributions do not cancel for $D$ odd.

\newsec{Comparing the index $\Tr (-1)^F$ of dual theories}

Obviously $\Tr (-1)^F$ must agree between dual theories; we here mention a few examples.  As mentioned in the introduction, any gauge theory that is dual to a  Wess-Zumino theory, without gauge fields, must have $\Tr (-1)^F=1$.  Indeed,  if the dual theory has fields $\Phi_i$ with real masses $m_i$ (suppressing flavor indices),
\eqn\kvm{\CL _{eff}\supset K(\Phi_i^\dagger e^{m_i\theta \bar\theta }\Phi_i)\qquad \rightarrow \qquad V_{eff}\supset m_i^2 {\partial K\over \partial |\Phi_i|^2}|\Phi _i|^2,}
with a unique SUSY vacuum at $\Phi _i=0$ for smooth, non-degenerate $K_{eff}(\Phi ^\dagger \Phi )$.

On the gauge theory side, $\Tr (-1)^F=1$ requires \indexconf\ $|k|-h+\half \sum _f T_2(r_f)=0$ (with $|k|$ replaced with $|k_c|$ for $|k|< |k_c|$).  For the example of $U(1)_k$ gauge theory with $N_\pm $ matter fields of charge $n_i=\pm 1$, we found $\Tr (-1)^F=|k|+\half N_++\half N_-$ (as in \indexgenn\ we replace $|k|$ with $|k_c|$ for $|k|\leq |k_c|$, and here $k_c=\half (N_+-N_-)$, see \kinfinity).  So the cases that can be dual to Wess-Zumino theories are the examples that we have mentioned.  One is $U(1)_k$, with $|k|=\half$ and a single matter field $Q$ of charge $|n_i|=1$, which is dual to a free chiral superfield, $X$.
Another is  $U(1)_0$ with $N_f=1$ flavor of fields,  $Q$ and $\tilde Q$, of charge $\pm 1$.   This theory is dual to $W=MX_+X_-$ \AharonyBX. The dualities for the two cases discussed above provide a basis for deriving other duality examples, see e.g.\ \DimofteJU.

 Let us explicitly verify that $\Tr (-1)^F$ matches for the duals of \AharonyBX, between $\CN =2$ $U(1)_0$ with $N_f$ flavors,  and the $U(1)_0^{N_f}/U(1)$ quiver, analogous to the $\CN =4$ mirror symmetry dual of \IntriligatorEX.  The $U(1)_0$ theory, for generic FI parameter and real masses has, according to \indexgen\ with $k=k_c=0$, $\Tr (-1)^F=N_f$.  To evaluate the index of the  $U(1)_0^{N_f}/U(1)$ mirror dual, it is convenient to consider first $U(1)^{N_f}$, and then mod out by the overall $U(1)$.  Integrating out the bifundamentals, the real masses induce $k_{mag}^{ab}= \half \vec n^a\cdot \vec n^b=C^{ab}$, where $\vec n^a$ is a vector with components $n_i^a$ the charges of bifundamental matter fields, and $C^{ab}$ is the Cartan matrix of the affine $SU(N_f)$ extended Dynkin diagram.  Note that $k_{mag}^{ab}$ has a zero eigenvalue, corresponding to the overall translation.  For example, for $N_f=2$ we get $k=\pmatrix{2&-2\cr -2&2}$, while for $N_f=3$ we get the Cartan matrix of affine $SU(3)$
\eqn\kmagthree{k_{mag}=\pmatrix{2&-1&-1\cr -1 &2&-1\cr -1&-1&2}.}
To get the $U(1)^{N_f-1}$ theory we need drop the zero eigenvalue, and rescale the remaining eigenvalues to have properly normalized $U(1)$ factors.  For example, for $N_f=2$ the eigenvalues are $0$ and $4$, but the relative $U(1)$ has $N_f=2$ (this case is self-dual), so we rescale to $k=2$.  In general, for all $N_f$, the index of the magnetic dual is given by
\eqn\indexmag{ \Tr (-1)^F_{mag}= {1\over N_f}\det ' k_{mag}=N_f,}
where $\det '$ means to omit the zero eigenvalue, and the factor of $1/N_f$ is to have properly normalized relative $U(1)$ charges.  The result \indexmag\ indeed agrees with \indexgen\ in the dual.

\newsec{$SU(2)_k$ gauge theory}

Recall first the case of $\CN =2$ $SU(2)_0$ theory, with $k=0$.   There is a classical Coulomb branch moduli space, where the real adjoint $\sigma _{adj}\sim {\rm diag}(\sigma , -\sigma)$ breaks $SU(2)\to U(1)$, and we impose $\sigma \geq 0$ since the  $SU(2)$ Weyl symmetry makes $\sigma$ and $-\sigma$ gauge equivalent.  The moduli space can be labeled by the chiral superfield $Y\sim e^{\sigma /g^2+ia}$ (analogous to $X$ for the low-energy $U(1)$).  For the theory without matter, the Coulomb branch  is  lifted by a 3d instanton (4d monopole) \AffleckAS\
and  the theory has no static vacuum.
\eqn\noflavor{SU(2)_0, \quad N_f=0: \qquad \hbox{runaway}\qquad W_{dyn}\sim {1\over Y}.}

For the $SU(2)_k$ theory with matter fields $Q_i$, $i=1\dots 2N_f$, we can form gauge invariant mesons $M_{ij}=Q_{ic}Q_{jd}\epsilon ^{cd}=-M_{ji}$ (with $c,d$ the usually-suppressed $SU(2)$ color indices).  The classical Higgs branch, for the theory with zero superpotential and real masses, is given by $\ev{M_{ij}}$ subject to the classical constraint that rank$(M)\leq 2$.  The $SU(2)_0$ theories with $2N_f$ fundamental matter fields $Q_i$, were analyzed in \AharonyBX, for vanishing real masses and integer $N_f$.

The result for $N_f=1$ is a quantum moduli space of SUSY vacua, merging the Higgs (with $M\equiv M_{12}$) and Coulomb branches:
\eqn\oneflavor{SU(2)_0, \qquad N_f=1: \qquad   \hbox{smooth SUSY }\ \CM _{\rm quantum}= \{ M, Y\  | \ MY=1\}.}
For $N_f=2$, the theory has a dual with chiral superfields $M_{ij}$ and $Y$, with
\eqn\twoflavor{SU(2)_0, \qquad N_f=2: \qquad \hbox{simple dual}\qquad W_{dual}=Y{\rm Pf} M ~.}
The $SU(2)_0$ gauge theory and its simple dual  \twoflavor\ flow to the same interacting SCFT at the origin.   For $N_f>2$, there is an interacting SCFT at the origin, with additional degrees of freedom beyond $M_{ij}$ and $Y$ \refs{\AharonyBX,\AharonyGP, \ARSW}. In this section, we generalize the above to allow for $k\neq 0$, real masses, half-integral $N_f$, and other matter representations.

For generic $k$ and real masses, there are isolated supersymmetric vacua, with a mass gap.  The index for the theory with $2N_f$ matter fields in the fundamental is found to be
\eqn\indexnfi{\Tr (-1)^F=|k| +N_f-1.}
More generally, for matter fields $Q_i$ in representations $r_i$ of $SU(2)$, the index is
\eqn\indexsuii{\Tr (-1)^F=|k|+\half \sum _i T_2(r_i)-1.}
Supersymmetry is dynamically broken for $|k|+N_f\leq 1$: SUSY is broken with a mass gap for $SU(2)_1$ with $N_f=0$ and $SU(2)_0$ with $N_f=1$ and generic real masses.        With non-generic real masses, and general $k$ and matter content, there can be Higgs and/or Coulomb branches, and interacting SCFTs.  The details will be given in the following subsections.

\subsec{Some generalities}

Consider generally $SU(2)_k$ with matter fields $Q_i$ in representations $r_i$ of $SU(2)$.  If the matter fields are given real masses $m_i$, they can be integrated out, and the low-energy theory is $SU(2)_{k_{eff}}$ with no matter, with  (writing $k=k_{SU(2)}$ here for clarity's sake)
\eqn\keffsuiig{k_{SU(2), eff}=k_{SU(2)}+\half \sum _iT_2(r_i)\sign (m_i),}
where  $T_2(r_i)=2(2I_i+1)I_i(I_i+1)/3$ is the quadratic index if representation $r_i$ is labeled by its $SU(2)$ isospin $I_i$ (i.e.\  $r_i$ is the $2I_i+1$ dimensional representation), with $T_2(r_{fund})=1$ for the fundamental.  The quantization condition on $k$, required for gauge invariance, is
\eqn\ksuiiq{k_{SU(2)}+\half \sum _i T_2(r_i)\in \Z.}
In particular, for $2N_f$ fundamentals \ksuiiq\ shows that
\eqn\kquant{k_{SU(2)}+N_f\in \Z.}

The semi-classical SUSY vacua satisfy
\eqn\classeqsuii{(\sigma ^a T^a_{r_i}+m_i{\bf 1}_{|r_i|}) Q_i=0, \qquad D^a\propto -{k_{eff}\over 2\pi}\sigma _{adj}^a +\sum _i Q_i^\dagger T^a Q_i=0,}
where $\sigma _{adj}^a=\Sigma ^a|$, $a=1, 2, 3$, are the real scalars in the vector multiplet.  When $\ev{\sigma ^a}\neq 0$, the gauge group is broken, $SU(2)\to U(1)$, and by an $SU(2)$ gauge rotation we can set $\sigma^1=\sigma^2=0$, $\sigma ^3= \sigma  \geq 0$. The low-energy theory  for $\sigma >0$ is $U(1)_{k_{U(1)}}$, with $k_{U(1)}$ rescaled to account for the normalization of the generators
\eqn\kreln{k_{U(1)}=2k_{SU(2)}\equiv 2k.}

More generally, for $\sigma > 0$, breaking $SU(2)_k\to U(1)_{2k}$, an  $SU(2)$ isospin $I_i$ rep matter field $Q_i$ decomposes into $2I_i+1$ $U(1)$ charged matter fields, $Q_i^{n_i}$, where $\half n_i\in \{I_i,\ I_i-1,\ \dots , -I_i\}$, with $n_i\in \Z$ the $U(1)$ charge.  These $U(1)$ charged matter fields have real masses as in \effmass, $m^{n_i}_i(\sigma)=m_i + n_i\sigma$.  So there is a massless component at $\sigma _{Q_i^{n_i}}=-m_i/n_i$, if this respects the $SU(2)$ requirement that $\sigma \geq 0$. The charge $n_i$ component can thus be massless if $\sign (n_i)=-\sign (m_i)$.  In the low-energy $U(1)_{2k}$ theory we define, as in \zetakeff,
\eqn\fsuii{F(\sigma)\equiv \zeta _{eff}+k_{U(1), eff}\sigma = 2k\sigma +\half \sum _i \sum _{{n_i\over 2}=-I_i}^{I_i} n_i |m_i + n_i\sigma|.}
This expression is valid only in the semiclassical region $e^2 |k| \ll \sigma$, where the signs of the gauginos real mass are determined by $\sigma$ and they do not contribute.
In particular, a fundamental matter field $Q$ with real mass $m$ contributes
\eqn\fsuiif{\Delta F(\sigma)_{fund}=\half (|m+\sigma |-|m-\sigma|)=\cases{ m & $\sigma >|m|$\cr \sigma \sign (m) & $\sigma < |m|.$}}
Note that $k_{U(1), eff}\to 2k$ for $\sigma \to \pm \infty$; comparing with \kinfinity,  here the quantity $k_c=0$, because $SU(2)$ representations are symmetric under $n_i\to -n_i$.  Because $k_c=0$, there will not be an analog of the case \indexgenn\ for the index of $SU(2)_k$ theories.

The vacua can be found by a semi-classical analysis in the low-energy $U(1)_{k_{U(1),eff}}$ theory, which is a good approximation for sufficiently large  $\sigma \gg e^2$, i.e.\  for $|m_i|\gg e^2$.  Corrections from instantons in the broken $SU(2)/U(1)$ need to be added to the low-energy $U(1)$ theory.

 The quantum vacua are as follows:

\lfm{1.} Higgs vacua, moduli spaces or isolated points.  At the origin of the Coulomb branch, $\sigma _a=0$, setting $m_i=0$, the fields $Q_i$ can have non-zero $\ev{Q_i}\in \CM _H$ solving \classeqsuii.  The moduli space $\CM _H $ is the same as for the corresponding 4d $\CN =1$ theory, which can be given as the gauge invariant composite chiral superfields, subject to classical constraints.  The space $\CM _H$ is unaffected by $k$ and unmodified by quantum effects (aside from the special case \oneflavor), and is generally singular at the origin.  The theory at $\sigma _a=Q_i=0$ is typically an interacting SCFT, which depends on the matter content and $k$.   For $m_i\neq 0$, there are Higgs branches where some $m_i+n_i\sigma =0$ in  \classeqsuii.   These can be analyzed in terms of the low-energy $U(1)_{k_{U(1)}}$ theory and, depending on the matter content and $m_i$, can be non-compact, compact, or isolated points.

\lfm{2.} Coulomb moduli space, when $k_{eff}=0$, setting $Q_i=0$ and $\sigma \neq 0$.   As we will see, this can occur in non-compact or compact regions.   The semi-classical Coulomb branches can be lifted by an instanton superpotential, in some cases, similar to \AffleckAS.

\lfm{3.} Topological vacua.  For $k_{eff}\neq 0$, when the matter fields have real masses and are integrated out, the low-energy $SU(2)_{k_{eff}}$ theory with no matter $|k_{eff}|-1$ topological vacua at $\sigma =0$.  For $\sigma \neq 0$, there can be $|k_{eff, U(1)}|$ additional topological vacua, at locations $\sigma _I$ where \fsuii\ vanishes.

The global symmetries for a general theory with matter fields $Q_i$ and real masses $m_i$ includes $U(1)_i$, which acts only on $Q_i$, and a $U(1)_R$ symmetry, with charges
\eqn\ntwogen{
\vbox{\offinterlineskip\tabskip=0pt
\halign{\strut\vrule#
%%%%%%%%%%%%%%%%%%
&~$#$~\hfil\vrule
&~$#$~\hfil\vrule
&~$#$~\hfil\vrule
&~$#$~\hfil\vrule
&~$#$~\hfil\vrule
&~$#$~\hfil%\vrule
&\vrule#
\cr
%%%%%%%%%%%%%%%%%
\noalign{\hrule}
&  & U(1)_j & U(1)_R \cr
\noalign{\hrule}
%%%%%%%%%%%%%%%%%%
&  Q_i        & \;\delta _{ij} &\;0\cr
&Y  &\; -K_j &\; -2+\sum _j K_j  \cr
}\hrule}}
$K_j$ can be computed in the low-energy $U(1)$ theory, where $Y\sim X_+$, with charges similar to \onengen\ obtained from the induced,   mixed Chern-Simons terms \kgaugefla\ between the low-energy $U(1)$ gauge group and the $U(1)$ global symmetries.  This (or, equivalently, the Callias index theorem \CalliasKG) gives
\eqn\kjcallias{K_j=\half \sum _{n=-2I_{r_j}}^{2I_{r_j}{}'}n \sign (n \sigma +m_j),}
where the ${}'$ is a reminder that the sum is only over even or odd $n$, depending on whether $I_{r_j}$ is integer or half-integer.  For example, for matter in the fundamental, $I=\half$,
\eqn\kjcalliasf{K_{fund}=\half \sign (\sigma +m)-\half \sign (-\sigma +m)=\Theta (\sigma -|m|).}
In general,  for $|\sigma |>|m_j|$, a Fermion in the $2I+1$ dimensional $SU(2)$ representation
has $K_j=I(I+1)$ for integer $I$, or $K_j=(I+\half)^2$ for half-integer $I$.  The results \noflavor\ and \oneflavor\ and \twoflavor\ are compatible with these charges.

A non-compact Coulomb branch exists semi-classically when $k_{eff}$ and $\zeta_{eff}$ vanish for $\sigma \to \infty$.  Using \fsuii, we find
\eqn\fsuiilim{k_{eff\infty}\equiv \lim_{\sigma \to \infty} k_{eff}= 2k, \qquad \zeta _{eff\infty}\equiv \lim_{\sigma \to \infty}\zeta _{eff}= \sum _i K_{i\infty}m_i ,}
where $K_{i\infty}$ is the index \kjcallias\ in the limit where $\sigma >|m_j|$ for all $m_j$,
\eqn\kjinfty{K_{i\infty}=\lim _{\sigma \to \infty} K_i=\half
\sum _{n=-2I_{r_i}}^{2I_{r_i}{}'}|n|.}
In particular, if all matter is in the fundamental, then $K_{i\infty}=1$ and \fsuiilim\ gives $\zeta _{eff\infty}=\sum _i m_i$.  There is a non-compact Coulomb branch only if $k=0$ and $\zeta _{eff, \infty}=0$.   The latter condition fits with the real mass assignment for the field $Y$, which follows from the $U(1)_i$ charge assignment of $Y$ in \ntwogen
\eqn\mygen{\lim _{\sigma \to \infty}m_Y=-\sum _i K_{i\infty}m_i=-\zeta _{eff\infty},}
so the non-compact Coulomb branch remains unlifted when $m_Y=0$.   When $k=m_Y=0$, the non-compact region of the Coulomb branch cannot be lifted by an instanton, if there is matter (unlike pure Yang-Mills \AffleckAS), since the $U(1)_i$ and $U(1)_R$ charges in \ntwogen\ are incompatible with a holomorphic superpotential if $K_{i,\infty}\neq 0$.  We will see examples where some compact semi-classical Coulomb branches are lifted by $W_{inst}\sim 1/Y$, which can happen in regions where all  $K_j=0$, as in \kjcalliasf\ for $\sigma< |m_j|$.

\subsec{The index for general $k$ and matter fields $Q_i$}

For generic $k$ and real masses, the theory has isolated vacua with a mass gap, so the index is finite and well-defined.  The index is then the sum of contributions from the $SU(2)_{k_{eff}}$ topological vacua at $\sigma =0$, and any Higgs or topological vacua at $\sigma \neq 0$:
\eqn\indexsuiiexp{\eqalign{\Tr (-1)^F&= \Tr (-1)^F|_{\sigma =0}+\sum _i \Tr (-1)^F|_{\sigma _{Q_i^{n_i}}}+\sum _{\sigma _I} \Tr (-1)^F |_{\sigma _I}\cr &=|k|-1+\half \sum _i T_2(r_i).}
}
The individual terms on the top line depend on the real parameters, while the sum of course does not. We here briefly illustrate that.  The $\sigma =0$ term is from the topological vacua of the low-energy $SU(2)_{k_{eff}}$ theory:
\eqn\indexorigin{\Tr (-1)^F|_{\sigma =0}=|k+\half \sum _i T_2(r_i)\sign (m_i)|-1}
The $\sigma _{Q_i^{n_i}}=-m_i/n_i$ and $\sigma _I$ terms in \indexsuiiexp\ are from the possible Higgs or topological vacua in the low-energy $U(1)_{k_{eff}}$ theory:
\eqn\indexiterm{\Tr (-1)^F|_{\sigma _{Q_i^{n_i}}}=n_i^2\Theta (-m_i/n_i)\Theta (n_i F(\sigma _{Q_i^{n_i}})),}
\eqn\indexiiterm{\Tr (-1)^F|_{\sigma _I}=\bigg| {dF(\sigma)\over d\sigma }\big| _{\sigma _I}\bigg| ,}
where $F(\sigma)$ is as in \fsuii\ and $\sigma _I$ are the topological vacua, where $F(\sigma _I)=0$.

The fact that \indexsuiiexp\ is independent of the values of $m_i$ can be proved by compactifying on an $\S^1$ and using holomorphy in the twisted chiral parameters, as mentioned in the introduction. Or it can be proved directly, much as in the $U(1)$ case, discussed in appendix D.  We can thus evaluate the index for a convenient choice of $m_i$.   We use the $P$ symmetry to take $k>0$, and choose all  $m_i>0$; in the end, we can replace  $k\to |k|$ for generality.  With this choice, the index comes entirely from the vacua at the origin  \indexorigin.  Indeed, using \fsuii\ here yields $F(\sigma )>0$ for all $\sigma >0$, so there are no vacua $\sigma _I\neq 0$ where $F(\sigma _I)=0$, and there are also no Higgs vacua, since only negative charge $n_i<0$ components $Q_i^{n_i}$ become massless, and $F(\sigma)$ has the wrong sign for $Q_i^{n_i}\neq 0$.  So the topological vacua at $\sigma =0$ yields the result \indexsuii, which is the complete, general answer.

If some $\sign (m_i) \neq  \sign (k)$, some Higgs or topological vacua compensate for the difference, leading again to \indexsuii.   As an illustration consider a theory with $2N_f$ fundamental flavors, with $k>0$ and all real masses $m_i=m<0$.  Then the vacua at the origin contribute
\eqn\indexo{\Tr (-1)^F|_{\sigma =0}=|k-N_f|-1.}
To look for the Higgs and topological vacua away from the origin, we consider $F(\sigma)$ \fsuii, which using \fsuiif\ here gives
\eqn\fsigmah{F(\sigma)= \cases{2k \sigma + 2N_f m & $\sigma >|m|$\cr 2(k-N_f)\sigma & $\sigma <|m|$}.}
The low-energy $U(1)$ theory has $2N_f$ massless matter fields, with $n_i=+1$, at $\sigma = |m|$. These  give  a $CP^{2N_f-1}$ compact Higgs branch if $F|_{\sigma =|m|}>0$,  and no SUSY vacua otherwise:
\eqn\indexha{\Tr (-1)^F|_{\sigma = |m|}=2N_f \Theta (k-N_f).}
Finally there are $2k$ topological vacua at $\sigma = N_f |m|/k$, if $N_f/k>1$:
\eqn\indexta{\Tr (-1)^F|_{\sigma = N_f|m|/k}=2k\Theta (N_f-k).}
Adding \indexo\ and \fsigmah\ and \indexha\ indeed gives the index \indexnfi, $\Tr (-1)^F=|k|+N_f-1$.  If we had taken $m>0$, all of these vacua would have instead been at the origin.

This illustrates a  point about decoupling matter fields via large $|m|$.  If the sign of $m$ is the same as the remaining $k_{eff}$, the additional vacuum or vacua associated with the heavy matter field remain at the origin, staying in the low-energy theory via increased $|k_{eff}|$.  If the sign of $m$ is opposite to the remaining $k_{eff}$, the additional vacuum or vacua associated with the heavy matter runs off to infinity, $\sigma _{I_{heavy}} = \CO(|m|)$, e.g.\ as in \indexha\ or \indexta.

\subsec{$SU(2)_k$ with $N_f=\half$, a single fundamental matter doublet, $Q$}
The condition \kquant\ requires $k\in \Z +\half$.  The global symmetry is $U(1)_Q\times U(1)_R$.
$Q$'s real mass $m$ is a background field for $U(1)_Q$.   By $P$ symmetry ($m\to -m$ and $k\to -k$), we could always restrict to $m>0$, or $k>0$, but we will not do so for the moment.
This theory does not have a Higgs branch at $\sigma =m=0$, because there is  no non-trivial $\ev{Q}\neq 0$ solution of the $SU(2)$ $D$ term equations \classeqsuii; this fits with the fact that no gauge invariant chiral operator can be formed from a single $Q$ field.

First consider the isolated Higgs and topological vacua for $m\neq 0$.  At $\sigma =0$, we can integrate out $Q$, to obtain a low-energy $SU(2)_{k+\half \sign (m)}$ theory without matter; so
\eqn\indexexq{\Tr (-1)^F|_{\sigma =0}=|k+\half \sign (m)|-1}
topological SUSY vacua, with a mass gap.   The cases $k=\pm \half$ will be discussed separately.

For $\sigma \neq 0$, breaking $SU(2)_k\to U(1)_{2k}$, the low-energy effective $U(1)$ theory has the $Q$ components $Q^\pm$, with mass $m_\pm = m\pm \sigma$. When $|m|\gg e^2$, we can analyze the theory semi-classically in the low-energy $U(1)$ theory, with \fsuii\ given by
\eqn\fsuiiex{F(\sigma)\equiv \zeta _{eff}+k_{eff}\sigma = 2k\sigma +\half |m+\sigma|-\half |m-\sigma| ~.}
 It follows from \indexiterm\ that
there is a Higgs vacuum at $\sigma _Q=|m|$ if $F(|m|)<0$ for $m>0$, or if $F(|m|)>0$ for $m<0$.  There are no topological vacua $\sigma _I\neq 0$ to contribute to \indexiiterm, since $F(\sigma)$ has no non-trivial zeros.  The total index is then given by the result \indexnfi\ for $N_f=\half$,
\eqn\indexextot{\Tr (-1)^F= |k|-\half.}

The cases $k=\pm \half$ dynamically break SUSY, fitting with $\Tr (-1)^F=0$ \indexextot.
The case $k=+\half \sign (m)$ leads to a mass gap, $k_{eff}=k+\half \sign (m)=\pm 1$, and none of the topological nor Higgs vacua exist in this case, since $|k_{eff}|=1$ and $F(|m|)$ has the wrong sign for $\ev{Q}\neq 0$.  The case $k=-\half \sign (m)$ is more involved, since $k_{eff}=0$, and thus a semi-classical Coulomb branch, for $\sigma \in [0, |m|]$, which gives $\CM ^{\rm S.C.} _{\rm Coulomb}\cong CP^1$ upon including the dual photon. The region
$\sigma >m$ is lifted by a  potential, since the low-energy $U(1)_{eff}$ theory has $k_{eff}=2k_{SU(2)} \neq 0$ (and $\zeta _{eff}=m$).

The semi-classical $\CM ^{\rm S.C.} _{\rm Coulomb}\cong CP^1$ for $k=-\half \sign (m)$ is lifted by an $SU(2)$ instanton superpotential, $W_{dyn}=1/Y$, where $Y$ is a local coordinate on the $CP^1$.  This $W_{dyn}$ is generated as in the $N_f=0$ case  \AffleckAS\ and is compatible with the symmetries \ntwogen, because the instanton does not have a $Q$ Fermion zero mode  for $\sigma <|m|$ (see \kjcalliasf). Unlike the $N_f=0$ case, here  there cannot be a runaway to infinite distance on the moduli space, because $CP^1$ is compact: $W_{dyn}$ pushes $Y$ to the end where semiclassically $\sigma \to |m|$ ($1/Y\to 0$) where there is a stable vacuum, with SUSY broken by the small non-zero vacuum energy from $W_{dyn}$ (and the global symmetries unbroken).

The case $k=\pm 3/2$ is also interesting, since \indexextot\ then gives $\Tr (-1)^F=1$.  For $k=3/2$ and $m>0$, the vacuum is at $\sigma =0$, while for $k=3/2$ and $m<0$ the vacuum is at $\sigma = \CO (|m|)$.

\subsec{$N_f=1$: matter fields $Q_1$ and $Q_2$. }

The global symmetry group is $SU(2)_F\times U(1)_Q\times U(1)_R$.   Consider first the theory with all masses set to zero.   For all $k$, there is a Higgs branch, with $\ev{M}$ arbitrary, where $M=Q_1 Q_2$ is an $SU(2)_F$ singlet, with $U(1)_Q$ charge equal to $2$.  For $k=0$, the quantum moduli space is the smooth space  $MY=1$ \oneflavor, while for $k\neq 0$ there is no Coulomb branch $Y$ modulus.  For $|k|=1$ and zero mass, as we will see via added matter and flowing down,  the theory at the origin is an IR-free field theory, consisting of the chiral superfield $M$, with $W=0$.  For $|k|>1$, the theory at $M=0$ is an SCFT labeled by $k$.

We now turn on real masses, $m_{1,2}$ for $Q_1$ and $Q_2$; $m_1+m_2$ is a  background gauge superfield of the $U(1)_Q$ symmetry, and $m_1-m_2$ is a background for $U(1) \subset SU(2)_F$.  Without loss of generality we can take $m_1 \ge |m_2|$.   The modulus $M$ has real mass $m_M=m_1+m_2$.  So if $m_1+m_2\neq 0$, the Higgs branch is lifted by $m_M\neq 0$.  For $m_1=-m_2$, $m_M=0$ and the Higgs branch is  unlifted, for all $k$.   For generic $m_1$ and $m_2$, there is a mass gap and \indexnfi\ gives $\Tr (-1)^F=|k|$.  In particular, for $k=0$, SUSY is broken for generic $m_i$.  Let us illustrate some aspects in more detail.

We consider first the semiclassical theory with
\eqn\semiclf{e^2 \ll |m_1|,\ |m_2| ~.}
The interesting points on the Coulomb branch are $\sigma=0,\ |m_1|,\ |m_2|$.
Integrating out the massive matter gives an $SU(2)_{k_{eff}}$ theory at $\sigma =0$, with, see \keffsuiig,
\eqn\keffex{k_{eff}=k+\half (1+\sign (m_2)).}
There can be semi-classical Coulomb vacua when $k_{eff}=0$, i.e.\  $k=0$ with $m_2<0$, or $k=-1$ with $m_2>0$.  If $k_{eff}\neq 0$, the IR theory at $\sigma =0$ has a mass gap,  with $|k_{eff}|-1$ SUSY vacua, and no SUSY vacua for $|k_{eff}|\leq 1$.  For $\sigma \neq 0$, we can use the low-energy $U(1)$ effective theory (with instanton corrections), with $F(\sigma)$ given by \fsuii\ and \fsuiif,
\eqn\zetakeffc{F(\sigma)=k_{eff}\sigma +\zeta _{eff}=\cases{
(2k +(1+\sign (m_2)))\sigma & $0<\sigma< |m_2|$ \cr
(2k+1)\sigma +m_2&$|m_2|<\sigma<m_1$ \cr
2k\sigma +m_1+m_2&$m_1 <\sigma $}
}

Consider the $k=0$ case.  The low energy theory around $\sigma=0$ is $SU(2)_{\half(1+ \sign(m_2))}$ with $N_f=0$, which does not have a SUSY vacuum.  It is clear from \zetakeffc\ that for $k=0$ there are no $\sigma _I\neq 0$ topological vacua, where $F(\sigma _I)=0$.  There are also no Higgs vacua: the massless field at $\sigma = m_1$ has charge $-1$, and $F(m_1)=m_1+m_2>0$ has the wrong sign for it to have an expectation value.  At $\sigma = |m_2|$, the massless field has charge  $-{\rm sign}(m_2)$ and the low-energy theory is $U(1)_{1+\half {\rm sign}(m_2)}$, and shifting $\sigma$ to vanish there leads to $\zeta = m_2+|m_2|$.  There is no Higgs vacuum there  since $\zeta$ either vanishes or has the wrong sign for the massless charged matter to get a Higgs vev.  In sum, the $k=0$ theory with generic real masses breaks SUSY.

Now consider $k=0$ for $m_1=\pm m_2$.  For $m_1=m_2$,  the low energy theory around $\sigma=m_1$ is $U(1)_1$ with $F(\sigma )=2m_1$ and  two massless chiral superfields with charge $-1$.  This theory breaks supersymmetry, since $F(\sigma)$ has the wrong sign for satisfying the $D$-term equations with nonzero charged matter vev.     For $m_1=-m_2$, on the other hand, the low energy theory around $\sigma=m_1$ is $U(1)_0$ with $\zeta =0$ and $N_f=1$ massless flavor of chiral superfields of charge $\pm 1$.  This low-energy theory has a moduli space of SUSY vacua with three  non-compact branches, two Coulomb and one Higgs, meeting at $\sigma =m_1$, i.e.\   it is the $W_{low, U(1)}=MX_+X_-$ theory.

We now consider nonperturbative corrections from $SU(2)$.  For $m_1+ m_2 \not=0$ the potential is everywhere nonzero, except perhaps at $\sigma=0$, where we know that $SU(2)_0$ has a
 runaway to large $\sigma$ and in $SU(2)_1$ supersymmetry is broken.  So the $m_1+m_2\neq 0$ theory breaks supersymmetry; we will momentarily give an independent derivation.

 Now consider the $SU(2)$ nonperturbative corrections for $m_1=-m_2$, where we have seen that the low-energy $U(1)$ theory near $\sigma = m_1$ is the $W_{low, U(1)}=MX_+X_-$ theory.   The low-energy $SU(2)_0$ theory $\sigma = 0$ gives a $W_{inst}$ correction, coinciding with that of \AffleckAS\ since the low-energy $SU(2)_0$ theory has no light matter: $W_{inst}=1/Y$.  Now the  Coulomb branch modulus $Y$ at $\sigma =0$ is related to the modulus $X_-$ at $\sigma = |m_2|$ by $YX_-=-1$; this is exactly as in  the identification in \AharonyBX\ in the context of Fig 2 there, written there as $V_{i+}V_{i+1-}=1$.  The instanton of \AffleckAS\ thus generates a term $W_{inst}=-X_-$ in the low-energy theory at $\sigma =m_1$, leading to the dual theory there (as also briefly discussed in \AharonyBX):
\eqn\effW{W=MX_+X_- - X_-~.}
The F-terms of \effW\ set $X_-=0$ and $MX_+=1$.  So the low energy theory has a smooth moduli space of vacua, $MX_+=1$, and therefore $\Tr (-1)^F$ is ill-defined.

Let us see how we can derive the same results in the limit opposite to \semiclf
\eqn\semicls{e^2 \gg |m_1|,\ |m_2| ~.}
We start with the smooth moduli space of $N_f=1$ with $m_i=0$, $MY=1$ \AharonyBX, and turn on the real masses $m_i$ as a small perturbation.  Since the global $SU(2)$ symmetry does not act on the low energy theory, only $m_1+m_2$ affects it.  In particular, if $m_1+m_2=0$ the real mass has no effect and we end up with the same smooth one dimensional moduli space of vacua $MY=1$.  This result is in accord with our conclusion in the opposite limit \semiclf.

For $m_1+m_2\neq 0$, there is a background expectation value for $U(1)_Q$, which leads to a real mass on the $MY=1$ moduli space of vacua, $m_M=-m_Y\propto m_1+m_2$, since the $U(1)_Q$ charge of $M$  is 2, and that of $Y$ is $-2$.  This breaks supersymmetry,  reminiscent of the quantum moduli space DSB models in 4d \refs{\IntriligatorPU, \IzawaPK}, with the classical supersymmetric vacuum at the origin eliminated by the quantum constraint.  Using the asymptotic form of the K\"ahler potential
\eqn\assyK{K\sim \cases{
|M| ={1\over |Y|} & $|Y| \to 0 $\cr
(\log(Y)+ \log (\bar Y))^2 &$|Y|\to \infty$}}
we learn that the potential is $V=-K(\bar Y e^{m_Y \theta\bar \theta}Y)|_{\theta ^2\bar\theta ^2}=m_Y^2 K_{Y\bar Y}|Y|^2$:
\eqn\assyV{V\sim \cases{
(m_1+m_2)^2|M| ={(m_1+m_2)^2\over |Y|} & $|Y| \to 0 $\cr
(m_1+m_2)^2 &$|Y|\to \infty$}}
and it is never zero.  Since, before turning on the real masses, the moduli space was smooth, and the symmetry associated with $m_1+m_2$ is everywhere broken, supersymmetry is necessarily broken when $m_1+m_2 \not=0$.  This breaking can be associated with one or several metastable states at finite $Y$.   The Fermion component of the $MY=1$ modulus remains massless, playing the role of the Goldstino.  For $Y\to \infty$ there is a classically massless pseudomodulus superpartner, seen from the classical flat direction in \assyV.  To determine whether the potential has a stable minimum, or a $Y\to \infty$ runaway, we now determine the leading quantum  correction to $K$ in \assyK\ for $Y\to \infty$.

The one-loop correction to the Coulomb branch metric can be written as in \refs{\SeibergNZ, \deBoerKR}
\eqn\oneloopm{ds^2={1\over 4}\left({1\over e^2}+ {s\over \sigma}\right) d\sigma ^2+\left({1\over e^2}+ {s\over \sigma}\right)^{-1}da^2}
where $s=N_f-3$ for $\CN =2$ SUSY $SU(2)$ with $2N_f$ doublets, see e.g.\ \refs{\DoreyKQ, \BuchbinderEM} and references therein.   The chiral superfield $\Phi = \log Y$ and the linear multiplet $\Sigma$ are related by a Legendre transform \refs{\HitchinEA, \deBoerKR, \AganagicUW}, as reviewed in appendix B, which here implies that
\eqn\metk{K_{Y\bar Y}|Y|^2= K''(\Phi +\bar \Phi)=\left({\partial (\Phi +\bar \Phi)\over \partial \Sigma}\right)^{-1}=\left({1\over e^2}+{s\over \sigma}\right)^{-1}.}
The potential $V=m_Y^2 K_{Y\bar Y}|Y|^2$ thus has slope with $\sign (dV/d\sigma)=\sign (s)$, pushing $\sigma$ toward the origin for $s>0$, or away from the origin for $s<0$.  In the present context, $SU(2)$ with $N_f=1$, $s=-2<0$, so the potential is a runaway, $|Y|\to \infty$.  For small $m$, we can regard this as a small correction to the $MY=1$ quantum moduli space.  Then $Y\to \infty$ means the runaway vacuum has $M\to 0$.

The result that the $m_1+m_2\neq 0$ theory breaks supersymmetry is in accord with our conclusion in the opposite limit \semiclf, and it shows that $SU(2)_1$ without matter breaks supersymmetry, consistent with $\Tr (-1)^F= |k|-1$, vanishing for $k=1$.  This is also compatible with our result  that $SU(2)_{\pm \half}$, with $N_f=\half$ breaks supersymmetry, since we can flow to that case by decoupling say $Q_2$, by taking $|m_2|\gg |m_1|$.

In conclusion, $SU(2)_0$ with $N_f=1$ and $m_1+m_2\not=0$ breaks supersymmetry. For $m_1+m_2=0$, there is  moduli space of SUSY vacua, $MY=1$, with K\"ahler potential \assyK.

\subsec{$N_f={3\over 2}$: matter fields $Q^{i=1, 2, 3}$, with $k\in \Z +\half$}

For zero mass, there is a Higgs branch labeled by $M_i=\epsilon _{ijk}Q^j Q^k$, in the $\bar 3$ of the global $SU(3)_F$.  The theory at the origin of this Higgs branch is an interacting SCFT, labeled by $k$, for $|k|>\half$ (it is perturbative for large $k$).  For $k=\pm \half$, on the other hand, the theory at the origin is a free field theory of the unconstrained chiral superfields $M_i$.  This will be justified in the following subsection, upon giving a real mass to RG flow from $N_f=2$, with $k=0$, to $N_f={3\over 2}$, with $|k|=\half$.  A check is that, for generic real masses, we get for the index $\Tr (-1)^F=|k|+\half$ SUSY vacua, so the $k=\pm \half$ theory has $\Tr (-1)^F=1$, matching that of a dual theory of  free chiral superfields  \indexconf.   The $SU(3)_F$ global symmetry has, in terms of the original $Q$ fields, background CS term $k_{SU(3)}\in \Z$.  The theory of the $M_i$ fields contributes $k_{SU(3)_F}\in\Z+\half $, so parity anomaly matching implies that the low-energy theory requires an added contribution $\Delta k_{SU(3)}=\half$, which can be induced by the RG flow, as in \ClossetVP, because the original $SU(2)$ theory anyway has parity violating $k_{SU(2)}\neq 0$.

We now consider the $SU(2)_\half$ theory, turning on various real masses $m_i$ for the matter fields.   The location of the vacua depend on the sign of the $m_i$.  For example, if all $m_i>0$, the SUSY vacuum contributing to $\Tr (-1)^F=1$ is at the origin, from the low-energy $SU(2)_{2}$ theory.  If, on the other hand, all $m_i=m<0$, the low-energy theory at the origin is $SU(2)_{-1}$,
which does not have a SUSY vacuum; the SUSY vacuum is at $\sigma = 3|m|$ and runs off to infinity as $|m|\to \infty$.  The $\sigma$ location of the vacuum is not evident in the dual description of the free $M_i$ fields, which gives a supersymmetric vacuum at $M_i=0$ regardless of the sign of the $m_i$.

Consider the $SU(2)_\half$ theory with a real mass $m_3$ for  $Q_3$, with $m_1=m_2=0$.
This gives real mass to $M_1$ and $M_2$, leaving $M=M^{12}=M_3$ massless, so there is a Higgs branch labeled by $\ev{M}$, and the low-energy theory near the origin is $SU(2)_{\half (1+\sign (m_3))}$ with $N_f=1$ massless flavor.   This low-energy theory is thus qualitatively different depending on the $\sign (m_3)$.  For $m_3>0$, the low energy $SU(2)_1$ has no Coulomb branch, and there are no other vacua for $\sigma \neq 0$.  For $m_3<0$, the low-energy theory at the origin is $SU(2)_0$ with $N_f=1$, so there is a classical Coulomb branch, with $0\leq |\sigma | \leq |m_3|$.  The low-energy $U(1)$ theory for $\sigma\neq 0$ has
\eqn\zetakefft{F(\sigma)= \sigma + \half  (|m_3+ \sigma| - |m_3-\sigma|)=\zeta _{eff}+k_{eff}\sigma}
so there is a compact Coulomb branch, where $F(\sigma)=0$, for $|\sigma |<|m_3|$; the Coulomb branch is lifted for $|\sigma|\geq m_3$, where $F(\sigma)=\sigma +m_3$.   Classically, there is thus a compact Coulomb branch which is topologically a $CP^1$.  The end near $\sigma \approx |m_3|$ is smooth, since the low-energy $U(1)$ theory near $\sigma = m_3$ is $U(1)_\half$ with a single charge $+1$ matter field, dual to a free-field theory.   The Coulomb modulus $Y$ has $R(Y)=0$ ($Q_{1,2}$ have a Fermion zero mode  \kjcalliasf\ $K_{1,2}=1$, while $Q_3$ has $K_3=0$), so $W_{dyn}(Y)=0$.  The low-energy $N_f=1$ theory has the smooth, quantum-deformed moduli space \oneflavor,
$MY=1$, with $Y\to \infty$ corresponding to $\sigma \approx |m _3|$.

In conclusion, the theory with finite $m_3$ has a smooth moduli space parameterized by $M=M^{12}$.  As $m_3 \to \infty$ the point $M=0$ is pushed to infinite distance.  In the dual description, this detailed structure is simply replaced with a free field $M$.   The K\"ahler potential information about the distance to, or location of $M=0$ is not immediately apparent in the dual description.

Now consider $SU(2)_\half$ for $m_{2,3}\neq 0$, with $m_1=0$.  If $m_2+m_3=0$, there is an unlifted Higgs branch moduli space of supersymmetric vacua, labeled by the massless field $M_1$.  For $m_2+m_3\neq 0$, the low energy theory has a mass gap.  For $m_2$ and $m_3$ both positive, the low-energy theory near the origin is $SU(2)_{{3\over 2}}$ with $N_f=\half$, which has $\Tr (-1)^F=1$ SUSY vacuum \indexextot.  For $m_2$ and $m_3$ both negative, the low-energy theory near the origin is $SU(2)_{-\half}$, which does not have a SUSY vacuum.  The location of the SUSY vacuum can be seen from the low-energy theory on a hypothetical Coulomb branch, using  \fsuii
\eqn\fsigmatwo{k_{eff}\sigma +\zeta _{eff}\equiv F(\sigma)=\sigma +\half (|m_2+\sigma|-|m_2-\sigma|)+\half (|m_3+\sigma |-|m_3-\sigma|).}
For $m_2$ and $m_3$ of opposite signs, the low-energy theory near the origin is $SU(2)_\half$, which again does not have a SUSY vacuum.   Taking the decoupling limit of large $m_{2,3}$, the SUSY vacuum is near the origin only if $m_2$ and $m_3$ are both positive, but in every other case the SUSY vacuum should decouple with large $m_{2,3}$.

Now consider $SU(2)_\half$ for $m_{1,2,3}$ all non-zero and generic.  The theory has a mass gap and $\Tr (-1)^F=1$ SUSY vacuum.   In terms of the low-energy theory of the $M_i$ fields, the real masses are $m_{M_i}\propto \sum _{j\neq i}m_j$, e.g.\ $M_3$ is massless if $m_1+m_2=0$, and there is a mass gap if all $m_{M_i}\propto \sum _{j\neq i}m_j \neq 0 $.  Note that the  SUSY vacuum must run off to infinite distance for $|m_1|\sim |m_2|\ll |m_3| \to \infty$, since we can there decouple $Q_3$ and connect to the low-energy $SU(2)_{\half (1+\sign (m_3))}$ effective theory, with $N_f=1$, which does not have a SUSY vacuum for $m_1+m_2\neq 0$.  Again, these K\"ahler metric details are less evident in the dual description in terms of the $M_i$ fields.

\subsec{$N_f=2$ flavors:  $Q^{i=1\dots 4}$, with $k\in \Z$.}

For $m_i=0$ and any $k$, there is a Higgs branch labeled by $M^{ij}=Q^iQ^j$, subject to the classical constraint ${\rm Pf}M=0$; the theory at $M^{ij}=0$, for all $k$, is an interacting SCFT.  The $k=0$ theory has the dual   \twoflavor, which matches the moduli space and the parity anomalies of the original $SU(2)_0$ theory \AharonyBX.    For generic real masses, the theory with general $k$ has index \indexnfi, $\Tr (-1)^F=|k|+1$.  So the $k=0$ theory has $\Tr (-1)^F=1$, which is an additional check of the duality of \twoflavor.

In the rest of this subsection, we consider the $k=0$ theory, with various non-zero real masses $m_i$ for the $Q^i$, flowing down to the theories considered in the previous subsections.   Without loss of generality, we can take $m_4>0$ and $m_4\geq |m_3|\geq |m_2|\geq |m_1|$.
In terms of the dual \twoflavor, the masses are
\eqn\dualmasses{m_{M^{ij}}=m_i+m_j, \qquad m_Y=-( m_1+m_2+m_3+m_4),}
as seen from matching the global symmetries.

Using \fsuiif, the low-energy theory on a hypothetical Coulomb branch has
\eqn\ftwof{F(\sigma)\equiv \zeta _{eff}+k_{eff}\sigma = \cases{m_1+m_2+m_3+m_4 & $\sigma >m_4$\cr m_1+m_2+m_3+\sigma & $m_4\geq \sigma \geq |m_3|$\cr
m_1+m_2+ \sigma (1+\sign (m_3)) &$ |m_3|\geq \sigma \geq |m_2|$\cr
m_1+\sigma (1+\sign m_3+\sign m_2) & $|m_2|\geq \sigma \geq |m_1|$\cr
\sigma (1+\sign (m_3)+\sign (m_2)+\sign (m_1)) & $|m_1|\geq \sigma \geq 0$.}}
The asymptotic $\sigma > m_4$ region of the Coulomb branch has $\zeta_{eff}=\sum_{i=1}^4 m_i$ and is lifted unless $\zeta _{eff}=0$;  this agrees with the dual theory \dualmasses, since $m_Y=-\zeta _{eff}$.

Consider first the case $m_4>0$, with $m_1=m_2=m_3=0$.  Using \ftwof, the Coulomb branch is always lifted, either by $\zeta _{eff}=m_4$ for $\sigma >m_4$, or by $k_{eff}=1$ for $\sigma <m_4$.  This fits with $m_Y=-m_4\neq 0$ in \dualmasses.  The Higgs branch moduli $M^{i4}$ are also lifted by $m_4\neq 0$.  The remaining massless moduli in the low energy theory are $M_k=\epsilon_{ijk} M^{[ij]}=\epsilon_{ijk}Q^{[i} Q^{j]}$ (with $i,j,k=1,2,3$) in the $\bar 3$ of the unbroken $SU(3)_F$.   In terms of the dual \twoflavor, the fields $M^{i4}$ and $Y$ are massive and can be integrated out, and the low-energy theory consists of the $M_i$ fields, with $W=0$.  This derives the dual of the $N_f={3\over 2}$ theory, discussed in the previous subsection, from the $N_f=2$ dual \twoflavor.  Note that integrating out $M^{i4}$ from the $N_f=2$ dual \twoflavor\ induces the background Chern-Simons term for the unbroken $SU(3)_F$ global symmetry, $\Delta k_{FF}=\half$, which we saw in the previous subsection was needed as an additional contribution in the low-energy theory of the $M_k$ fields, to match the $SU(3)_F$ parity anomaly $k_{FF}\in \Z$ of the original $Q$ fields.   Note also that the $SU(2)_0$ theory with $N_f=2$ and the $SU(2)_\half$ theory with $N_f={3\over 2}$ have the same index, $\Tr (-1)^F=1$, so no vacua need to move out or in from infinity in the $m_4\to \infty$ decoupling limit.

We now consider the $SU(2)_0$ theory with $m_{3,4}\neq 0$, with $m_{1,2}=0$.   For general $m_3+m_4\neq 0$, only $M^{12}$ remains massless in \dualmasses; the remaining $M^{ij}$, and $Y$ become massive.  In particular, \ftwof\ shows that the $\sigma \to \infty$ region of the Coulomb branch is lifted by $\zeta = m_3+m_4$, if non-zero.  For $m_3+m_4=0$, on the other hand, $Y$ and $M^{34}$ become massless.  More generally, $M^{12}$, $M^{34}$ and $Y$ are all massless when $m_3+m_4=m_1+m_2=0$.   Setting the other, massive fields to zero in the superpotential \twoflavor, the low-energy superpotential for the light fields in the dual theory in these cases is
\eqn\wlowex{W_{low}=YM^{12}M^{34},}
which looks similar to the $W=MX_+X_-$ dual of $N_f=1$ SQED -- it flows to a nontrivial fixed point, which is often referred to as the ``$XYZ$-theory.''

The dual \wlowex\ in this case seems at odds with the semi-classical picture of the moduli space, from the electric variables and \ftwof.  The semi-classical picture on the electric side suggests that there are two distinguished points on the Coulomb branch, $\sigma = |m_2|$ and $\sigma = m_4$, where the two Higgs branches have their roots.  These points separate the Coulomb branch to three distinct branches: $0\le \sigma <|m_2|$,  $|m_2|\le  \sigma \le m_4$ and $\sigma >m_4$.   If we consider $|m_i|\gg e^2$, the low-energy theory near both $\sigma = |m_2|$ and $\sigma = m_4$ is a copy of $N_f=1$ SQED, so the situation looks similar to \wex.

In fact, instanton effects modify this semi-classical picture, and the result is perfectly compatible with \wlowex.  First, an instanton leads to $W_{dyn}\neq 0$ in the region  $\sigma <|m_3|$, as in \AffleckAS, since there the instanton does not have any matter Fermion zero modes (see \kjcalliasf).  Next, the Coulomb branch region $m_4>\sigma > |m_3|$ gets quantum-merged, as in \oneflavor, with the $M^{12}$ Higgs branch; this is because the instanton there has precisely two
two matter Fermion zero modes (for the fields $Q_1$ and $Q_2$).  The upshot is perfectly compatible with the dual \wlowex: the $M^{12}$ and $M^{34}$ branches do intersect each other,  despite their semi-classical separation by distance $\Delta \sigma = |m_4|-|m_3|$, because the separation becomes part of the $M^{12}$ Higgs branch.    The non-compact $Y$ Coulomb branch also intersects this point, and there is an interacting SCFT there.

Now consider the case where $m_3$ and $m_4$ are non-zero, with $m_3+m_4\neq 0$, for $m_1=m_2=0$.   There is the unlifted $M^{12}$ Higgs branch at $\sigma =0$.  For $\sign (m_3)=-\sign (m_4)=-1$, the low-energy theory is $SU(2)_{k=0}$, and there is the associated unlifted Coulomb branch $|m_3|\geq \sigma \geq 0$, where $F(\sigma )=0$ in \ftwof; this region gets quantum-merged together with the $M^{12}$ Higgs branch, as in \oneflavor.  The other regions of the Coulomb branches are lifted by $F(\sigma )\neq 0$ in \ftwof.  As seen from the dual \twoflavor, the low-energy theory consists of the IR-free field modulus $M^{12}$.  In terms of \wlowex, the fields $Y$ and $M^{34}$ are set to zero by their real masses $m_{M^{34}}=-m_Y= m_3+m_4$.   For $m_3=m_4$, the low-energy theory at $\sigma = m_4$ is SQED with two massless fields of charge $-1$, but we see from \ftwof\ that $F(\sigma=m_4)>0$, which is the wrong sign for obtaining a compact $CP^1$ Higgs branch there.

For $m_3$ and $m_4$ large, the low-energy theory is $SU(2)_{\half (1+\sign (m_3))}$ with $N_f=1$ light flavor.   According to \indexnfi, this low-energy theory has $\Tr (-1)^F=0$ for $m_3<0$, or $\Tr (-1)^F=1$ for $m_3>0$.  In sum, upon turning on small $m_1$ and $m_2$ to lift all moduli, the supersymmetric vacuum of the $N_f=2$ theory should either run off to infinity and decouple, if $m_3<0$, or remain near the origin for $m_3>0$.  Again, these K\"ahler metric details are less evident in the $W=Y{\rm Pf}M$ dual description.

\subsec{$\CN =2$ $SU(2)_k$ with matter $Q$ in a triplet}

One can consider matter in general representations, with no analog of the 4d asymptotic freedom restriction on the matter representation. Here we just briefly mention one example.
Consider the theory of a single matter field $Q$ in the adjoint representation.  The index \indexsuii\ gives\foot{This theory differs from $\CN =3$ SYM, which has a {\it complex} mass superpotential term, with $m_C=k$, see e.g.\ \KapustinHA.  Since $m_C\neq 0$, the adjoint of $\CN =3$ SYM cannot be given a real mass, and so the theories need not have the same index.  Indeed, the $\CN =3$ SYM theory has $|k'_{\CN =3}|=|k'|-{3\over 2}h$ in \indexnomat, so it has $\Tr (-1)^F=|k|-2$. The $\CN =4$ case has $k=0$, so the index is ill-defined.}  $\Tr (-1)^F=|k|+1$ when $Q$ is given a real mass.   The theory with massless $Q$ has a moduli space, where $M=Q^2$ has an expectation value, breaking $SU(2)_k\to U(1)_{2k}$, with no matter.  The non-compact $M$ moduli space means that $\Tr (-1)^F$ is well-defined only if the adjoint is given a real mass $m_Q\neq 0$.

The index gives a simple check of the duality in  \JafferisNS\ between $SU(2)_1$, with an adjoint and a free field theory with field $M=Q^2$, plus a topological sector.  If not for the topological sector, there would be a mismatch.  Upon giving $Q$ a real mass, the index is $\Tr (-1)^F=2$, whereas an IR-free field $M$ contributes $\Tr (-1)^F=1$ for $m_M\neq 0$.  The index matching works upon including the tensor product with the topological sector $U(1)_2$, which is the
low-energy theory left unbroken by $\ev{M}$ for $m_Q=0$, and which indeed has $\Tr (-1)^F=2$.

\newsec{Preliminary aspects of $SU(N_c)$ and $U(N_c)$}

We here briefly discuss, in parallel,  gauge groups $G=SU(N_c)$ and $G=U(N_c)$, with $N_+$ matter fields $Q_f\in {\bf N_c}$ and $N_-$ matter fields $\tilde Q_{\tilde f}\in \bar{\bf N}_c$.    We can turn on real masses $m_f^g$ and $m_{\tilde f}^{\tilde g}$ for the matter fields, and Chern-Simons term $k$.  For the case of $U(N_c)$, we can also add an FI term ${\cal L}\supset -\int d^4\theta {\zeta \over 2\pi}\Tr V$.   (In this case we can also have different Chern-Simons coefficients for $SU(N_c)$ and $U(1)$, but we will not do it here.) The classical vacua satisfy \eqn\classeq{\eqalign{
&\left(\sigma_c^{c'}\delta_{f}^{f'}+ \delta_c^{c'}m_{f}^{f'}\right) Q^{c f} =0\cr
&\left(-\sigma_c^{c'}\delta_{\tilde f}^{\tilde f'} + \delta_c^{c'}m_{\tilde f}^{\tilde f'}\right) \tilde Q_{c' \tilde f} =0\cr
& D_c^{c'}\propto -{k\over 2\pi} \sigma_c^{c'} -{\zeta \over 2\pi} \delta _c^{c'}+ Q^{\dagger}_{cf}  Q^{c' f}- \tilde Q^{\dagger c'\tilde f}  \tilde Q_{c \tilde f} =0 ~~.}}
For $G=SU(N_c)$, we set $\zeta =0$ and relax the RHS of the last line in \classeq\ to be proportional to $\delta _c^{c'}$.   We can always choose the real Coulomb moduli $\sigma _c^{c'}=\sigma _c\delta _c^{c'}$ to be diagonal by a gauge rotation; the off-diagonal components are eaten by the massive gauge fields in the breaking to the Cartan subgroup, $U(N_c)\to U(1)^{N_c}$ or $SU(N_c)\to U(1)^{N_c-1}$.    We can restrict the $\sigma _c$ to the Weyl chamber $\sigma _1\geq \sigma _2\geq \dots \geq \sigma _{N_c}$.

The semi-classical vacua are obtained via \classeq\ with $k$ and $\zeta$ replaced with the one-loop exact quantum corrected expressions, much as in \zetakeff.  One then adds in the effects of instanton corrections.   Consider, for example, $U(N_c)_k$, with massless matter, expanded around $\sigma _c^{c'}\approx \sigma \delta _{c}^{c'}$, with $\sigma$ large.  Then $U(N_c)$ is approximately unbroken, and the matter fields are massive and can be integrated out leading to
\eqn\kuninf{k_{eff}(\sigma \to \pm \infty)= k\pm k_c, \qquad k_c\equiv \half (N_+-N_-)}
as in the $U(1)$ case \kinfinity.  For $k\neq \mp k_c$, this region of the Coulomb branch is lifted by $k_{eff}\neq 0$.  For $k=\mp k_c$, there is a non-compact Coulomb branch, which is partially lifted by instantons.

When the $\sigma _c$ all differ, we are out along the Coulomb branch, $G\to U(1)^r$, and the theory can be  approximately analyzed in terms of the low-energy $U(1)^r$ gauge theory, with the matter charges given by its $G$-rep weight vector.    We then need to add instanton effects in the broken $SU(2)$ subgroups of $G$.   In particular, each low-energy $U(1)$ gauge factor yields an approximate $U(1)_J$ global topological symmetry, so there is an approximate $\prod _{i=1}^r U(1)_{J_i}$ global symmetry.  Instantons explicitly break $N_c-1$ of these global $U(1)$ factors, so $U(N_c)$ has a single  $U(1)_J$ global symmetry, while $SU(N_c)$ has none.
The Coulomb branch can be described either in terms of real, linear multiplets $\Sigma_c$, in the $U(1)^r$ Cartan subalgebra, or dualized to chiral superfields.    In $U(N_c)$ we can have $X_{i\pm}\sim e^{\pm (\sigma _i/e_{eff, i}^2+i a_i)}$, with charges  $\pm 1$ under the erstwhile global $U(1)_{J_i}$ charge of the low-energy $U(1)^r$ Abelian theory.  In $SU(N_c)$ it is better to use $Y_i\sim X_{i+1,-}X_{i,+}$.

As discussed in \refs{\AharonyBX, \deBoerKR}, instantons indeed generate superpotential terms which lift most of the Coulomb branch moduli.  For $SU(N_c)$, a single Coulomb branch chiral superfield, $Y$, remains unlifted.  For $U(N_c)$, two Coulomb branch chiral superfields, $X_\pm$, remain unlifted.   It will be soon useful to recall here how this works.  Semi-classically, $X_{i+}\sim 1/X_{i-}$ and the $SU(2)$ subgroups of $G$ have $Y_i\sim X_{i+1,-}X_{i,+}\sim 1/X_{i+1,+}X_{i,-}$, $i=1\dots N_c-1$, with global $U(1)$ charges  which are similar to \ntwogen, but with the Callias index giving Fermion zero modes, $K_j\neq 0$, only for the $Y_{i_0}$ instanton factor which has $\sigma _{i_0+1}>0>\sigma _{i_0}$.  Instantons in these $SU(2)$ subgroups of $G$ generate the dynamical superpotential \refs{\AffleckAS, \ARSW}
\eqn\winst{W_{dyn} =\sum _i  {1\over Y_i}~.}
 The effect of this superpotential, is to set $\sigma_2=\cdots =\sigma _{N_c-1}=0$.
For $SU(N_c)$, there is a single, unlifted Coulomb modulus, $Y$,  given by
\eqn\ygenis{Y=\prod _{i=1}^{N_c-1} Y_i.}
For $U(N_c)$, there are two unlifted Coulomb moduli, $X_+\sim X_{1+}$ and $X_-\sim X_{N_c-}$, and we can replace $Y\to X_+X_-$.

We recall, for example, that $SU(N_c)_0$ and $U(N_c)_0$, with $N_f=N_c$ flavors, has a nice dual description in terms of the unlifted Coulomb moduli, together with the Higgs branch moduli,  with superpotential \AharonyBX
\eqn\wsun{SU(N_c)_0:\qquad \hbox{with $N_f=N_c$}:\qquad W=-Y(\det M-B\tilde B),}
\eqn\wun{U(N_c)_0:\qquad \hbox{with $N_f=N_c$}: \qquad W=-X_+X_-\det M.}
 The quantum description for $N_f<N_c$ follows from \wsun\ or \wun, upon adding complex mass terms and integrating out flavors.   For example, the theories with $N_f=N_c-1$ have quantum moduli space of supersymmetric vacua, with a constraint that follows from \wsun\ or \wun\ upon adding $W=m_{N_c}M_{N_cN_c}$ and integrating out the massive flavors:
 \eqn\wsunx{SU(N_c)_0\qquad\hbox{with $N_f=N_c-1$}: \qquad Y\det M=1.}
 \eqn\wunx{U(N_c)_0\qquad\hbox{with $N_f=N_c-1$:}\qquad X_+X_-\det M=1.}
 For $N_f<N_c-1$, this yields a runaway superpotential  for $Y$ or $X_+X_-$, so the theory with zero masses has no stable vacuum; upon adding real masses, they break supersymmetry.    The quantum description for $N_f>N_c$ requires additional degrees of freedom, e.g.\ those of the Aharony dual \AharonyGP.

We now consider these theories with $k\neq 0$, and more general matter content. Consider the case of $N_+$ matter fields $Q$ in the fundamental and $N_-$ matter fields $\tilde Q$ in the anti-fundamental.  For $k=0$, the quantum numbers are similar to those in \AharonyGP, with $N_f$ there replaced with $N_f\equiv \half (N_++N_-)$ for the $U(1)_A$ and $U(1)_R$ charges.  For $SU(N_c)_0$,
\eqn\sunch{
\vbox{\offinterlineskip\tabskip=0pt
\halign{\strut\vrule#
%%%%%%%%%%%%%%%%%%
&~$#$~\hfil\vrule
&~$#$~\hfil\vrule
&~$#$~\hfil\vrule
&~$#$~\hfil\vrule
&~$#$~\hfil\vrule
&~$#$~\hfil%\vrule
&\vrule#
\cr
%%%%%%%%%%%%%%%%%
\noalign{\hrule}
&  SU(N_c)_0 &  SU(N_+) &  SU(N_-) &  U(1)_A& U(1)_R  \cr
\noalign{\hrule}
%%%%%%%%%%%%%%%%%%
&  Q^i        & \; {\bf  N_+}   &\; {\bf  1}  &\; 1    &\;0 \cr
& \tilde Q_{\tilde i} & \; {\bf  1}   & \; {\bf  \bar N_-} & \; 1   & \;0   \cr
\noalign{\hrule}
%%%%%%%%%%%%%%%%%%
& M^i_{\tilde i}=Q^i \tilde Q_{\tilde i} & \; {\bf  N_+}  & \; {\bf  \bar N_-}       & \;2 & \; 0    \cr
&Y & \; {\bf  1} & \; {\bf  1} &\; -2N_f &\;  2(N_f-N_c+1) \cr
}\hrule}}
For $U(N_c)_0$, there is the extra $U(1)_J$ symmetry, and we replace $Y\to X_+X_-$
\eqn\unch{
\vbox{\offinterlineskip\tabskip=0pt
\halign{\strut\vrule#
%%%%%%%%%%%%%%%%%%
&~$#$~\hfil\vrule
&~$#$~\hfil\vrule
&~$#$~\hfil\vrule
&~$#$~\hfil\vrule
&~$#$~\hfil\vrule
&~$#$~\hfil%\vrule
&\vrule#
\cr
%%%%%%%%%%%%%%%%%
\noalign{\hrule}
&U(N_c)_0  &  SU(N_+) &  SU(N_-) &  U(1)_A& U(1)_R &  U(1)_J &\cr
\noalign{\hrule}
%%%%%%%%%%%%%%%%%%
&  Q^i        & \; {\bf  N_+}   &\; {\bf  1}  &\; 1    &\;0& \quad 0   &\cr
& \tilde Q_{\tilde i} & \; {\bf  1}   & \; {\bf  \bar N_-} & \; 1   & \;0 &  \quad 0   &\cr
\noalign{\hrule}
%%%%%%%%%%%%%%%%%%
& M^i_{\tilde i}=Q^i \tilde Q_{\tilde i} & \; {\bf  N_+}  & \; {\bf  \bar N_-}       & \;2 & \; 0  &   \quad  0   &\cr
&X_\pm & \; {\bf  1} & \; {\bf  1} &\; -N_f &\;  N_f -N_c+1 & \; \pm 1 &\cr
}\hrule}}
There is no analog of superpotentials \wsun\ or \wun\ if $N_+\neq N_-$, since the  holomorphic $SU(N_\pm)$ singlet object $\det M$ can only be formed from  $M^i_{\tilde i}$  if $N_+=N_-$.

Consider the general $N_\pm$ case, with all matter fields massless, $m_f^{f'}=m_{\tilde f}^{\tilde {f'}}=0$ in \classeq.   There is a non-compact, classical Higgs branch $\CM _H^{cl}$ moduli space of solutions of \classeq, at $\sigma _c^{c'}=0$, as long as $N_+N_-\neq 0$.  Taking $N_+\geq N_- >0$, on $\CM _H^{cl}$ the gauge group is broken as $SU(N_c)\to SU(N_c-N_-)$, or $U(N_c)\to U(N_c-N_-)$.  So the $SU(N_c)$ gauge group is fully broken if $N_-\geq N_c-1$, and the $U(N_c)$ gauge group is fully broken if $N_-\geq N_c$.   In the cases where the gauge group is fully broken, there is a quantum Higgs branch moduli space of SUSY vacua, $\CM_H\cong \CM _H^{cl}$.  When the gauge group is not fully broken, the unbroken gauge group will break supersymmetry if $|k|<N_c-N_-$, with a runaway Coulomb branch in the $k=0$ case.

\newsec{From Giveon-Kutasov to Aharony}

Aharony duality \AharonyGP\ is unusual in that the dual Lagrangian contains terms explicitly involving the dual theory's monopole operators. One might reasonably question whether or not this is a legitimate theory: is one allowed to include monopole operators in the defining Lagrangian of a theory?  We here show how to UV-complete this theory to one without the monopole operators in the Lagrangian.  We do this by showing that Aharony duality can be derived via an RG-flow, as an IR-consequence of Given-Kutasov duality in the UV. The monopole operators in the dual Lagrangian are in this way UV-completed to a more-conventional theory without them.

In the following subsections, we briefly review these dualities in greater detail, and then show how to derive Aharony duality as a consequence of Given-Kutasov duality.

\subsec{Review of Aharony duality}

 The electric theory is a $U(N_c)_0$ gauge theory, with $N_f$ flavors in the fundamental and anti-fundamental, $Q^{cf}$ and $\tilde Q_{c\tilde f}$.  The gauge and global charges, including those of the composite  meson $M=Q\tilde Q$ and the monopole operators $X_\pm$,  are as in \unch\ with $N_+=N_-=N_f$.
The dual theory \AharonyGP\ is a $U(n_c=N_f-N_c)_0 $ gauge theory with quarks and anti-quarks $q$ and $\tilde q$, and the elementary fields $M$, $ X_\pm$, with quantum numbers
\eqn\Aham{
\vbox{\offinterlineskip\tabskip=0pt
\halign{\strut\vrule#
%%%%%%%%%%%%%%%%%%
&~$#$~\hfil\vrule
&~$#$~\hfil\vrule
&~$#$~\hfil
&~$#$~\hfil%\vrule
&~$#$\hfil
&~$#$\hfil
&~$#$\hfil
&\vrule#
\cr
%%%%%%%%%%%%%%%%%%
\noalign{\hrule}
&  &  U(n_c)_0 & SU(N_f)&   SU(N_f)& U(1)_A & U(1)_J & U(1)_R &\cr
\noalign{\hrule}
%%%%%%%%%%%%%%%%%%
&  q^{\tilde c}_f        & \; {\bf {n_c}}     & \; {\bf \b N_f}    &\; {\bf  1} & \quad -1  &    \quad  0 &  \quad  1   &\cr
& \tilde q_{\tilde c}^{\tilde f}                & \; {\bf \b { n_c} }    & \; {\bf 1}   & \; {\bf N_f}   & \quad -1  &    \quad  0 &   \quad 1   &\cr
%%%%%%%%%%%%%%%%%%
& M  ^f_{\tilde f}             & \; {\bf  1}     & \; {\bf N_f}   & \; {\bf \b N_f}   & \quad 2  &    \quad  0 &  \quad  0   &\cr
& X_\pm              & \; {\bf  1}     & \; {\bf 1}   & \; {\bf 1}   & \quad -N_f  &    \quad \pm 1 &  \quad  N_f-N_c+1   &\cr}
\hrule}}
The dual theory has the superpotential \AharonyGP\
\eqn\supm{W=M q\tilde q + X_+\tilde X_- + X_-\tilde X_+ ~.}
where $\tilde X_\pm$ are the monopole operators of the magnetic theory. It should be stressed that, due to the explicit appearance of the magnetic operators $\tilde X_\pm$ in the Lagrangian, the magnetic theory is not a standard weakly coupled renormalizable Lagrangian.  At weak coupling the last two terms in \supm\ are high dimension operators.

\subsec{Giveon-Kutasov Duality}
 The electric theory is $U(N_c)_k$ theory with $N_f$ quarks.  The quantum numbers are as in \unch\ with $N_+=N_-=N_f$ except that, for $k\neq 0$, we should omit  the operators $X_\pm$ which are not gauge invariant (as in \onengen, they get charge $\sim k$ from the overall $U(1)\subset U(N_c)$).
 \eqn\GKe{
\vbox{\offinterlineskip\tabskip=0pt
\halign{\strut\vrule#
%%%%%%%%%%%%%%%%%%
&~$#$~\hfil\vrule
&~$#$~\hfil\vrule
&~$#$~\hfil
&~$#$~\hfil%\vrule
&~$#$\hfil
&~$#$\hfil
&~$#$\hfil
&\vrule#
\cr
%%%%%%%%%%%%%%%%%
\noalign{\hrule}
&  &  U(N_c)_k & SU(N_f)&   SU(N_f)& U(1)_A & U(1)_J & U(1)_R &\cr
\noalign{\hrule}
%%%%%%%%%%%%%%%%%%
&  {Q^{cf}}         & \; {\bf  N_c}     & \; {\bf N_f}    &\; {\bf 1} & \quad 1  &    \quad  0 &  \quad  0   &\cr
& \tilde Q_{c \tilde f}                & \; {\bf \b N_c}     & \; {\bf 1}   & \; {\bf\b N_f}   & \quad 1  &    \quad  0 &   \quad 0   &\cr
\noalign{\hrule}
%%%%%%%%%%%%%%%%%%
& M^f_{\tilde f}               & \; {\bf  1}     & \; {\bf N_f}   & \; {\bf \b N_f}   & \quad 2  &    \quad  0 &  \quad  0   &\cr}
\hrule}}
The dual theory \GiveonZN\ has gauge group $U(n_c)_{-k}$, with $n_c= N_f + |k|-N_c$ and matter
\eqn\GKm{
\vbox{\offinterlineskip\tabskip=0pt
\halign{\strut\vrule#
%%%%%%%%%%%%%%%%%%
&~$#$~\hfil\vrule
&~$#$~\hfil\vrule
&~$#$~\hfil
&~$#$~\hfil%\vrule
&~$#$\hfil
&~$#$\hfil
&~$#$\hfil
&\vrule#
\cr
%%%%%%%%%%%%%%%%%%
\noalign{\hrule}
&  &  U(n_c)_{-k} & SU(N_f)&   SU(N_f)& U(1)_A & U(1)_J & U(1)_R &\cr
\noalign{\hrule}
%%%%%%%%%%%%%%%%%%
&  {q^c_f}         & \; {\bf {n_c}}     & \; {\bf \b N_f}    &\; {\bf  1} & \quad -1  &    \quad  0 &  \quad  1   &\cr
& \tilde q_{c\tilde f}                & \; {\bf \b { n_c} }    & \; {\bf 1}   & \; {\bf N_f}   & \quad -1  &    \quad  0 &   \quad 1   &\cr
%%%%%%%%%%%%%%%%%%
& M ^f_{\tilde f}              & \; {\bf  1}     & \; {\bf N_f}   & \; {\bf \b N_f}   & \quad 2  &    \quad  0 &  \quad  0   &\cr}
\hrule}}
The dual has the superpotential \GiveonZN
\eqn\GKsupm{W=M q\tilde q ~.}

As a warmup, we outline how Giveon-Kutasov duality can be derived from Aharony duality.   Consider starting with an electric $U(N_c)_0$ with $N_f+|k|$ flavors, and turn on equal mass $m$ for $|k|$ flavors, with $\sign (k)=\sign (m)$.   In the  $U(N_f+|k|-N_c)_0$ Aharony dual of the original electric theory, the real masses map to real masses $-m$ for $|k|$ flavors, real masses for $X_\pm$, and real masses for the $(N_f+|k|)^2-N_f^2$ components of the singlets $M$ with flavor indices for the massive electric flavors.  The masses follow from the $U(1)_A$ assignments in \Aham.   Integrating out the massive flavors on both sides, the electric theory flows in the IR to $U(N_c)_k$ with $N_f$ flavors, and the Aharony dual flows in the IR to the $U(N_f+|k|-N_c)_{-k}$ Giveon-Kutasov dual.  The peculiar $\tilde X_\pm$ of the UV Aharony dual \supm\ decouple in the IR limit, together with the elementary fields $X_\pm$, which got real masses.

\subsec{Turning on a real mass for one flavor in $U(N_c)_k$}

We start with the electric side \GKe\ of Giveon-Kutasov duality, $U(N_c)_k$ theory with $N_f$ quark flavors, $Q^f$ and $\tilde Q_{\tilde f}$, and consider turning on real masses.  The classical vacua satisfy \classeq, where we set the FI term $\zeta =0$.  For simplicity, consider the case where we give a non-zero real mass to only one flavor, giving the same real mass to $Q^{N_f}$ and $\tilde Q_{N_f}$, \
\eqn\massmatrix{m_f^{f'} =m \delta_{f N_f} \delta^{f' N_f} \qquad ; \qquad \tilde m_f^{f'} =m \delta_{\tilde f N_f} \delta^{\tilde f' N_f} ~~}
and we take $m>0$; the results for $m<0$ are similar.  The solutions of \classeq\ are\foot{A closely related discussion appeared in \Shamirthesis.}:
\item{1.} $\sigma=0$, with $Q^{cN_f}=\tilde Q_{cN_f}=0$.  The theory here has a Higgs branch for the remaining $N_f-1$ massless flavors.  Integrating out the massive flavor, the low-energy theory at the root of the Higgs branch is a $U(N_c)_{k+1}$ gauge theory with $N_f-1$ flavors.  The case $k=-1$, where there is a Coulomb branch, will be discussed separately below.
\item{2.} $\sigma_{N_c}^{N_c} \approx -m$, with the other entries of $\sigma$ vanish, breaking the gauge symmetry as $U(N_c) \to U(N_c-1)\times U(1)$.  The unbroken $U(N_c-1)$ factor is similar to the previous case: it has $N_f-1$ massless flavors and one massive flavor, which leads to $U(N_c-1)_{k+1}$ with $N_f-1$ massless flavors.  The unbroken $U(1)$ gauge factor has $N_f$ fields of charge $\pm 1$, from the $c=N_c$ color component of the original $Q^f$ and $\tilde Q_{\tilde f}$.  Of these $N_f$ charged flavors, $N_f-1$ have real masses $\pm \sigma _{N_c}^{N_c}$,
 so they can be integrated out and their one-loop corrections to $\zeta _{eff}$ and $k_{eff}$ in \zetakeff\ cancel.  The $U(1)$ charge $+1$ matter field $Q^{N_c, N_f}$ has real mass $\sigma _{N_c}^{N_c}+m\approx 0$, so it is light and kept in the low-energy theory.  The $U(1)$ charge $-1$ matter field $\tilde Q_{N_c, N_f}$ has real mass $m-\sigma _{N_c}^{N_c}\approx 2m$, so it is integrated out and then \zetakeff\ gives $\zeta \to \zeta +m$, $k\to k+\half$.  It is natural to shift $\sigma _{N_c}^{N_c}$ by $\delta \sigma =m$ to vanish in the vacuum; as in \ginva, this additionally shifts $\zeta \to \zeta +km$.  In sum, in addition to $U(N_c-1)_{k+1}$ with $N_f$ flavors, there is a low-energy, decoupled $U(1)_{k+\half}$ factor, with a single light field $Q=Q^{N_c, N_f}$, of charge $+1$ and
    \eqn\lowzetaa{\zeta = m(k+1) ~.}
We analyze this low-energy $U(1)_{k+\half}$ theory as in the discussion following eqn.    \condos, with $k_{low}=k+\half$ and $\zeta$ given by \lowzetaa.   According to \indexgen, this $U(1)_{k+\half}$ theory generically has $\Tr (-1)^F=|k+\half |+\half$ SUSY vacua, but the topological vacua do not count, since they are far from $\sigma _{N_c}^{N_c}\approx -m$, so outside of our low-energy effective theory; only the Higgs vacua are in the low-energy theory.  For generic $k<-1$, the effective FI term $F(\sigma _Q)<0$ and there is no supersymmetric vacuum.  For $k>0$, there is an isolated supersymmetric vacuum, with a mass gap, on the Higgs branch, with $Q\sim  \sqrt{ \zeta}$.  The $U(1)$ gauge field and charged matter $Q$ are massive, so the remaining low energy theory is $U(N_c-1)_{k+1}$ with $N_f-1$ flavors.  Again, $k=-1$ is a special case, giving $\zeta_{low}=0$, $k_{low}=+\half$, which we discuss separately below.
\item{3.}  $\sigma_{N_c}^{N_c} \approx m$, with the other entries of $\sigma$ vanish. The analysis here is similar to the previous case with $Q\leftrightarrow \tilde Q$.  The low-energy theory is $U(N_c-1)_{k+1}\times U(1)_{k+\half}$, where the $U(1)$ has a single light field, of charge $-1$, with $\zeta =-m(k+1)$.  For generic $k$, there is a supersymmetric vacuum on the Higgs branch if $k>0$, with $\tilde Q^{N_c, N_f}\sim \sqrt{-\zeta}$, and no supersymmetric vacuum if $k<-1$.
\item{4.} The only nonzero entries of $\sigma$ are $\sigma_{N_c}^{N_c} =-\sigma_{N_c-1}^{N_c-1}=m$, breaking $U(N_c)_k\to U(N_c-2)_{k+1}\times U(1)_{k+\half} \times U(1)_{k+\half}$.
This is similar to the above two cases: the $U(N_c-2)$ factor has $N_f-1$ massless flavors, and the $U(1)_{k+\half}$ each have a single charged field, $N_f=\half$.  Again, for $k<-1$ these $U(1)$ factors do not give supersymmetric vacua within the low-energy theory, while for $k>0$ they give a single supersymmetric vacua, with only $U(N_c-2)_{k+1}$ with $N_f-1$ flavors remaining light.

\bigskip
\noindent
Next, consider the special case $k=-1$ and follow the four cases mentioned above:
\item{1.} The  low energy theory is $U(N_c)_{0}$ with $N_f-1$ flavors and a Coulomb branch.
\item{2.} The $U(1)$ factor is $U(1)_{-\half} $ with a single charge $+1$ chiral superfield and no FI-term (see \lowzetaa).  This theory is dual at low energies to a single chiral superfield $U_+$ (the reason for the subscript will be clear below), representing  half of the Coulomb branch of this chiral Abelian theory.  The $U_+$ chiral superfield must couple to the $U(N_c-1)_0$ fields and in particular, to the operator $X_-$ associated with the $U(1) \subset U(N_c-1)$.  The original $U(N_c)$ theory had a single topological symmetry under which $U_+$ and $X_-$ have opposite charges. Therefore, effects in the broken high energy $U(N_c)$ theory must break the separate topological symmetries of the two factors $U(N_c-1)\times U(1)$.  In particular, monopole configurations that acts as $3d$ instantons must do the job and produce a superpotential as in  \winst
    \eqn\firstdyS{W\sim U_+X_-~.}
    In conclusion, the low energy theory is $U(N_c-1)_0$ with $N_f-1$ flavors and a chiral superfield $U_+$ with the superpotential \firstdyS.
\item{3.} This case is similar to the previous one.  The low energy theory is $U(N_c-1)_0$ with $N_f-1$ flavors and a chiral superfield $U_-$.  In addition, it has a superpotential
    \eqn\seconddyS{W\sim U_-X_+~,}
    where $X_+$ is constructed out of the $U(N_c-1)$ fields.
\item{4.} Repeating the discussion in the two previous cases, we find a  $U(N_c-2)_0$ theory with $N_f-1$ flavors and two chiral superfields $U_\pm$ with the superpotential
    \eqn\thirddyS{W\sim U_+ X_- + U_-X_+~.}

\subsec{Flowing by real masses from a pair of dual Giveon-Kutasov theories.}

The previous discussion can be interpreted as flowing from the electric Giveon-Kutasov theory.  We see that the answers depend on the value of $k$.  It is straightforward to repeat this analysis in the magnetic theory $U(n_c=N_f+|k| - N_c)_{-k}$ with $N_f$ flavors, chiral superfields $M$ and the superpotential \GKsupm.  The electric real masses \massmatrix\ map, according to the global symmetries, to real masses in the dual for $q_{N_f}$, $\tilde q^{N_f}$, $M^f_{\tilde N_f}$, and $M^{N_f}_{\tilde f}$. The discussion is similar to the above so we will not repeat it in detail.  Again, there are four kinds of vacua to consider and the answers depend on if $k>0$, $k<-1$, or $k=-1$.  The main point is that the four kinds of vacua in the electric theory and the analogous four kinds of vacua in the magnetic theory are mapped as
\eqn\vacuama{\eqalign{
&1\leftrightarrow 4\cr
&2\leftrightarrow 3}}

In particular, starting with $U(N_c)_{-1}$ with $N_f$ flavors on the electric side, vacuum 1 gives a low-energy $U(N_c)_0$ theory with $N_f-1$ flavors.  In the magnetic dual, we start with $U(N_f+1-N_c)_{1}$ and flow in vacuum 4 to $U(N_f-1-N_c)_0$ theory with the superpotential \supm, where the $X_+\tilde X_-+X_-\tilde X_+$ terms come from the analog of \thirddyS\ in the magnetic dual, so $U_\pm \to X_\pm$ and $X_\pm \to \tilde X_\pm$.

We learn several lessons from this analysis.  First, flowing from a Giveon-Kutasov dual pair we find a pair of Aharony duals.  This gives additional evidence for the two kinds of dualities.  Furthermore, we see where the mysterious fields $X_\pm$ in Aharony duality come from, and how to interpret the superpotential \thirddyS, which couples to complicated operators $U_\pm$.
We also derived a new duality (which can be derived trivially from Aharony duality) from the $2\leftrightarrow 3$ vacua duality exchange:  The electric theory is a $U(N_c)_0$ theory with $N_f$ flavors, a singlet $U_+$ and a superpotential \firstdyS\ $U_+X_-$.  The magnetic theory is a $U(N_f-N_c)_0$ theory with $N_f$ flavors, singlets $M$, a singlet $\tilde X_+$, and a superpotential $qM\tilde q + U_-\tilde X_+$ where $U_-$ is constructed out of the magnetic gauge fields.
Finally, we can extend this analysis and flow in the entire chain of dualities by turning on positive and negative real masses.  The main subtlety is how to flow to $k=0$, where additional fields show up.  Our discussion shows how to do it.

\bigskip
\noindent {\bf Acknowledgments:}

We would like to thank Antonio Amariti, Cyril Closset, Thomas Dumitrescu, Guido Festuccia, Davide Gaiotto, Daniel Jafferis, Hans Jockers, Zohar Komargodski, John McGreevy, Shlomo Razamat, Brian Willett, and Edward Witten for discussions.  We especially thank Ofer Aharony and David Tong for helpful discussions, correspondence, participation in some of this work, and thoughtful comments about the manuscript.   We would also like to thank Jihye Seo and McGill University for the ``N=2 Jeometry And ApplicationZ"
workshop, where this collaboration was initiated. The research of NS is supported
in part by DOE grant DE-FG02-90ER40542.  The research of KI was supported in part by DOE-FG03-97ER40546.  KI also thanks the IHES, CdF, Jussieu U., and the ENS for hospitality and support in Paris in August 2012, while some of this work was done.

\appendix{A}{Notation and additional details, review}

We here set our notation.  There are some crucial relative signs and normalizations, between the Chern-Simons term, the Fayet-Iliopoulos term, and the sign of their shifts when matter is integrated out, which must be properly fixed to obtain consistent results.

\subsec{Susy algebra and conventions}

We use the conventions of \WessCP, reduced to 3d, e.g.\ the metric is\foot{In Euclidean space, $iS_M\to -S_E$, e.g.\ introducing an $i$ factor in the Chern-Simons action.} $\eta _{\mu \nu}=diag(-++)$. The supersymmetry algebra is
\eqn\zenop{\{Q_\alpha, \bar Q_\beta\} = 2\gamma^\mu_{\alpha\beta} P_\mu + 2i \epsilon_{\alpha\beta} Z.}
The 3d $\gamma ^\mu$ matrices satisfy  $(\gamma ^\mu)_\alpha {}^\beta (\gamma ^\nu)_\beta  {}^\lambda = \eta ^{\mu \nu}\delta _\alpha {}^\lambda +\epsilon ^{\mu \nu \rho}(\gamma _\rho )_\alpha{}^\lambda$. We choose the spinor basis\foot{ Another sometimes-convenient basis is one that diagonalizes  $(\gamma ^0)_\alpha {}^\beta$, and the $\half [\gamma _1, \gamma _2]$ angular momentum contribution, $(\gamma ^{\mu =0,1,2})_\alpha {}^\beta = (i\sigma _3, \sigma _2, \sigma _1)$.} that diagonalizes $\gamma ^0_{\alpha \beta}$:
\eqn\gammawb{\gamma^{\mu=0,1,2}_{\alpha\beta}= -{\bf 1}, \sigma^1, \sigma^3,}
which are  the $(\sigma ^\mu)_{\alpha \dot \beta}$ of \WessCP, reduced along the $x^{\mu =2}$ direction, and the sign of $\gamma ^0$ fits with positive energy, $P^0=-P_0\geq 0$.

Indices are raised, lowered, and contracted with $\epsilon ^{\alpha \beta}$ or $\epsilon _{\alpha \beta}$ (as in appendix B of \DumitrescuIU), e.g. $\psi \chi \equiv \epsilon ^{\alpha \beta}\psi _\alpha \chi _\beta$, and $\psi \bar \chi \equiv \epsilon ^{\alpha \beta}\psi _\alpha \bar \chi _\beta$.  In this notation\foot{The notation of \GatesNR, where indices are contracted with $C^{\alpha \beta}=i\epsilon ^{\alpha \beta}$ nicely eliminates the minus sign and $i$ above, but we will use the above convention to remain closer to \WessCP.}, $\overline{(\psi \chi)}= \epsilon ^{\alpha \beta }\bar \chi _\beta \bar \phi _\alpha=-\bar \chi \bar \phi$, and $i\theta \bar \theta$ and $\bar \theta \gamma ^\mu \theta$ are real, with e.g. $(\theta \bar \theta)^2=\half \theta ^2\bar \theta ^2$. Supercharges $Q_\alpha$ are represented in superspace by differential operators $\CQ _\alpha$ which act on superfields $\CO$ similar to \WessCP
\eqn\diffopi{\delta _{\zeta}\CO = i[\zeta Q-\bar \zeta \bar Q, \CO]=(\zeta \CQ -\bar \zeta \bar \CQ)\CO,}
where the supercharges can be written as
\eqn\qare{\CQ _\alpha = {\partial \over \partial \theta ^\alpha}-i\gamma ^\mu _{\alpha \beta}\bar \theta ^\beta \partial _\mu , \qquad \bar \CQ _\alpha = -{\partial \over \partial \bar \theta ^\alpha }+i\theta ^{\beta} \gamma ^\mu _{\beta\alpha}\partial _\mu.}
The superspace derivatives, anti-commuting with \qare\ are
\eqn\dare{D_\alpha =  {\partial \over \partial \theta ^\alpha}+i\gamma ^\mu _{\alpha \beta }\bar \theta ^\beta \partial _\mu , \qquad \bar D_\alpha = -{\partial \over \partial \bar \theta ^\alpha }-i\theta ^\beta \gamma ^\mu _{\beta \alpha} \partial _\mu.}

\subsec{Gauge fields and Lagrangians}

A $U(1)$ gauge theory has vector multiplet (in Wess-Zumino gauge \WessCP) $V=-i\theta\bar \theta \sigma -\theta \gamma ^\mu \bar \theta A_\mu +i \theta ^2\bar \lambda -i\bar \theta ^2\theta \lambda +\half \theta ^2\bar\theta ^2 D$. The gauge field strength is in the real linear multiplet
\eqn\sigmais{\eqalign{\Sigma &\equiv   -{i\over 2}\epsilon ^{\alpha \beta} \overline{D}_\alpha D_\beta V\equiv 2\pi {\cal J}_J
=\sigma +\theta \bar \lambda +\bar \theta \lambda + \half \theta \gamma ^\mu  \bar \theta F^{\nu \rho }\epsilon _{ \mu \nu\rho }+\cr &+i\theta \bar \theta D+{i\over 2}\bar \theta ^2 \theta \gamma ^\mu \partial _\mu \lambda -{i\over 2}  \theta ^2\bar \theta \gamma ^\mu \partial _\mu \bar \lambda +{1\over 4} \theta ^2\bar \theta ^2 \partial ^2 \sigma ,}}
with $D^2\Sigma = \bar D^2\Sigma =0$.  Then \diffopi\ gives, for example, the photino variation
\eqn\lamvar{i\{ Q_\alpha, \lambda _\beta\}= \gamma ^\mu _{\alpha \beta} (-i\partial _\mu \sigma +
\half  F^{\nu \rho }\epsilon _{ \mu \nu\rho })+i\epsilon _{\alpha\beta}D.}
${\cal J}_J$ is the current superfield for the $U(1)_J$ global symmetry, with topological current, $j_J^\mu \equiv {1\over 2\pi} \epsilon ^{\mu \nu \rho}\partial _\nu A_\rho$, so $j_J^0={F_{12}\over 2\pi} \equiv {B\over 2\pi}$.  The gauge kinetic, CS, and FI terms are\foot{ Our sign convention is that integrating out a Fermion of real mass $m>0$ induces $\Delta k>0$.  The following are odd under parity and time reversal: $k$, $m_i$, $\sigma$.  Under charge conjugation ${\cal C}$, $V\to -V$, or equivalently we can leave $V$ alone and take $n_i\to -n_i$; other quantities are ${\cal C}$ even. The following quantities all have dimensions of mass: $e^2$, $\Sigma$, $\zeta$; while $k$ and $V$ are dimensionless.}
\eqn\lgauge{\CL _{gauge}=\int d^4 \theta \left(-{1\over e^2} \Sigma ^2- {k\over 4\pi} \Sigma V-\FI V\right)}
where $e$ is the gauge coupling constant.  The terms involving the gauge field include
\eqn\lgaugec{\CL _{\rm gauge}\supset -{1\over 4e^2} F_{\mu \nu}F^{\mu \nu}+{k\over 4\pi}\epsilon ^{\mu \nu\rho}A_\mu \partial _\nu A_\rho + A_\mu j_{matter}^\mu,}
so the $A_\mu$ equations of motion are
\eqn\eom{-{1\over e^2}\partial _\nu F^{\nu \mu}=j_{matter}^\mu +k j_J^\mu\equiv j^\mu _{Gauss}. }
The CS term $k$ thus imparts electric charge to states carrying $U(1)_J$ topological charge, as in \Gaussk.  For states with $q_{Gauss}=0$, as in \eleck, $q_{matter}=-kq_J$. Vortex states with $q_J\neq 0$ acquire electric charge $-k_{eff}q_J$, from \eleck\ in the effective theory.  The consistency condition \consk, $k_{eff}\in \Z$, then follows from the Dirac quantization condition.

For $U(1)_k$ with chiral superfields $Q_i$ of charges $n_i$, the matter has classical terms
\eqn\matterk{\CL _{kin}=\sum _{i=1}^N \int d^4\theta Q^\dagger _i e^{n_i V+im_i\theta \bar \theta} Q_i.}
The auxiliary $D$ component of the $U(1)$ gauge multiplet appears in the Lagrangian as
\eqn\Dterms{\CL _{cl}\supset {1\over 2 e^2}D^2+{D\over 2}\left(\sum _i n_i |Q_i|^2-{k\over 2\pi}\sigma -\FI \right).}
Solving for $\ev{D}$,
\eqn\Deq{\ev{D}=-{e^2\over 2}(\sum_i n_i|Q_i|^2 - {\zeta \over 2\pi}- {k \over 2\pi} \sigma ),}
the classical potential of the system is
\eqn\classpot{V_{cl}= {e^2 \over 32\pi ^2}\left( \sum_i 2\pi n_i |Q_i|^2 -\zeta -k \sigma\right)^2 + \sum_i (m_i + n_i\sigma)^2 |Q_i|^2~~.}

Quantum corrections modify $\zeta$ and $k$ in \classpot, as in \claspot.  More generally, including the gauge kinetic term, the quantum corrections   to \lgauge\ are of the form
\eqn\lgaugec{\CL _{\rm gauge, eff }=-\int d^4\theta \left( f(\Sigma) +{1\over 4\pi} (k_{eff} \Sigma V + 2 \zeta _{eff} V)\right)~,}
where $k_{eff}$ and $\zeta_{eff}$ are one loop exact, as in  \zetakeff.
The function $f(\Sigma)$ in \lgaugec\ is given at one loop by
\eqn\fis{f(\Sigma) = {1\over e^2} \Sigma ^2+2s\Sigma \log (\Sigma /e^2) + \cdots ~,}
with $s=\half \sum _i n_i^2$ for the $U(1)$ theory.  The non-Abelian case is similar, see  \refs{\SeibergNZ, \deBoerKR}, with negative contributions to $s$ in \fis\ from integrating out the massive, non-Cartan gauge fields on the Coulomb branch, e.g.\  $s=N_f-3$ for $SU(2)$ with $N_f$ flavors.

For a non-Abelian gauge theory, the $\CN =2$ SUSY Chern-Simons term is
\eqn\snonab{S_{CS}^{\CN =2}={k\over 4\pi}\int \Tr (AdA + {2\over 3}A^2-\bar \lambda \lambda +2D\sigma)}
(see e.g.\ \refs{\ZupnikRY\IvanovFN\GatesQN-\GaiottoQI}\ for $S_{CS}^{\CN=2}$ written in superspace).
If the gauge group is $U(N_c)$ we can also add a FI term $\CL \supset -\int d^4\theta {\zeta \over 2\pi}\Tr V$.  For the $\CN =2$ theory with matter chiral superfields in representations $r_f$ of the gauge group $G$, the quantization condition is $k+\half \sum _f T_2(r_r)\in \Z$, with $T_2(r_f)=1$ e.g.\ for a fundamental or anti-fundamental of $SU(N_c)$ or $U(N_c)$.  Integrating out massive matter shifts $k_{eff}=k+\half \sum _f T_2(r_f) \sign (m_f(\phi))$.  For $k_{eff}=0$, there is a Coulomb branch with $G\to U(1)^r$; when $k_{eff}\neq 0$, the Coulomb branch is lifted.

\subsec{Vortices and their zero modes}

We here review and collect some formulae for vortices in 3d, ${\cal N}=2$ SUSY, $U(1)_k$ Chern-Simons-Maxwell theory with matter.  See e.g.\ \refs{\LeeEQ, \LeePM}\ for more details.

BPS field configurations are annihilated by $Q_-$ and $\bar Q_+$, and anti-BPS configurations are annihilated by $Q_+$ and $\bar Q_-$, with $Q_\pm=\half (Q_1\pm i Q_2)$ and $\bar Q_\pm = \half (\bar Q_1\pm i\bar Q_2)$ \Qpmd.  In Wess-Zumino gauge, the variations of the photino, with $\lambda _\pm = \lambda _1\pm i \lambda _2$, are
\eqn\photiv{\eqalign{
&\{Q_-, \lambda_-\}= F_{ \bar z t } + \partial_ {\bar z} \sigma   \cr
&\{Q_-, \lambda_+\} = F_{z\bar z} +  \partial_t \sigma -iD }\qquad \eqalign{
&\{Q_+, \lambda_+\}= F_{ z t } - \partial_ {z} \sigma   \cr
&\{Q_+, \lambda_-\} = F_{z\bar z} +  \partial_t \sigma + iD }}
with $F_{z\bar z}=iF_{xy}$ since $z=x+iy$.
The complex conjugate equations are obtained via $Q_\pm \to \bar Q_\mp$, $\bar \lambda _\pm \to \lambda _\mp$, $z\leftrightarrow \bar z.$  The variation of a Fermions in chiral multiplets $Q_i=\phi _i +\theta \psi _i +\theta ^2 F_i$, with charges $n_i$,  for mass $m_i=0$,  are (with $D_\mu \to  \partial _\mu -i n_i A_\mu$ and $D_z\equiv \half (D_x-iD_y)$)
\eqn\Fermionv{\eqalign{\{\bar Q_+, \psi_{i+}\}&= D_ z \phi_i,\cr
\{\bar Q_+, \psi_{i-}\} &= iD_t \phi_i +  n_i\sigma \phi_i, \cr
\{Q_-, \psi_{i+}\}&= F_i,}\qquad \eqalign{\{\bar Q_-, \psi_{i-}\}&= D_ {\bar z} \phi_i,\cr
\{\bar Q_-, \psi_{i+}\} &= -iD_t \phi_i +  n_i\sigma \phi_i, \cr
\{Q_+, \psi_{i-}\}&= F_i.} }
with similar complex conjugate equations for the Fermions in the anti-chiral multiplets.

 The BPS equations obtained from \photiv\ and \Fermionv\ are thus
\eqn\bpst{\partial _t \sigma =0, \qquad F_{zt}+\partial _z\sigma =0, \qquad i\partial _t \phi _i+n_i(\sigma +A_0)\phi _i=0,}
\eqn\finaleq{F_{z\bar z }-iD=0,}
\eqn\mattereq{D_z \phi_i=F_i=0.}
Equations \bpst\ can be solved in a static configuration, setting all $\partial _t\to 0$ and $A_0=-\sigma$.  Integrating \finaleq\ over the spatial plane yields
\eqn\fxyint{2\pi q_J=-i \int d^2z  F_{z\bar z}= \int d^2 z D\equiv {\zeta e^2 \over 4\pi} R_{core}^2,}
where $R_{core}$ is the length scale of the vortex core, where the $\phi _I\approx 0$ and $\sigma \approx 0$ in \Deq.  The signs in \fxyint\ show that \finaleq\ is only consistent if $Z=q_J\zeta>0$.  Likewise, anti-BPS states have $z\leftrightarrow \bar z$ in \finaleq\ and \mattereq, and the analog of \fxyint\  requires $Z=q_J\zeta <0$.

\appendix{B}{Coulomb branch and K\"ahler form nonrenormalization}

Consider first an Abelian gauge theory.  The massless modes along the Coulomb branch are the microscopic gauge fields $V_r$, $r=1,...,n$.  They can be described by the linear superfields $\Sigma_r=-{i\over 2}\bar DD V_r$, and the effective Lagrangian is given by a real function $f$:
\eqn\lineeff{\CL _{eff}=-\int d^4\theta f(\Sigma_r).}
The kinetic terms for $\sigma_r=\Sigma_r|_{\theta=\bar \theta=0}$ is determined from the metric
\eqn\sigmak{ds^2=f^{rs}d \sigma_rd\sigma_s, \qquad \hbox{with}\qquad f^{rs}\equiv \partial^r \partial^s f(\sigma_c) .}
In a sensible theory, the Hessian $f^{rs}>0$ is a positive definite matrix.

It is convenient to dualize the linear multiplets $\Sigma_r$  \refs{\HitchinEA, \deBoerKR}.  Locally, this is achieved by adding Lagrange multiplier chiral superfields $U^r$ and studying the Lagrangian\foot{We use
Lorentzian signature; in Euclidean space the second term has an imaginary coefficient.}
\eqn\lineeffr{\int d^4\theta \Big(-f(\Sigma_r) + (U^r +\bar U ^r){\Sigma _r\over 2\pi}\Big) ~.}
Where the bottom component of $U^r|$ has
\eqn\reimU{ (U^r+\bar U^r)|=2\phi ^r, \qquad (U^r-\bar U^r)|=2ia^r,}
so e.g. $U^r+\bar U^r\supset -2\theta\gamma ^\mu \bar \theta \partial _\mu a^r$.  The normalization of the added term in \lineeffr\  is such that $a^r$ have $2\pi$ periodicity, given that $\CJ _r=\Sigma _r/2\pi$, the conserved current for a $U(1)_{J_r}$ global symmetry, has integral charges $q_{J_r}=\int d^2 x \half \epsilon ^{0\nu \rho}F_{\nu \rho}/2\pi$ \qjis. The shift symmetries
\eqn\shiftsy{U^r\to U^r + i \alpha^r,}
with real $\alpha^r$, shifting $a^r \to a^r + \alpha^r$ in \reimU, reflect the conserved $U(1)_{J_r}$ currents $\Sigma_r/2\pi$.

 Since the low energy theory is IR free, we can use the semiclassical approximation and integrate out $\Sigma_r$ by a Legendre transform
\eqn\SigmaUr{U^r+\bar U^r=2\pi \partial_{\Sigma_r} f(\Sigma_t)|_{\Sigma_t=\Sigma_t^U}~,}
where $\Sigma_t^{U}$ is the solution of \SigmaUr, and we find the action
\eqn\lineeffr{\eqalign{
&\CL _{eff}=\int d^4\theta K \cr
&K=K(U^r+\bar U^r)= -f(\Sigma_r^U)+{1\over 2 \pi} (U^r+\bar U^r)\Sigma_r^U~.}}
The Legendre transform gives
\eqn\Kfrel{\partial_{U^r} K=\partial _{\bar U^r}K= {1\over 2\pi} \Sigma_r^U}
and the Hessian of $K$,
\eqn\HessK{K_{rs}=\partial_{U^r} \partial_{U^s} K= {1\over 2\pi} \partial_{U^s}\Sigma_r^U~,}
gives the inverse matrix of the Hessian of $f$ in \sigmak
\eqn\inverse{K^{rs} =(2\pi )^2  f^{rs} ~.}
This determines the K\"ahler metric,
\eqn\Kahlermetric{ds^2= 2K_{r\bar s} dU^r d\bar U^{\bar s}=2K_{rs}(d\phi^r d\phi^s + da^r da^s).}
The Legendre transform breaks down at the vanishing eigenvalues of $f^{rs}$ or $K_{rs}$, where there are singularities on the Coulomb branch.

To illustrate the above procedure, consider a single Coulomb variable with $f(\Sigma)\approx \Sigma ^2/e^2$ for large $\sigma$.  Then \SigmaUr\ and \lineeffr\ give $K\approx {e^2\over 16\pi ^2} (U+\bar U)^2$, with
\eqn\exapprox{\phi \approx 2\pi \sigma /e^2, \qquad \epsilon _{\mu \nu \rho}\partial ^\rho a \approx {2\pi \over e^2} F_{\mu \nu},  \qquad X=e^{\phi +ia}\approx e^{2\pi \sigma /e^2+ia}.}

 The coordinates $a^r$ are natural because they are associated with the $U(1)^n$ isometry \shiftsy\ of the space, and $\phi^r$ are related to them by the complex structure.
Similarly, the coordinates $\sigma_r$ are natural because they are related by supersymmetry to the field strengths and conserved currents.  For large $\sigma$, \exapprox\ gives $\phi \approx  2\pi \sigma /e^2$.  But it is important that $\sigma_r$ and $a^r$ reside in different superfields.

Using the general  K\"ahler metric \Kahlermetric, for any $f(\Sigma _r)$ in \lineeffr, the K\"ahler form is
\eqn\Kahlerform{\Omega = {i\over 2} K_{r\bar r} dU^r \wedge d\bar U^{\bar r}=K_{rs}d\phi^r \wedge da^s= {1\over 4\pi} d\sigma_r \wedge da^r~,}
where we used the above properties of the Legendre transform. In this context,  this result was derived in \AganagicUW, where it was emphasized that the $f$ independence of  \Kahlerform\ means that the K\"ahler form on the Coulomb branch is given exactly by its classical value.

This conclusion has several consequences and it can be extended in several directions. First, consider a Coulomb branch of a non-Abelian theory.  Because of magnetic monopoles, the semiclassical $\Sigma _r$ cannot be used beyond perturbation theory.  In fact, as in \AffleckAS, one must first dualize to the variables $U^r$ and then the shift symmetry \shiftsy\ is violated.  Hence, the K\"ahler form can be renormalized non-perturbatively.

It turns out that in many situations we can still prove such a non-renormalization result.  Consider for example an $SU(2)$ gauge theory.  The theory has a one dimensional Coulomb branch and it is invariant under a global $U(1)$ symmetry that semiclassically acts on it as in \shiftsy.  Depending on the number of flavors this symmetry is an R-symmetry or an ordinary symmetry.  Therefore, the K\"ahler potential must be of the form $K(U+\bar U)$.  Note that this is true even in the $N_f=0$ theory, where a superpotential for $U$ is generated non-perturbatively.  Now, we can run the discussion above in reverse, define $\Sigma$ using the Legendre transform and show that the K\"ahler form is given exactly by $d\sigma \wedge da$.

More generally, whenever the theory has an $n$ dimensional branch with $n$ shift symmetries \shiftsy\ (which could be accompanied with an R-transformation), the K\"ahler potential has the form $K(U^r+\bar U^r)$.  Then we can define $\Sigma_r $ as in \Kfrel, and conclude that the K\"ahler form has the simple form \Kahlerform.

This discussion breaks down at the singularities of the Coulomb branch. They occur when the Hessian $K_{rs}$ has vanishing or divergent eigenvalues.  Comparing with known examples we can find one of the following situations:
\item{1.} At some finite point, $\sigma_r=\sigma^{(0)}_r$ the K\"ahler metric $K_{rs}$ has a vanishing eigenvalue.  This means that in terms of $U^r$ the space ends.  Nevertheless, the moduli space of vacua can continue beyond this point.  Then, a new set of chiral superfields $\tilde U^r$ are needed on the other side.  An example of this situation arises in a $U(1)$ gauge theory with several flavors with real masses $m_i$.  Here, the Coulomb branch of the moduli space of vacua is parameterized by $-\infty < \sigma <+\infty$.  It includes various regions $m_i < \sigma <m_{i+1}$ parameterized by different chiral superfields $U_i$ (see, e.g.\ Fig.\ 1).
\item{2.} The classical Coulomb branch ends at some finite $\sigma_r=\sigma^{(0)}_r$, but in the quantum theory $\sigma_r$ is extended beyond this point.  This is the case in an $SU(2)$ gauge theory with $N_f=1$.  The moduli space is parameterized by a meson $M$, which is related to $U$ through $M e^U=1$.

In the discussion above, we ignored the fact that the functions $f$ and $K$ could depend on various parameters.  In particular, locations of singularities on the Coulomb branch can depend on real masses $m_i$, or  Fayet-Iliopoulos parameters.  Such parameters can affect the Coulomb branch because they can be regarded as background linear multiplet superfields.  It is then easy to argue that whenever the discussion above applies (i.e.\ when the theory has a $U(1)^n$ isometry so that the K\"ahler potential depends only on $U^r+\bar U^r$) the locations of the singularities is not corrected from the semiclassical values.  The singularities are at $\Sigma=\Sigma_i$, where $\Sigma_i$ is a background linear superfield, which accounts for the real masses and FI parameters. The quantum corrections to the linear superfields $\Sigma_i$ are very restrictive, the only ambiguity is in shifts by some linear superfields, and the ambiguity by such shifts is easily determined semiclassically.

This conclusion has important consequences.  It is often the case that the theory has such a $U(1)^n$ global symmetry so that we can apply the result above, and the Coulomb branch has singularities.  There can then be BPS states that wrap the regions in the Coulomb branch between these singularities.  The BPS masses of such Skyrmions are determined by the K\"ahler form.  Since the K\"ahler form is not renormalized, and its singularities are easily determined in the classical approximation, we can easily determine the masses of the BPS states.  This result was found in Abelian theories in \AganagicUW, and here we extend it also to many non-Abelian theories.

\appendix{C}{Comments about modes with logarithmically divergent norm}

The purpose of this appendix is to clarify a confusing issue in the study of vortices in theories with non-minimal matter, where it was found, starting with \WardIJ, that there are non-normalizable zero modes.  A typical example is the ``semi-local vortex" of the Abelian Higgs model with several fields \VachaspatiDZ; it occurs both for vortex particles in 3d and for vortex strings in 4d; see e.g.\ \refs{ \HananyHP\AuzziFS\ShifmanDR\HananyEA \OlmezAU- \KoroteevRB}.  Various prescriptions have been given for handling the non-normalizable modes -- either to regulate the divergence, e.g.\ by putting the theory on compact space, or to remove them, e.g.\ by giving them mass. We here interpret these modes and explain how to treat them, without regulators or removing them.

For concreteness, consider a 3d $U(1)$ gauge theory with two complex scalar fields $\phi_i$ of charge $+1$, and with FI term $\zeta >0$.   This system describes the bosonic sector of our SQED with two $Q$s with charge $+1$ (section 4.2).  The low energy dynamics of this system is the $CP^1$ model.  Therefore, if our space is $\R^2$, the global $SU(2)$ symmetry of the system is spontaneously broken to $U(1)$.  If our space is compact, e.g.\ a sphere or torus, then the symmetry is not broken.

More concretely, in finite volume $V=L^2$ the system has a zero mode associated with a global $SU(2)$ rotation on the doublet $\phi_i$.  The norm of this constant mode is $L^2$ and, since it is finite, this mode is dynamical.  As the volume of the compact space becomes large, the norm of this mode diverges like $L^2$, and it is frozen in the infinite volume limit.  Correspondingly, the Hilbert space breaks into superselection sectors labeled by the boundary conditions on $CP^1$.  Up to a global $SU(2)$ rotation, we will take the boundary conditions at infinity to be
\eqn\boucond{\phi_2 \to 0 \qquad ; \qquad |\phi_1| \to \sqrt{\zeta \over 2\pi}~.}

The configurations fall into sectors labeled by the vortex number \qjis
\eqn\vort{q_J=-{i\over 2\pi} \int d^2 z F_{z\bar z} ~,}
where $z$ and $\bar z$ parameterize space.  The BPS static configurations with $q_J>0$ are determined by a meromorphic function of $z$ \refs{\WardIJ, \LeeseFN}
\eqn\wzpa{w(z) = {\phi_2\over \phi_1}= \sum_{I=1}^{q_J} {\rho_I \over z-z_I} ~,}
where for simplicity we assume that all the poles are simple.  When $|\rho_I|$ are much smaller than the separation between $z_I$ this configuration can be interpreted as $q_J$ vortices each centered around $z_I$.  $|\rho_I|$ corresponds to the scale size of the $I$'th vortex and $\arg(\rho_I)$ is the orientation in the target space. The moduli space of these vortices and the metric on that space are analyzed in detail in \LeeseFN.  It was shown in \refs{\VachaspatiDZ, \HindmarshYY, \LeeseFN}
that these vortices smoothly interpolate between the ANO vortices and the $CP^1$ Skyrmions.

The large $z$ behavior of \wzpa\ is
\eqn\wzpal{w(z) \to {\sum_I\rho_I \over z} + \CO(1/z^2) ~.}
The mode $\rho= \sum_I\rho_I$ is interesting.  Unlike all other modes in the system, which have finite norm, the norm of $\rho$ is logarithmically divergent.  (In the finite volume system this mode has finite norm proportional to $\log L$.)  Therefore, it is not dynamical in the infinite volume system and it is frozen, as noted in \WardIJ.

We have seen two modes that should be frozen.  The constant mode associated with $SU(2)$ rotations has quadratically divergent norm and $\rho$, whose norm is logarithmically divergent.  The constant mode is visible already in the sector of the theory with $q_J=0$ and its physics leads to the superselection sectors there and the corresponding spontaneous symmetry breaking.  But the physical interpretation of $\rho$ appears to be mysterious.  In particular, if the mode of the vortices $\rho$ is not treated as a collective coordinate, then the spectrum of the vortices appears to be a continuum labeled by $\rho$.  This looks quite unphysical, and indeed we will now argue that this conclusion is wrong.

We claim that,  as with the constant mode, $\rho$ is a superselection parameter already for the   $q_J=0$ sector.  Consider e.g.\ a single, real, massless scalar field $\varphi$, with boundary conditions $\lim_{|z| \to \infty }\varphi =0$.  Naively, we can set up the quantization of this system in a standard way.   However, consider a smooth configuration of $\varphi$ at some fixed time $t=t_0$
\eqn\varphiz{\varphi(z,\bar z, t=t_0) = {\rm Re}\ {\rho \over z} + \CO(1/|z|^2) }
with some complex $\rho$.   Setting the initial conditions \varphiz\ with $\partial_t\varphi(z, \bar z, t=t_0) =0$ the configuration has finite energy and it spreads out.  But since ${\rm Re}{\rho \over z} $ satisfies the static equations of motion, no matter how much time we wait, the large $|z|$ behavior of $\varphi$ is always as in \varphiz.  The value of $\rho$ can be measured at infinity, and it does not change.

We see that the configurations of the classical system are labeled by the complex parameter $\rho$ and its value cannot change with time.  This means that in the quantum theory $\rho$ labels different superselection sectors in the Hilbert space.  Indeed, starting with a state with some value of $\rho$ and acting on it with any local operators, and letting the system evolve for any finite time, we cannot find states with any different value of $\rho$.

Normally, we focus only on the states with $\rho=0$.  These states include the standard vacuum and we can take its energy to be zero.  The states with nonzero $\rho$ are unusual.  Their energy is bounded below by zero.  But there is no state with zero energy among them.  These states are necessarily time dependent.

Now we can move to the states with nonzero $q_J$.  Here we find static states with any $\rho$.  The states with nonzero $q_J$ can be created out of states with $q_J=0$ by acting on them with monopole operators.  Such local operators cannot change the value of $\rho$.

We conclude that the Hilbert space breaks into superselection sectors labeled by the continuous complex parameter $\rho$. If we want to focus on the standard vacuum with $\rho=0$ we can limit the discussion of vortices to vortices with $\rho=0$.  Vortices with nonzero $\rho$ are excitations of states with $q_J=0$ with nonzero $\rho$.

The discussion above of the Abelian Higgs model is easily generalized to $N+1$ complex charged fields $\phi_i$.  Here the low energy theory is the $CP^N$ model.  It has $N$ modes with quadratically divergent norm associated with the boundary values $\lim_{|z|\to \infty }\phi_i$ (with fixed $\sum_i|\phi_i|^2$).  They are frozen when the system is in $\R^3$ signaling the spontaneous breaking of $SU(N+1) \to SU(N)$.  In addition, there are $N$ logarithmically divergent modes associated with the $1\over z$ fall-off of the scalar fields at infinity.  As with our single $\rho$ in the example above, they are superselection sectors, and are present both with and without vortices.

We now consider the Fermionic zero modes.  Again, some of them can have logarithmically divergent norms.  On a compact space, such modes have finite norm and they should be quantized.  But when the system is on $\R^3$ they are frozen.  A simple way to handle them, which is clearly motivated by our discussion above of the bosonic modes, is the following.  Start with the system in a compact space, where the quantization leads to a big Hilbert space.  Next, take the infinite volume limit and then this big Hilbert space breaks to distinct superselection sectors.  Doing that with the Fermions, these different superselection sectors have different charges. Clearly, there are no transitions between states in distinct superselection sectors and we can focus only on one of them.

The new point is the identification of all these superselection sectors, associated both with the bosonic and with the Fermionic modes, also when $q_J=0$ and there are no solitons.  Therefore, solitons with different values of $\rho$, and different non-normalizable Fermi zero modes, are not merely in different sectors of the one body problem:  they are in different sectors of the whole quantum field theory.  In each superselection sector of the whole second-quantized system, we have solitons only with a fixed value of $\rho$, and with some charges associated with the Fermion zero modes.

We end this appendix by pointing out that a similar phenomenon can take place in four dimensions.  In addition to its zero mode, a massless real scalar field $\varphi$ can have asymptotic behavior at spatial infinity analogous to \varphiz, specified by a real parameter $\rho$
\eqn\varphizf{\varphi(\vec x, t=t_0) = {\rho \over |\vec x|} + \CO(1/|\vec x|^2). }
With such asymptotics, the energy is finite, the $\rho$-term satisfies the local equations of motion, and the norm of the $\rho$-mode is (linearly) divergent.  Hence, $\rho $ leads to a new superselection parameter.

\appendix{D}{Derivation of $\Tr (-1)^F$ for $U(1)_k$ with matter}

We here derive the index  \indexgen\ and \indexgenn, by explicitly verifying that the sum  \indexch\ is independent of the real parameters $m_i$ and $\sigma$, and then evaluating it for a convenient choice.  Phase transitions can occur when, upon varying $m_i$ and $\sigma$, two (or more) $\sigma$ vacua locations collide.  The Higgs vacua can exist at $\sigma _{Q_i}$, the locations where the slope, $k_{eff}(\sigma)$, of $F(\sigma)$ can change.  Topological vacua are at $\sigma _I$, where $F(\sigma)=0$.  Varying the real parameters, two  $\sigma _{Q_i}$ vacua can cross each other, or a $\sigma _{Q_i}$ can cross a $\sigma _I$; two $\sigma _I$ cannot cross each other, since the equation for $\sigma _I$ is piecewise linear.  We here verify that, although the individual terms can vary, the sum \indexch\ is unchanged by all such vacuum crossings.

As we vary the real parameters, the number of Higgs vacua at some $\sigma _{Q_i}$ can only change if its $s_i$ in \higgsmaybe\ changes, which only happens if $\sigma _{Q_i}$ crosses a $\sigma _I$ where $F(\sigma _{Q_i})$ passes through a zero.  Likewise, the number of topological vacua at some $\sigma _I$ can only change if $|k_{eff}(\sigma _I)|$ changes, which also only happens if $\sigma _I$ crosses a $\sigma _{Q_i}$.  We verify that these changes always cancel in the total $\Tr (-1)^F$. The dependence of \higgsn\ and \indexcn\ on $\sign (\zeta)$ illustrates this general result.

Consider first the case of two crossing $\sigma _{Q_i}$ and $\sigma _{Q_i'}$, at the point, where $\sigma _{Q_i}=\sigma_{Q_i'}$.  If $\sign (n_i)=\sign ({n_i'})$, then $s_i=s_{i'}$.  For $s_i=s_{i'}=1$,
there is a compact space of Higgs-vacua $\CM _H$ at  $\sigma _{Q_i}=\sigma _{Q_{i'}}$, and continuity of $\Tr (-1)^F$ requires that these vacua contribute $\Tr (-1)^F|_{\CM _H}= n_i^2+n_{i'}^2$, to equal those of the Higgs vacua at $\sigma _{Q_i}$ and $\sigma _{Q_i'}$ when they are separated. For $n_i=n_{i'}=1$, the compact space of vacua is $\CM _H\cong CP^1$, which contributes to the index according to its Euler character, $\Tr (-1)^F|_{\CM _H}=\chi (\CM _H)=2$.
More generally, for $n_i=n_{i'}$, $\CM _H$ is a $Z_{n_i}$ orbifold of $CP^1$, which gives $\Tr (-1)^F|_{\CM _H}=2n_i^2$.  These cases indeed agree with  $n_i^2+n_{i'}^2$.  For $n_i\neq n_{i'}$, the space $\CM _H$ is a singular weighted projective space and the contribution to $\Tr (-1)^F$ is better computed via the UV $U(1)$ linear sigma model, which must give $\Tr (-1)^F|_{\CM _H} =n_i^2+n_{i'}^2$.  If $\sign (n_i)=-\sign ({n_i'})$, then $s_i\neq s_{i'}$:  only one contributes, both before and after the crossing.  At the crossing point, there is a non-compact Higgs branch, so $\Tr (-1)^F$ is ill-defined; elsewhere, the index \indexch\ is well defined, and unchanged by the crossing.

Now consider the transition where a $\sigma _{Q_i}$ hits a $\sigma _I$, i.e.\  where $F(\sigma _{Q_I})$ changes sign.   The effect depends on the relative sign of the slopes \kzetajumps\ $k_{eff,i}^\pm$ of $F(\sigma)$ on either side of $\sigma _{Q_i}$. If $\sign (k_{eff,i}^-) = \sign (F(\sigma _{Q_i}))$, there is a $\sigma _I$ topological vacuum at $\sigma _I<\sigma _{Q_i}$, and if $\sign (k_{eff,i}^+) =-\sign (F(\sigma _{Q_i}))$, there is a $\sigma _I$ vacuum at $\sigma _I>\sigma _{Q_i}$.   So if $k_{eff, i}^\pm$ have the same sign, the transition is that a single $\sigma _I$ crosses over to the other side of $\sigma _{Q_i}$.  If $k_{eff,i}^\pm$ are of opposite sign, the transition is instead that two topological vacua $\sigma _I$ and $\sigma _{I'}$ pair create or annihilate when they meet at $\sigma _{Q_i}$.

 Consider first the same sign case, $\sign (k^+ _{eff,i})=\sign (k^-_{eff, i})\equiv \sign (k_{eff, i})$, and suppose that initially $\sigma _I<\sigma _{Q_i}$, and after the crossing that $\sigma _{I}'>\sigma _{Q_{i}}'$;  the reversed situation is analogous.
So initially $k_{eff}(\sigma _I)=k_{eff,i}^-$, with $\sign (F(\sigma _{Q_i}))=\sign (k_{eff,i})$, and afterwards $k_{eff}(\sigma _I')=k_{eff,i}^+$, with $\sign (F(\sigma _{Q_i}'))=-\sign (k_{eff,i})$.
Using  \kzetajumps,
\eqn\kjump{k_{eff}(\sigma _I')=k_{eff,i}^+=k_{eff,i}^-+n_i^2\sign (n_i),}
so the number of topological vacua counted in \indexch\ changes after the crossing, by
\eqn\dindexc{\Delta \Tr (-1)^F_{\rm topo}=|k_{eff,i}^- + n_i^2\sign (n_i)|-|k_{eff,i}^-| =n_i^2\sign (n_i k_{eff,i}).}
The number of Higgs vacua counted in \indexch\ also changes after the crossing, by
\eqn\dindexh{\Delta \Tr (-1)^F_{\rm Higgs}=n_i^2 (s_{i'}-s_i)=n_i^2 \left(\Theta (-n_i k_{eff,i})- \Theta (n_i k_{eff,i})\right).   }
The changes in \dindexc\ and \dindexh\ indeed cancel: the total number of vacua remains constant, though some of the Higgs vacua can become topological vacua, or visa versa.

Now consider the case of $\sign (k^-_{eff,i})=-\sign (k^+_{eff,i})$, where topological vacua can pair create or annihilate at $\sigma _{Q_i}$.  Consider the annihilation process; the  reverse process is analogous.  Initially, $\sign (k_{eff,i}^-)=\sign (F(\sigma _{Q_i}))=-\sign (k_{eff,i}^+)$, so there are $\sigma _I$ and $\sigma _{I'}$ vacua on either side of $\sigma _{Q_i}$, which then annihilate  when $F(\sigma _{Q_i})$ changes sign. The number of topological  vacua counted in \indexch\ thus changes by
\eqn\dindexcm{\Delta \Tr (-1)^F_{\rm topo}=-|k_{eff,i}^-+n_i^2\sign (n_i)|-|k_{eff,i}^-|=n_i^2 \sign (n_ik_{eff,i}^-).}
The number of Higgs vacua also changes, and using \higgsmaybe\ here gives
\eqn\dindexhm{\Delta \Tr (-1)^F_{\rm Higgs}=n_i^2 (s_{i'}-s_i)=n_i^2 \left(\Theta (-n_i k_{eff,i}^-)- \Theta (n_i k_{eff,i}^-)\right).   }
Again, the changes \dindexcm\ and \dindexhm\ cancel.

We have verified that \indexch\ is invariant under all $m_i$ and $\zeta$ deformations, and can thus evaluate \indexch\ by a convenient $m_i$, $\zeta$ choice. Note from \kinfinity\ that $F(\sigma)$ has asymptotic slopes of the same (opposite) sign if  $|k|>|k_c|$ ($|k|<|k_c|$), and hence an even (odd) number of topological vacua.  By the $\sigma _I$ annihilation process, we can always adjust $\zeta$  such that $F(\sigma)$ has only one zero for $|k|>|k_c|$, or no zeros for $|k|<|k_c|$.

We first consider $|k|>|k_c|$.  We take $\sign m_i=\sign n_i$, so that $\sigma _{Q_i}=-m_i/n_i<0$ for all fields, and choose $\zeta$ such that there is a single topological  vacuum, at $\sigma _I=0$,   $F(\sigma =0)=\zeta _{eff}=\zeta +\half \sum _i n_i |m_i|=0$.  The topological contribution to \indexch\ is then
\eqn\indexci{\Tr (-1)^F_{\rm topo }=|k_{eff}(\sigma =0)|=|k|+\half \sum _i n_i^2  \sign (n_ik).}
For the Higgs contributions, note that $\sign F(\sigma _{Q_i})=-\sign k$, so
\eqn\indexhi{\Tr (-1)^F_{\rm Higgs}=\sum _i  s_i n_i ^2 =\half \sum _i n_i^2 (1-\sign (n_ik)).}
Adding \indexci\ and \indexhi\ gives the result \indexgen\ for the total index.

We now consider the $|k|<|k_c|$ case, and take $\zeta$ such that there are no topological vacua, $F(\sigma )\neq 0$, by taking $|\zeta|$ sufficiently large with $\sign \zeta = \sign k_c$, so
$\sign F(\sigma)=\sign k_c$ everywhere.  Then \indexch\ yields the result \indexgenn:
\eqn\indexgenx{\Tr (-1)^F=\Tr (-1)^F_{\rm Higgs}=\half \sum _i n_i^2 (1+\sign n_i k_c)=|k_c|+\half \sum _i n_i^2.}

\listrefs
\end